\documentclass[prx,twocolumn]{revtex4-2}
\usepackage{mathrsfs}
\usepackage{amsmath}
\usepackage{amssymb}
\usepackage{graphicx}
\usepackage{braket}
\usepackage{hyperref}
\usepackage[usenames,dvipsnames]{xcolor}
\usepackage{stmaryrd}

\newcommand{\dblbrace}[1]{\llbracket #1\rrbracket}

\hypersetup{
    colorlinks,
    citecolor=blue,
    linkcolor=blue,
    urlcolor=blue
}

\newcommand{\SUPPcMPS}{1}
\newcommand{\SUPPFisher}{2}
\newcommand{\SUPPQFisher}{3}
\newcommand{\SUPPMixedState}{4}
\newcommand{\SUPPSimulation}{5}
\newcommand{\SUPPTimeDependent}{6}
\newcommand{\SUPPHeisenberg}{7}
\newcommand{\SUPPDefinitions}{8}

\begin{document}
\title{
Unifying Speed Limit, Thermodynamic Uncertainty Relation and Heisenberg Principle via Bulk-Boundary Correspondence
}
\author{Yoshihiko Hasegawa${}^1$}
\email{hasegawa@biom.t.u-tokyo.ac.jp}

\affiliation{${}^1$Department of Information and Communication Engineering, Graduate
School of Information Science and Technology, The University of Tokyo,
Tokyo 113-8656, Japan}
\date{\today}
\begin{abstract}
The bulk-boundary correspondence provides a guiding principle for tackling strongly correlated and coupled systems.
In the present work, we apply the concept of the bulk-boundary correspondence to thermodynamic bounds described by classical and quantum Markov processes.
Using the continuous matrix product state, we convert
a Markov process to a quantum field, such that
jump events in the Markov process are represented by the creation of particles in the quantum field. 
Introducing the time evolution of the continuous matrix product state,
we apply the geometric bound to its time evolution. 
We find that the geometric bound reduces to the speed limit relation
when we represent the bound in terms of the system quantity,
whereas the same bound reduces to the thermodynamic uncertainty relation
when expressed based on quantities of the quantum field. 
Our results show that the speed limit and thermodynamic uncertainty
relations are two aspects of the same geometric bound.
\end{abstract}
\maketitle

\section*{Introduction}

The bulk-boundary correspondence is a guiding principle for solving complex and strongly coupled systems
\cite{Bousso:2002:HoloReview,Ammon:2015:HolographyBook,Baggioli:2019:HoloBook}. 
The main idea of the bulk-boundary correspondence is that the information on the bulk of a system is encoded in its boundary. 
In particular, a system that is complex with no apparent approaches for solving problems can be mapped to a different system that becomes simpler to tackle.
By using the bulk-boundary correspondence, a strongly correlated quantum field theory (conformal field theory; CFT) is 
mapped to classical gravity (anti-de Sitter space; AdS)
at one dimension higher, where
physical quantities in the boundary are evaluated via those in
the bulk space \cite{Policastro:2001:HoloViscosity,Ryu:2006:EntEnt,Hartnoll:2007:HoloNernst},
which is referred to as the AdS/CFT correspondence.

In the present manuscript,
we consider quantum and stochastic thermodynamics \cite{Seifert:2012:FTReview,VandenBroeck:2015:Review,Funo:2018:QFT,Manzano:2021:QThermo}.
They are associated with
quantities such as heat, work and entropy that can be defined based on a stochastic trajectory.
Stochastic and quantum thermodynamic systems exhibit behaviors that occur far from equilibrium and are described by correlated and coupled Markov processes.
This fact leads us to consider that the bulk-boundary correspondence
might play a fundamental role in stochastic and quantum thermodynamics. 
Recently, Refs.~\cite{Verstraete:2010:cMPS,Osborne:2010:Holography} proposed
the continuous matrix product state representation
that enables realization of the bulk-boundary correspondence in Markov processes. 
The continuous matrix product state relates a
Markov process to the quantum field,
with the Markov process and the quantum field corresponding to the boundary and the bulk, respectively. 
Using the continuous matrix product state, 
we can investigate properties of a quantum field from
the point of view of the corresponding Markov process. 
In contrast, we can study a Markov process 
by mapping it to a quantum field and unveiling its properties. 
Indeed, the continuous matrix product state has been employed in the thermodynamics of trajectory,
where it has been used to investigate phase transitions and the role of gauge symmetry
in classical and quantum Markov processes
\cite{Garrahan:2010:QJ,Lesanovsky:2013:PhaseTrans,Garrahan:2016:cMPS}.
Moreover, we have recently employed the continuous matrix product state to derive 
quantum thermodynamic uncertainty relations
\cite{Hasegawa:2020:QTURPRL,Hasegawa:2021:QTURLEPRL,Hasegawa:2021:FPTTUR}. 

In the present paper, we use the bulk-boundary correspondence to 
examine thermodynamic bounds, such as thermodynamic uncertainty relations \cite{Barato:2015:UncRel,Gingrich:2016:TUP,Garrahan:2017:TUR,Dechant:2018:TUR,Terlizzi:2019:KUR,Hasegawa:2019:CRI,Hasegawa:2019:FTUR,Vu:2019:UTURPRE,Dechant:2020:FRIPNAS,Vo:2020:TURCSLPRE,Koyuk:2020:TUR,Pietzonka:2021:PendulumTURPRL,Erker:2017:QClockTUR,Brandner:2018:Transport,Carollo:2019:QuantumLDP,Liu:2019:QTUR,Guarnieri:2019:QTURPRR,Saryal:2019:TUR,Hasegawa:2020:QTURPRL,Hasegawa:2020:TUROQS,Sacchi:2021:BosonicTUR,Kalaee:2021:QTURPRE,Monnai:2022:QTUR}
(see \cite{Horowitz:2019:TURReview} for a review)
and (quantum and classical) speed limit relations  \cite{Mandelstam:1945:QSL,Margolus:1998:QSL,Deffner:2010:GenClausius,Taddei:2013:QSL,DelCampo:2013:OpenQSL,Deffner:2013:DrivenQSL,Pires:2016:GQSL,OConnor:2021:ActionSL,Shiraishi:2018:SpeedLimit,Ito:2018:InfoGeo,Ito:2020:TimeTURPRX,Nicholson:2020:TIUncRel,Vu:2021:GeomBound} (see \cite{Deffner:2017:QSLReview} for a review).
The speed limit relation concerns a trade-off relation between the
speed of time evolution and thermodynamic costs,
and was first introduced in quantum dynamics \cite{Mandelstam:1945:QSL,Margolus:1998:QSL,Deffner:2010:GenClausius,Taddei:2013:QSL,DelCampo:2013:OpenQSL,Deffner:2013:DrivenQSL,Pires:2016:GQSL}.
Recently, the concept has been generalized to classical Markov processes as well \cite{Shiraishi:2018:SpeedLimit,Ito:2018:InfoGeo,Ito:2020:TimeTURPRX,Nicholson:2020:TIUncRel,Vu:2021:GeomBound}. 
It states that faster time evolution should be accompanied by
higher thermodynamic costs, such as dynamical activity and energy.
The thermodynamic uncertainty relation gives the fundamental limit for
the precision of thermodynamic machines and states that higher precision can only be achieved
at the expense of higher thermodynamic costs. 
Thermodynamic uncertainty relations have become important not only from a theoretical point of view but also from a practical standpoint,
such as the estimation of entropy production from measurements \cite{Li:2019:EPInference,Manikandan:2019:InferEPPRL,Vu:2020:EPInferPRE,Otsubo:2020:EPInferPRE,Roldan:2021:EPInfer}.
As noted above, the continuous matrix product state
has been applied to classical and quantum Markov processes.
These approaches use the quantum field representation for analyses but its time evolution has not been explicitly incorporated.
In the present manuscript, we introduce a time evolution operator into the continuous matrix product state.
The space of the continuous matrix product state is one dimension higher than
that of the original Markov process, and the original Markov process exists
at the boundary and thus it is referred to as \textit{bulk}.
We apply the concept of the geometric speed limit inequality
to the bulk space to derive speed limits [Eqs.~\eqref{eq:Lps_bound} and \eqref{eq:Ldo_bound}] and thermodynamic uncertainty relations [Eqs.~\eqref{eq:cTUR} and \eqref{eq:qTUR}]. In the resulting speed limit relations, the distances between the initial and the final states are bounded from above by terms comprising classical or quantum dynamical activities. 
In the case of the thermodynamic uncertainty relations obtained in this work, we show that the precision of an observable that counts the number of jumps is bounded from below by costs composed of classical or quantum dynamical activities. 
We establish a duality relation in that
the speed limit and the thermodynamic uncertainty relation can be 
understood as two different aspects of the geometric speed limit inequality. 
Specifically, when we bound the geometric inequality with the quantities in
the Markov process, the inequality reduces to classical
and quantum speed limits [Eqs.~\eqref{eq:Lps_bound} and \eqref{eq:Ldo_bound}]. 
In contrast, the geometric inequality becomes the thermodynamic uncertainty relations [Eqs.~\eqref{eq:cTUR} and \eqref{eq:qTUR}] when we bound the geometric inequality with the quantities in
the quantum field. 
This duality is demonstrated for both classical and quantum Markov processes. 
We also consider the Heisenberg uncertainty relation in the bulk space to show that the Heisenberg uncertainty relation reduces to the thermodynamic uncertainty relation in the Markov process.

\section*{Results}

\subsection*{Continuous matrix product state}
Let us consider a quantum Markov process described by a Lindblad equation. 
Classical Markov processes are included in quantum Markov processes
as particular cases (see Eq.~\eqref{eq:master_eq_def}). 
Let $\rho(s)$ be a density operator of the system at time $s$. 
We assume that $\rho(s)$ is governed by the time-independent
Lindblad equation:
\begin{equation}
\frac{d}{ds}\rho(s)=\mathfrak{L}(\rho(s))=-i\left[H_{\mathrm{sys}},\rho(s)\right]+\sum_{m=1}^{M}\mathcal{D}(\rho(s),L_{m}),
\label{eq:Lindblad_def}
\end{equation}where $\mathfrak{L}(\bullet)$ is a Lindblad super-operator,
$H_\mathrm{sys}$ is the system Hamiltonian, $\mathcal{D}(\rho,L)\equiv L\rho L^{\dagger}-\{L^{\dagger}L,\rho\}/2$
with $L_m$ being the $m$th jump operator (there are $M$ jump operators, $\{L_1,L_2,\ldots,L_M\}$),
$[\bullet,\bullet]$ is the commutator and
$\{\bullet,\bullet\}$ is the anti-commutator. 
Here, we assume that $H_\mathrm{sys}$ and $L_m$ are time-independent. 
Suppose that the dynamics starts at $s=0$ and ends at
$s = \tau$ ($\tau > 0$).
When we apply a continuous measurement to the Lindblad equation, 
we obtain a record of jump events, given by
\begin{equation}
    \Gamma \equiv [(s_{1},m_{1}),(s_{2},m_{2}),\ldots,(s_{K},m_{K})],
    \label{eq:Gamma_def}
\end{equation}
where $K$ is the number of jump events and $s_k$ and $m_k \in \{1,2,\ldots,M\}$ specify the time and type of the $k$th jump event, respectively. 
The record of these jump events $\Gamma$ is termed the \textit{trajectory}. 
For a given trajectory, $\rho(s)$ is governed by a quantum Markov process referred to as
the stochastic Schr{\"o}dinger equation. 
By averaging all possible measurements in the stochastic Schr{\"o}dinger equation, we can recover the original Lindblad equation [Eq.~\eqref{eq:Lindblad_def}].

\begin{figure*}
\includegraphics[width=18cm]{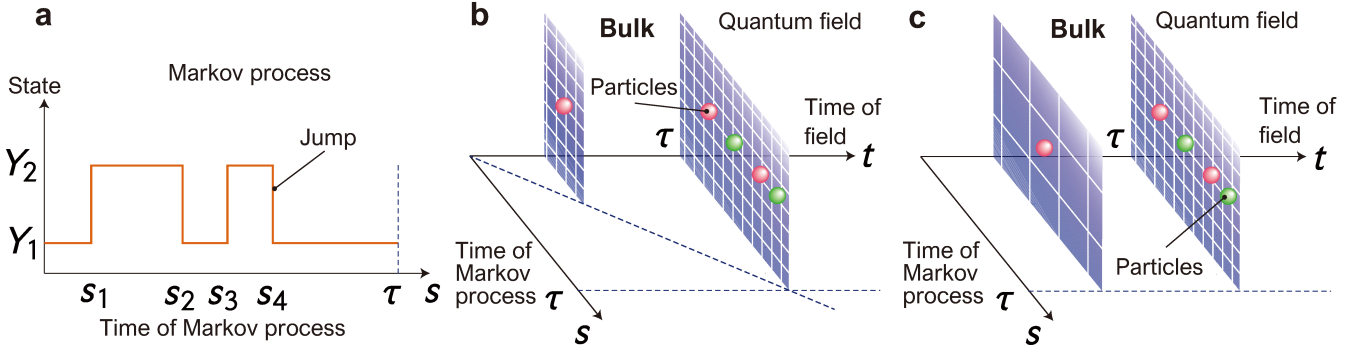} 
\caption{
\textbf{Bulk-boundary correspondence in a Markov process.}
\textbf{a} Trajectory of the Markov process
as a function of $s$ within the time interval $[0,\tau]$.
$Y_1$ and $Y_2$ denote states of the Markov process,
and $s_k$ is the time stamp of the $k$th jump event. 
\textbf{b} Bulk space corresponding to the Markov process of \textbf{a},
generated by Eq.~\eqref{eq:cMPS_def2}. 
The record of jump events is represented by particle creation
in the quantum field. 
The boundary at $t = \tau$ represents the Markov process of \textbf{a}.
The axis of $t$ specifies the time evolution of the quantum field. 
\textbf{c} Bulk space corresponding to the Markov process of \textbf{a},
generated by Eq.~\eqref{eq:Psi_U_def}. In contrast to \textbf{b}, the space is scaled so that
the quantum field is defined for $s \in [0,\tau]$ for all $t \in [0,\tau]$.
\label{fig:holography}}
\end{figure*}

We now consider the bulk-boundary correspondence
in the continuous measurement of the Lindblad equation.
The bulk-boundary correspondence relates a
Markov process to the quantum field, and  
this correspondence is possible through a representation
known as the continuous matrix product state \cite{Verstraete:2010:cMPS,Osborne:2010:Holography}. 
When we apply the continuous measurement to Eq.~\eqref{eq:Lindblad_def},
we obtain a trajectory $\Gamma$ [Eq.~\eqref{eq:Gamma_def}].
The quantum field that records the trajectory is defined as
\begin{equation}
    \ket{\Gamma}\equiv \phi_{m_{K}}^{\dagger}(s_{K})\cdots\phi_{m_{2}}^{\dagger}(s_{2})\phi_{m_{1}}^{\dagger}(s_{1})\ket{\mathrm{vac}},
    \label{eq:cMPS_traj_def}
\end{equation}where
$\phi_m(s)$ is a field operator having the canonical commutation relation $[\phi_m(s),\phi_{m^\prime}^\dagger(s')]=\delta_{mm^\prime}\delta(s-s')$;
$\phi^\dagger_m(s)$ creates a particle of type $m$ at $s$ and 
$\ket{\mathrm{vac}}$ is a vacuum state.
The time evolution of the measurement record and the state of the principal system
can be represented by the continuous matrix product state:
\begin{equation}
    \ket{\Phi(t)}=\mathfrak{U}(t;H_{\mathrm{sys}},\{L_{m}\})\ket{\Phi(0)},
    \label{eq:cMPS_def2}
\end{equation}
where $\mathfrak{U}(t;H_{\mathrm{sys}},\{L_{m}\})$ is an operator parametrized by $t$ and the operators $H_\mathrm{sys}$ and $\{L_m\}$:
\begin{align}
    \mathfrak{U}(t;H_{\mathrm{sys}},\{L_{m}\})\equiv\mathbb{T}\exp\biggl[-i\int_{0}^{t}ds\,[H_{\mathrm{sys}}\otimes\mathbb{I}_{\mathrm{fld}}\nonumber\\\,\,\,+\sum_{m}\left(iL_{m}\otimes\phi_{m}^{\dagger}(s)-iL_{m}^{\dagger}\otimes\phi_{m}(s)\right)]\biggr].
    \label{eq:Ufrak_def}
\end{align}
Here
the initial state is $\ket{\Phi(0)}=\ket{\psi(0)}\otimes\ket{\mathrm{vac}}$
with $\ket{\psi(0)}$ being the initial state in the system,
$\mathbb{T}$ is the time ordering operator
and $\mathbb{I}_\mathrm{fld}$ is the identity operator in the field.
$\ket{\Phi(t)}$ records the jump events within
the interval $0\le s \le t$. 
Figure~\ref{fig:holography} shows an intuitive illustration of the bulk-boundary
correspondence in Markov processes. 
Figure~\ref{fig:holography}(a) shows an example of a Markov process,
where the horizontal and vertical axes denote the time $s$ and
the state of the Markov process, respectively. 
By using the bulk-boundary correspondence, 
all information concerning measurement is recorded by creating particles in the quantum field by applying $\phi_{m}^\dagger(s)$ to $\ket{\mathrm{vac}}$. 
The bulk-boundary correspondence maps the system to 
a quantum field that is one dimension higher than the original one,
as depicted in Fig.~\ref{fig:holography}(b). 
In Fig.~\ref{fig:holography}(b),
the original time evolution of the Markov process is shown by the $s$ axis while the extra dimension $t$ in the bulk space represents the time evolution of the continuous matrix product state. 
In Fig.~\ref{fig:holography}(b), the boundary at $t=\tau$
represents the original Markov process, and thus
the space of Fig.~\ref{fig:holography}(b) is the bulk space. 
Any information that can be obtained from the original Markov process
can be derived from Eq.~\eqref{eq:cMPS_def2}.
Let us define
\begin{equation}
    \rho^{\Phi}_\mathrm{sys}(t)\equiv \mathrm{Tr}_{\mathrm{fld}}[\ket{\Phi(t)}\bra{\Phi(t)}],
    \label{eq:rho_from_Phi}
\end{equation}where $\mathrm{Tr}_\mathrm{fld}$ is the trace with respect to
the field.
$\rho^{\Phi}_\mathrm{sys}(t)$ satisfies $\rho^{\Phi}_\mathrm{sys}(t) = \rho(t)$,
where $\rho(t)$ is the density matrix in Eq.~\eqref{eq:Lindblad_def}.
The quantum field that encodes all information about
the jump events is given by 
\begin{equation}
    \rho^{\Phi}_\mathrm{fld}(t)\equiv \mathrm{Tr}_{\mathrm{sys}}[\ket{\Phi(t)}\bra{\Phi(t)}],
    \label{eq:varrho_fld_def}
\end{equation}
where $\mathrm{Tr}_\mathrm{sys}$ is the trace with respect to
the system. 
See Supplementary Note \SUPPcMPS{} for details of the
continuous matrix product state.

\subsection*{Scaled quantum field}

We now consider the time evolution of the continuous matrix product state. 
Since 
the operator defined in 
Eq.~\eqref{eq:Ufrak_def} is already a unitary operator, it seems satisfactory to
employ it as its time-evolution operator.
However, such an approach appears to be problematic, as explained below.
We will be interested in the fidelity between two continuous matrix product states at
different times,
$\braket{\Phi(t_2)|\Phi(t_1)}$ for $t_1 \ne t_2$.
However, since the integration ranges for $\ket{\Phi(t_1)}$ and $\ket{\Phi(t_2)}$ 
are different, as indicated by Eqs.~\eqref{eq:cMPS_def2} and \eqref{eq:Ufrak_def}, it is not possible to evaluate the fidelity (Fig.~\ref{fig:holography}(b)).
In the present work, instead of using 
the continuous matrix product state defined by Eq.~\eqref{eq:cMPS_def2},
we employ the scaled representation:
\begin{align}
    \ket{\Psi(t)}=\mathfrak{U}\left(\tau;\frac{t}{\tau}H_{\mathrm{sys}},\left\{ \sqrt{\frac{t}{\tau}}L_{m}\right\} \right)\ket{\Psi(0)},
    \label{eq:Psi_U_def}
\end{align}
where $\ket{\Psi(0)} \equiv \ket{\psi(0)}\otimes \ket{\mathrm{vac}}$.
Here, we use $\ket{\Phi(t)}$ and $\ket{\Psi(t)}$ to represent the genuine [Eq.~\eqref{eq:cMPS_def2}] and the scaled [Eq.~\eqref{eq:Psi_U_def}] continuous matrix product state representations, respectively. 
Since $\ket{\Psi(t)}$ and $\ket{\Phi(t)}$ are different states, 
we show justification for using $\ket{\Psi(t)}$ instead of $\ket{\Phi(t)}$ as follows.
Let us define
\begin{equation}
    \rho^{\Psi}_\mathrm{sys}(t)\equiv \mathrm{Tr}_{\mathrm{fld}}[\ket{\Psi(t)}\bra{\Psi(t)}].
    \label{eq:rho_from_Psi}
\end{equation}
In Eq.~\eqref{eq:Psi_U_def}, $H_\mathrm{sys}$ and $L_m$ are scaled by
$t/\tau$ and $\sqrt{t/\tau}$, respectively,
leading to
the Lindblad equation
$\partial_{s}\rho(s)=(t/\tau)\mathfrak{L}(\rho(s))$, which is the same as Eq.~\eqref{eq:Lindblad_def} except for 
its time scale; the scaled operators yield the dynamics,
which is $t/\tau$ times as fast as the original dynamics. 
Due to the scaling, the integration range in Eq.~\eqref{eq:Psi_U_def}
is the same for all $t \in [0,\tau]$, making evaluation
of the fidelity at different times possible. 
Moreover, the system state (i.e., the state of the original Markov process) can be
obtained by both $\ket{\Psi(t)}$ and $\ket{\Phi(t)}$:
\begin{equation}
    \rho(t)=\rho_{\mathrm{sys}}^{\Psi}(t)=\rho_{\mathrm{sys}}^{\Phi}(t),
    \label{eq:rho_Psi_Phi}
\end{equation}
where $\rho(t)$ is the density operator in the Lindblad equation \eqref{eq:Lindblad_def}.
Equation~\eqref{eq:rho_Psi_Phi} shows that, with respect to the state of the system,
$\ket{\Phi(t)}$ and $\ket{\Psi(t)}$ provide the state consistent with Eq.~ \eqref{eq:Lindblad_def}.

It is helpful to assess the difference between $\ket{\Phi(t)}$ and $\ket{\Psi(t)}$
with respect to a field observable.
Let $\rho^\Psi_\mathrm{fld}(t)$ be a density operator in the field:
\begin{equation}
    \rho^\Psi_\mathrm{fld}(t) \equiv \mathrm{Tr}_\mathrm{sys}[\ket{\Psi(t)}\bra{\Psi(t)}].
    \label{eq:rho_fld_def}
\end{equation}
In general, 
we cannot use $\ket{\Psi(t)}$ instead of $\ket{\Phi(t)}$ for a
general measurement in the quantum field. 
However, if we are interested in the number of jump events,
$\ket{\Phi(t)}$ and $\ket{\Psi(t)}$ yield the same 
statistics since $\ket{\Psi(t)}$ is based on dynamics
that are exactly the same as $\ket{\Phi(t)}$ except for the time scale. 
Since the jump events are recorded in the field as the creation of
particles, information of the jump events can be obtained by 
measuring the field with the number operator:
\begin{align}
    \mathcal{N}_m \equiv \int_{0}^{\tau}\phi_{m}^{\dagger}(s)\phi_{m}(s)ds,
    \label{eq:numop_def}
\end{align}
which counts the number of $m$th jumps during $[0,\tau]$. 
When we are interested in the state of the system (the state of the original Markov process) and
the number of jump events, $\ket{\Psi(t)}$ and $\ket{\Phi(t)}$
provide exactly the same information. 
This property justifies the use of $\ket{\Psi(t)}$ in place of $\ket{\Phi(t)}$.

Thus far, our focus has been on the number operator $\mathcal{N}_m$ alone, but more general observables can be considered. 
The number operator
[Eq.~\eqref{eq:numop_def}]
admits the
spectral decomposition:
\begin{align}
    \mathcal{N}_{m}=\sum_{n_{m}=0}n_{m}\Pi_{n_{m}},
    \label{eq:Nm_decomp_def}
\end{align}
where the eigenvalue $n_m$ denotes the number of $m$th jumps within $[0,\tau]$ and 
$\Pi_{n_m}$ is its corresponding projector.
The first-level generalization of $\mathcal{N}_m$ is
\begin{align}
    \mathcal{N}_{m}^{\circ}\equiv\sum_{n_{m}=0}\eta_{m}(n_{m})\Pi_{n_{m}},
    \label{eq:Nm1_decomp_def}
\end{align}
where $\eta_m(n)$ is a real function satisfying $\eta_m(0)=0$. 
Thus, $\mathcal{N}^\circ_m$ is a generalization of $\mathcal{N}_m$ as $\eta_m(n)=n$ recovers $\mathcal{N}_m$ in Eq.~\eqref{eq:Nm_decomp_def}. 
The second-level generalization would be
\begin{align}
    \mathcal{N}_{m}^{\bullet}\equiv\sum_{n_{m}=0}\xi_{m}(n_{m})\Pi_{n_{m}},
    \label{eq:Nm2_decomp_def}
\end{align}
where $\xi_m(n)$ is an arbitrary real function. 
$\mathcal{N}_m^{\bullet}$ is the most general form of observable that commutes with $\mathcal{N}_m$. 
Note that $\mathcal{N}_m^\circ$ and $\mathcal{N}_m^{\bullet}$ can also be used for the scaled representation $\ket{\Psi(t)}$
instead of $\ket{\Phi(t)}$ (see the Methods section).

Figures~\ref{fig:holography}(b) and (c) depict the bulk spaces corresponding to $\ket{\Phi(t)}$ and $\ket{\Psi(t)}$, respectively. 
In Fig.~\ref{fig:holography}(c), we see that $\ket{\Psi(t)}$ is defined for $s \in [0,\tau]$, where the 
scaling factor of the space depends on $t$. 
In contrast, in the case of Fig.~\ref{fig:holography}(b),
the quantum field is defined for $s\in [0,t]$ while the scaling factor does 
not depend on $t$.

\subsection*{Geometric bound in probability space\label{se:cgeom}}

The previous section introduced the time evolution
of the continuous matrix product state. 
In this section, we consider the geometric properties
of its time evolution.
These geometric properties have been
extensively employed in the quantum speed limit \cite{Deffner:2017:QSLReview}.
We first consider a space of classical probability and 
then move to a space of the quantum state in the next section. 

Let us consider a classical Markov process with $N_S$ states $\{Y_1,Y_2,\ldots,Y_{N_S}\}$.
The dynamics of the Markov process is governed by a classical Markov process:
\begin{equation}
\frac{d}{ds}P(\nu,s) = \sum_\mu W_{\nu \mu} P(\mu,s),
    \label{eq:master_eq_def}
\end{equation}where $P(\nu,s)$ is the probability of being $Y_\nu$
at time $s$ and $W_{\nu\mu}$ is the transition rate from $Y_\mu$ to $Y_\nu$ with $W_{\mu\mu} = - \sum_{\nu \ne \mu} W_{\nu \mu}$. 
Taking $H_\mathrm{sys} = 0$, $L_{\nu\mu}=\sqrt{W_{\nu\mu}}\ket{Y_{\nu}}\bra{Y_{\mu}}$ and $\rho(t)=\mathrm{diag}\left([P(\nu,t)]_{\nu}\right)$
in Eq.~\eqref{eq:Lindblad_def},
the Lindblad equation is reduced to the corresponding classical Markov process,
where $\{\ket{Y_{1}},\ket{Y_{2}},\cdots,\ket{Y_{N_{S}}}\}$ constitutes an orthonormal basis with
each $\ket{Y_\nu}$ corresponding to $Y_\nu$. 
Here, the index of the jump operator should be mapped as
$L_m \to L_{\nu\mu}$ by mapping $m\to (\nu,\mu)$.
Therefore, the $m$th jump in Eq.~\eqref{eq:Lindblad_def} corresponds to the jump
from $Y_\mu$ to $Y_\nu$ in Eq.~\eqref{eq:master_eq_def}.
Using the continuous matrix product state, the probability of measuring a trajectory $\Gamma$ and $Y_\nu$ at the end time is
\begin{align}
    \mathcal{P}(\Gamma,\nu,t)\equiv\braket{\Psi(t)|(\ket{Y_{\nu}}\bra{Y_{\nu}}\otimes\ket{\Gamma}\bra{\Gamma})|\Psi(t)}.
    \label{eq:path_prob_def}
\end{align}

Let us consider the time evolution of the continuous matrix product state.
Its time evolution corresponds to the $t$ axis
in Fig.~\ref{fig:holography}(c). 
Applying the projector $\ket{Y_\nu}\bra{Y_\nu}\otimes\ket{\Gamma}\bra{\Gamma}$,
we can consider the time evolution of the probability distribution 
$\mathcal{P}(\Gamma,\nu,t)$ as a function of $t$. 
For such a time-evolving probability distribution, 
by using Ref.~\cite{Wootters:1981:StatDist},
the following relation holds:
\begin{equation}
    \frac{1}{2}\int_{t_{1}}^{t_{2}}dt\,\sqrt{\mathcal{I}(t)}\ge\mathcal{L}_{P}\left(\mathcal{P}(\Gamma,\nu,t_{1}),\mathcal{P}(\Gamma,\nu,t_{2})\right),
    \label{eq:classical_bound}
\end{equation}
where  $\mathcal{I}(t)$ is the classical Fisher information defined by
\begin{equation}
    \mathcal{I}(t)\equiv\sum_{\Gamma,\nu}\mathcal{P}(\Gamma,\nu,t)\left(-\frac{\partial^{2}}{\partial t^{2}}\ln\mathcal{P}(\Gamma,\nu,t)\right),
    \label{eq:CFI_def}
\end{equation}
and $\mathcal{L}_P$ is the Bhattacharyya angle:
\begin{align}
    \mathcal{L}_{P}\left(p_{1}(x),p_{2}(x)\right)\equiv\arccos\left[\mathrm{Bhat}\left(p_{1}(x),p_{2}(x)\right)\right].
    \label{eq:LP_def}
\end{align}
In Eq.~\eqref{eq:LP_def}, $\mathrm{Bhat}(p_1(x),p_2(x))$ is the Bhattacharyya coefficient:
\begin{equation}
    \mathrm{Bhat}\left(p_{1}(x),p_{2}(x)\right)\equiv\sum_{x}\sqrt{p_{1}(x)p_{2}(x)}.
    \label{eq:Batt_def}
\end{equation}Here $p_1(x)$ and $p_2(x)$
are arbitrary probability distributions,
and Eq.~\eqref{eq:LP_def} quantifies the distance between the two probability distributions. 
Equation~\eqref{eq:classical_bound} was used in Refs.~\cite{Ito:2018:InfoGeo,Ito:2020:TimeTURPRX}
to obtain thermodynamic trade-off relations in
classical Markov processes. 
Note that the probability state in Refs.~\cite{Ito:2018:InfoGeo,Ito:2020:TimeTURPRX} is
the actual state.
This corresponds to $P(\nu,s)$, whose time evolution is
the $s$ axis in Fig.~\ref{fig:holography}(c). 
The state considered herein concerns the path probability space $\mathcal{P}(\Gamma,\nu,t)$, whose time evolution is shown by
the $t$ axis in Fig.~\ref{fig:holography}(c).
A straightforward calculation shows that $\mathcal{I}(t)$ can be written as
\begin{equation}
    \mathcal{I}(t)=\frac{\mathcal{A}(t)}{t^{2}},
    \label{eq:Ic_and_A_relation}
\end{equation}with $\mathcal{A}(t)$ being the dynamical activity \cite{Maes:2020:FrenesyPR}:
\begin{equation}
    \mathcal{A}(t)\equiv\int_{0}^{t}ds\sum_{\nu,\mu,\nu\ne\mu}P(\mu,s)W_{\nu\mu}.
    \label{eq:activity_def}
\end{equation}
$\mathcal{A}(t)$ quantifies the average number of jumps
within $[0,t]$ (see Supplementary Note \SUPPFisher{}).

The Bhattacharyya coefficient
satisfies the monotonicity with respect to any classical channel.
Using the monotonicity and Eq.~\eqref{eq:Ic_and_A_relation}, we can write (see Methods)
\begin{equation}
    \frac{1}{2}\int_{0}^{\tau}\frac{\sqrt{\mathcal{A}(t)}}{t}dt\ge\mathcal{L}_{P}(P(\nu,0),P(\nu,\tau)).
    \label{eq:Lps_bound}
\end{equation}
Equation~\eqref{eq:Lps_bound} is the first result of this paper, showing that the distance between the initial and final probability distributions in a classical Markov process has an upper bound
comprising the dynamical activity $\mathcal{A}(t)$. 
Equation~\eqref{eq:Lps_bound} is reminiscent of the classical speed limit obtained in Ref.~\cite{Shiraishi:2018:SpeedLimit}.
The bound in Ref.~\cite{Shiraishi:2018:SpeedLimit}
compared the initial and final probability distributions
by means of the total variation distance. 
Equation~\eqref{eq:Lps_bound} is a direct classical analog of the 
geometric quantum speed limit \cite{Taddei:2013:QSL}.

In Eq.~\eqref{eq:Lps_bound}, we obtained the lower bound for the right hand side in terms of
the quantity in the system ($P(\nu,s)$ in the Markov process).
We next obtain a lower bound using
the quantity in the quantum field, which
leads to a classical thermodynamic uncertainty relation.
We notice that the right hand side of Eq.~\eqref{eq:classical_bound}
can be bounded from below by the distance between
$\mathcal{P}(\Gamma,t_1)$ and $\mathcal{P}(\Gamma,t_2)$,
where $\mathcal{P}(\Gamma, t) \equiv \sum_\nu \mathcal{P}(\Gamma,\nu,t)$.
However, in general, obtaining $\mathcal{P}(\Gamma,t)$ 
requires a large amount of measurement that is impractical.
Thus, as an alternative, we use a time-integrated observable and 
bound the right hand side of Eq.~\eqref{eq:classical_bound} with the statistics of the time-integrated observable. 
Consider the observable in the continuous measurement of 
the Lindblad equation [Eq.~\eqref{eq:Lindblad_def}]:
\begin{equation}
    \mathfrak{C}(\Gamma)\equiv \sum_{m}\alpha_{m}\mathfrak{N}_{m}(\Gamma),
    \label{eq:C_def}
\end{equation}
where $\mathfrak{N}_m(\Gamma)$ counts the number of $m$th jumps in
a given trajectory $\Gamma$, and $\alpha_m$ is a real parameter
defining the weight of the $m$th jump. 
The Hermitian observable corresponding to Eq.~\eqref{eq:C_def}
in the quantum field is written by
\begin{equation}
    \mathcal{C}\equiv\sum_{m}\alpha_{m}\mathcal{N}_{m},
    \label{eq:counting_obs_def}
\end{equation}
where $\mathcal{N}_m$ is the number operator defined in Eq.~\eqref{eq:numop_def}.
Equation~\eqref{eq:counting_obs_def} is the weighted sum of jump events during
the time interval $[0,\tau]$.
Let us define
\begin{align}
    \braket{\mathcal{C}}_{t}&\equiv\mathrm{Tr}_{\mathrm{fld}}\left[\rho_{\mathrm{fld}}^{\Psi}(t)\mathcal{C}\right],\label{eq:C_braket_def}\\
    \dblbrace{\mathcal{C}}_{t}&\equiv\sqrt{\braket{\mathcal{C}^{2}}_{t}-\braket{\mathcal{C}}_{t}^{2}},
    \label{eq:C_dblbrace_def}
\end{align}
where $\rho_{\mathrm{fld}}^{\Psi}(t)$ is defined in Eq.~\eqref{eq:rho_fld_def}.
$\braket{\mathcal{C}}_t$ and $\dblbrace{\mathcal{C}}_t$
correspond to
the mean and standard deviation of the number of 
jump events during the time interval $[0,t]$
in the original Markov process.
In Eqs.~\eqref{eq:Nm1_decomp_def} and \eqref{eq:Nm2_decomp_def}, we have defined $\mathcal{N}_m^\circ$ and $\mathcal{N}_m^{\bullet}$, the generalization of the number operator $\mathcal{N}_m$. We also define generalizations of $\mathcal{C}$ as follows: 
\begin{align}
    \mathcal{C}^{\circ}\equiv\sum_{m}\alpha_{m}\mathcal{N}_{m}^{\circ},\,\,\,\mathcal{C}^{\bullet}\equiv\sum_{m}\alpha_{m}\mathcal{N}^{\bullet}.
    \label{eq:Cbullet_Ccirc_def}
\end{align}
Relations that hold for $\mathcal{C}^{\bullet}$ should be satisfied by $\mathcal{C}^\circ$, and those that hold for $\mathcal{C}^\circ$ should also be satisfied by $\mathcal{C}$
(see the Methods section).

Applying the inequality relation for the Bhattacharyya coefficient to Eq.~\eqref{eq:classical_bound}, 
we obtain a thermodynamic uncertainty relation for $0\le t_1 < t_2\le \tau$ (see the Methods section for details):
\begin{equation}
    \left(\frac{\dblbrace{\mathcal{C}^{\bullet}}_{t_{2}}+\dblbrace{\mathcal{C}^{\bullet}}_{t_{1}}}{\braket{\mathcal{C}^{\bullet}}_{t_{2}}-\braket{\mathcal{C}^{\bullet}}_{t_{1}}}\right)^{2}\ge\frac{1}{\tan\left[\frac{1}{2}\int_{t_{1}}^{t_{2}}\frac{\sqrt{\mathcal{A}(t)}}{t}dt\right]^{2}},
    \label{eq:cTUR}
\end{equation}
which holds for $(1/2)\int_{t_{1}}^{t_{2}}\sqrt{\mathcal{A}(t)}/t\,dt\le\pi/2$. 
Equation~\eqref{eq:cTUR} is the second result of this paper and holds for an arbitrary time-independent
classical Markov process. 
In Refs.~\cite{Hasegawa:2020:TUROQS,Hasegawa:2021:QTURLEPRL},
we derived thermodynamic uncertainty relations that hold for arbitrary
classical Markov chains.
However, the thermodynamic cost terms in Refs.~\cite{Hasegawa:2020:TUROQS,Hasegawa:2021:QTURLEPRL} are
not thermodynamic quantities, whereas
the thermodynamic cost in Eq.~\eqref{eq:cTUR} is
the dynamical activity. 
Let us employ $t_1 = 0$ and $t_2 = \tau$ in Eq.~\eqref{eq:cTUR}.
Since there is no jump for $t=0$, $\braket{\mathcal{C}^\circ}_{t=0} = 0$ and $\dblbrace{\mathcal{C}^\circ}_{t=0} = 0$, and we obtain
\begin{equation}
    \frac{\dblbrace{\mathcal{C}^{\circ}}_{\tau}^{2}}{\braket{\mathcal{C}^{\circ}}_{\tau}^{2}}\ge\frac{1}{\tan\left[\frac{1}{2}\int_{0}^{\tau}\frac{\sqrt{\mathcal{A}(t)}}{t}dt\right]^{2}},
    \label{eq:cTUR2}
\end{equation}
which holds for $(1/2)\int_{t_{1}}^{t_{2}}\sqrt{\mathcal{A}(t)}/t\,dt\le\pi/2$. 
Equations~\eqref{eq:cTUR} and \eqref{eq:cTUR2} are previously unknown relations.
Note that Eqs.~\eqref{eq:cTUR} and \eqref{eq:cTUR2} should hold for $\mathcal{C}$ defined by Eq.~\eqref{eq:counting_obs_def},
since $\mathcal{C}^{\circ}$ and $\mathcal{C}^\bullet$ are generalizations of $\mathcal{C}$. 
In addition, Eq.~\eqref{eq:cTUR}
can derive known classical thermodynamic uncertainty relations, as shown below. 
Let $\varepsilon$ be a sufficiently small parameter. 
Considering $t_1 = \tau - \varepsilon$ and $t_2 = \tau$ in Eq.~\eqref{eq:cTUR}, 
Eq.~\eqref{eq:cTUR} reduces to (see the Methods section for details)
\begin{equation}
    \frac{\dblbrace{\mathcal{C}^{\bullet}}_{\tau}^{2}}{\tau^{2}\left(\partial_{\tau}\braket{\mathcal{C}^{\bullet}}_{\tau}\right)^{2}}\ge\frac{1}{\mathcal{A}(\tau)}.
    \label{eq:cTUR_dCdt}
\end{equation}
Equation~\eqref{eq:cTUR_dCdt} is equivalent to the bound in Ref.~\cite{Terlizzi:2019:KUR}.
Both Eqs.~\eqref{eq:cTUR2} and \eqref{eq:cTUR_dCdt}
hold for an arbitrary time-independent Markov process, 
but the denominator
in the left hand side of Eq.~\eqref{eq:cTUR_dCdt}
is not the time-integrated observable but rather the time derivative of its average value. 
The left hand side of Eq.~\eqref{eq:cTUR2}
can be
defined through the time-integrated observable $\braket{\mathcal{C}}_\tau$,
and so can be interpreted as the precision. 
For the steady state condition, 
Eq.~\eqref{eq:cTUR_dCdt} reduces to
\begin{equation}
    \frac{\dblbrace{\mathcal{C}}_{\tau}^{2}}{\braket{\mathcal{C}}_{\tau}^{2}}\ge\frac{1}{\mathcal{A}(\tau)},
    \label{eq:conv_cTUR}
\end{equation}which is the thermodynamic uncertainty relation derived in Ref.~\cite{Garrahan:2017:TUR,Terlizzi:2019:KUR}.
Therefore,  Eq.~\eqref{eq:cTUR} is a generalization of
the well-known classical bounds.

\subsection*{Geometric bound in quantum space\label{sec:qgeom}}
Thus far, we have considered the classical probability space.
We now move to the quantum space and obtain the geometric
bound for the continuous matrix product state. 
We consider the time evolution of $\ket{\Psi(t)}$,
which is induced by the unitary in Eq.~\eqref{eq:Psi_U_def}. 
We analyze the dynamics through the 
quantum speed limit \cite{Deffner:2017:QSLReview}. 
Similar to Eq.~\eqref{eq:classical_bound},
the bound for the fidelity is given by the relation: \cite{Uhlmann:1992:BuresGeodesic,Taddei:2013:QSL}
\begin{equation}
    \frac{1}{2}\int_{t_{1}}^{t_{2}}dt\,\sqrt{\mathcal{J}(t)}\ge\mathcal{L}_{D}(\ket{\Psi(t_{1})},\ket{\Psi(t_{2})}),
    \label{eq:quantum_bound}
\end{equation}
where $\mathcal{J}(t)$ is the \textit{quantum}
Fisher information \cite{Meyer:2021:QFI}
\begin{equation}
    \mathcal{J}(t)\equiv4\left[\braket{\partial_{t}\Psi(t)|\partial_{t}\Psi(t)}-\left|\braket{\partial_{t}\Psi(t)|\Psi(t)}\right|^{2}\right],
    \label{eq:QFI_def}
\end{equation}
and $\mathcal{L}_D$ is the Bures angle defined by
\begin{align}
    \mathcal{L}_{D}(\rho_{1},\rho_{2})\equiv\arccos\left[\sqrt{\mathrm{Fid}(\rho_{1},\rho_{2})}\right],
    \label{eq:LD_def}
\end{align}
with $\mathrm{Fid}(\rho_1,\rho_2)$ being the quantum fidelity: \cite{Nielsen:2011:QuantumInfoBook}
\begin{equation}
    \mathrm{Fid}(\rho_{1},\rho_{2})\equiv\left(\mathrm{Tr}\sqrt{\sqrt{\rho_{1}}\rho_{2}\sqrt{\rho_{2}}}\right)^{2}.
    \label{eq:fidelity_def}
\end{equation}
Here, $\rho_1$ and $\rho_2$ are arbitrary density operators and 
the fidelity satisfies $0\le\mathrm{Fid}(\rho_{1},\rho_{2})\le1$. 
Since $\ket{\Psi(t)}$ is a pure state, the fidelity reduces to
$\mathrm{Fid}(\ket{\Psi(t_{1})},\ket{\Psi(t_{2})})=|\braket{\Psi(t_{2})|\Psi(t_{1})}|^{2}$. 
$\mathcal{L}_D$ quantifies the distance between two density operators and is widely employed in quantum speed limits \cite{Deffner:2017:QSLReview}. 
Equation~\eqref{eq:quantum_bound} is also commonly used in
the quantum speed limit \cite{Deffner:2017:QSLReview}.
The quantum Fisher information $\mathcal{J}(t)$
can be computed using the two-sided Lindblad equation
introduced in Ref.~\cite{Gammelmark:2014:QCRB} (see Supplementary Note \SUPPQFisher{}).

For the classical case, the Fisher information $\mathcal{I}(t)$ reduces to the dynamical activity $\mathcal{A}(t)$ [Eq.~\eqref{eq:activity_def}].
However, it is difficult to represent the quantum Fisher information $\mathcal{J}(t)$ by a well-known physical quantity.
Therefore, from Eq.~\eqref{eq:Ic_and_A_relation},
we may define the quantum generalization of the dynamical activity by
\begin{equation}
    \mathcal{B}(t)\equiv t^{2}\mathcal{J}(t),
    \label{eq:B_activity_def}
\end{equation}where
the classical Fisher information $\mathcal{I}(t)$ in Eq.~\eqref{eq:Ic_and_A_relation}
is replaced with the quantum counterpart. 
In the present manuscript, we refer to $\mathcal{B}(t)$ as the \textit{quantum} dynamical activity. 

The fidelity obeys the monotonicity relation
with respect to any completely positive and trace-preserving map \cite{Nielsen:2011:QuantumInfoBook}. Using the monotonicity, we obtain (see the Methods section for details)
\begin{equation}
    \frac{1}{2}\int_{0}^{\tau}\frac{\sqrt{\mathcal{B}(t)}}{t}dt\ge\mathcal{L}_{D}(\rho(0),\rho(\tau)).
    \label{eq:Ldo_bound}
\end{equation}
Equation~\eqref{eq:Ldo_bound} is a continuous measurement case of the quantum speed limit reported in Ref.~\cite{Taddei:2013:QSL}.
Regarding a quantum speed limit in open quantum dynamics,
Ref.~\cite{DelCampo:2013:OpenQSL}
considered a Lindblad dynamics and employed the relative purity as a distance measure.
Equation~\eqref{eq:Ldo_bound} itself can be derived from Eq.~\eqref{eq:quantum_bound} via 
the monotonicity of the quantum fidelity.
Although there are infinitely many ways to describe open quantum dynamics through purification, 
we will show that the quantum dynamical activity $\mathcal{B}(\tau)$
in Eq.~\eqref{eq:Ldo_bound} plays a central role in a quantum thermodynamic uncertainty relation derived as follows. 
The speed limit relations derived in Eqs.~\eqref{eq:Lps_bound} and \eqref{eq:Ldo_bound} do not explicitly include time $\tau$. However, by rearranging terms, we can obtain  lower bounds for the evolution time $\tau$ (see the Methods section).

Next, we consider a quantum thermodynamic uncertainty relation that follows directly from Eq.~\eqref{eq:quantum_bound}.
Again, we consider the observables $\mathcal{C}$, $\mathcal{C}^\circ$ and $\mathcal{C}^{\bullet}$.
Similar to the classical case [Eq.~\eqref{eq:cTUR}], we obtain the thermodynamic uncertainty relation for $0 \le t_1 < t_2 \le \tau$ (see the Methods section for details):
\begin{equation}
\left(\frac{\dblbrace{\mathcal{C}^{\bullet}}_{t_{2}}+\dblbrace{\mathcal{C}^{\bullet}}_{t_{1}}}{\braket{\mathcal{C}^{\bullet}}_{t_{2}}-\braket{\mathcal{C}^{\bullet}}_{t_{1}}}\right)^{2}\ge\frac{1}{\tan\left[\frac{1}{2}\int_{t_{1}}^{t_{2}}\frac{\sqrt{\mathcal{B}(t)}}{t}dt\right]^{2}},
    \label{eq:qTUR}
\end{equation}
which holds for $(1/2)\int_{t_{1}}^{t_{2}}\sqrt{\mathcal{B}(t)}/t\,dt\le\pi/2$. 
This relation is a quantum analog of Eq.~\eqref{eq:cTUR}
and constitutes the third result of this manuscript.
Equation~\eqref{eq:qTUR} holds for arbitrary time-independent quantum Markov processes. 
Similar to Eqs.~\eqref{eq:Lps_bound} and \eqref{eq:cTUR},
the quantum dynamical activity $\mathcal{B}(\tau)$
plays a fundamental role in both Eqs.~\eqref{eq:Ldo_bound} and \eqref{eq:qTUR},
indicating that $\mathcal{B}(t)$ is a physically important quantity. 
Although we previously derived thermodynamic uncertainty relations that hold for arbitrary
quantum Markov chains in Refs.~\cite{Hasegawa:2020:TUROQS,Hasegawa:2021:QTURLEPRL},
the thermodynamic cost terms in Refs.~\cite{Hasegawa:2020:TUROQS,Hasegawa:2021:QTURLEPRL} are
not thermodynamic quantities as in the classical case. 
Since Eq.~\eqref{eq:qTUR} is the same as Eq.~\eqref{eq:cTUR} except that $\mathcal{A}(t)$ is replaced by $\mathcal{B}(t)$, we can obtain quantum counterparts of Eqs.~\eqref{eq:cTUR2}--\eqref{eq:conv_cTUR} in the same manner.
Equation~\eqref{eq:cTUR2} with $\mathcal{A}(t)$ replaced by $\mathcal{B}(t)$
is a quantum thermodynamic uncertainty relation that holds for arbitrary time-independent quantum Markov processes. 
In particular, 
Eq.~\eqref{eq:conv_cTUR} with $\mathcal{A}(t)$ replaced by $\mathcal{B}(t)$ is equivalent to the quantum thermodynamic uncertainty relation derived in
Ref.~\cite{Hasegawa:2020:QTURPRL},
which was derived using the quantum Cram\'er-Rao inequality.
In Ref.~\cite{Hasegawa:2020:QTURPRL},
we calculated $\mathcal{B}(\tau)$ for $\tau \to \infty$
to show that $\mathcal{B}(\tau)$ 
is given by a sum of the classical dynamical activity and
the coherent contribution, which is induced by the effective Hamiltonian.

In speed limit and thermodynamic uncertainty relations, the bounds require the condition $(1/2)\int_{t_{1}}^{t_{2}}\sqrt{\mathcal{A}(t)}/t\,dt\le\pi/2$ (classical) or $(1/2)\int_{t_{1}}^{t_{2}}\sqrt{\mathcal{B}(t)}/t\,dt\le\pi/2$ (quantum). It is helpful here to examine a physical meaning of the conditions. 
When the system is in a steady state, the dynamical activity is  $\mathcal{A}(t) = \mathfrak{a} t$, where $\mathfrak{a}$ is a proportionality coefficient. 
Consequently, $(1/2)\int_{0}^{\tau}\sqrt{\mathcal{A}(t)}/t\,dt=\sqrt{\mathfrak{a}t}$, which transforms the constraint into $\tau\le\pi^{2}/(4\mathfrak{a})$. 
Therefore, physically, the conditions can be identified as the constraint for $\tau$, demonstrating that the predictive power of the bounds is limited to a prescribed time determined by the system's dynamics. 
This limitation on $\tau$ can be ascribed to the geometric speed limit relations. In Eqs.~\eqref{eq:classical_bound} and \eqref{eq:quantum_bound}, the range of values for the left hand side is $[0,\infty)$ while that for the right hand side is $[0,\pi/2]$.
Therefore, although the geometric speed limit relations hold for $\tau \to \infty$, predictive power is lost for finite time values.

The derivations above assume the initially pure state $\rho(0)=\ket{\psi(0)}\bra{\psi(0)}$.
Using the purification, we can show that the speed limit and thermodynamic uncertainty relations hold for an initially mixed state (see Supplementary Note \SUPPMixedState{}).
Thus far, we have been concerned with theoretical aspects of the bounds. We numerically test the speed limits and the thermodynamic uncertainty relations and verify the bounds (see Supplementary Note \SUPPSimulation{}).

\section*{Discussion}

\begin{figure*}
\includegraphics[width=16cm]{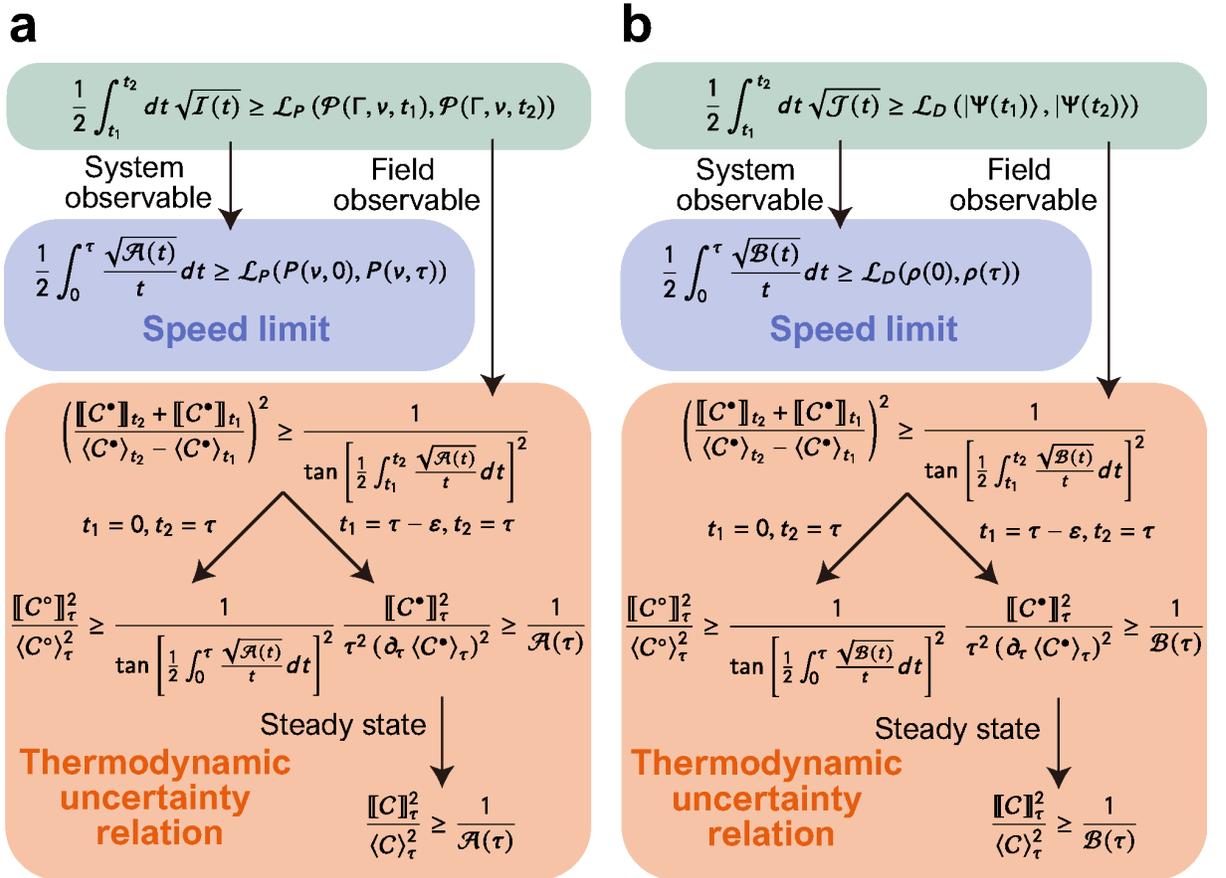} 
\caption{
\textbf{Relation of obtained inequalities.}
\textbf{a}
When we bound the right-hand side of Eq.~\eqref{eq:classical_bound} by the system and field quantities, 
we obtain classical speed limits and classical thermodynamic uncertainty relations, respectively. 
\textbf{b}
Similar relation for the quantum case. 
In \textbf{a} and \textbf{b}, 
variable and function definitions are presented in Supplementary Note \SUPPDefinitions{}. 
\label{fig:relation}}
\end{figure*}

The results represented by Eqs.~\eqref{eq:Lps_bound}, \eqref{eq:cTUR}, \eqref{eq:Ldo_bound} and \eqref{eq:qTUR} show that the speed limits and the thermodynamic uncertainty relations can be 
understood as two different aspects of Eqs.~\eqref{eq:classical_bound} and \eqref{eq:quantum_bound}. 
When we bound the right hand sides of Eqs.~\eqref{eq:classical_bound} and \eqref{eq:quantum_bound} with the quantities in
the principal system, that is, the probability distribution $P(\nu,s)$ or the density operator $\rho(s)$, the inequalities reduce to the classical
and quantum speed limits expressed by Eqs.~\eqref{eq:Lps_bound} and \eqref{eq:Ldo_bound}, respectively.
On the other hand, when we bound the right hand sides of
Eqs.~\eqref{eq:classical_bound} and \eqref{eq:quantum_bound}
with the field quantity, $\braket{\mathcal{C}^{\bullet}}_t$ and $\dblbrace{\mathcal{C}^{\bullet}}_t$,
we obtain the classical and quantum thermodynamic uncertainty relations, expressed by
Eqs.~\eqref{eq:cTUR} and \eqref{eq:qTUR}, respectively.
Therefore, the speed limit and the thermodynamic uncertainty relations can be derived from the common ancestral relation. 
Figure~\ref{fig:relation} shows an intuitive illustration of the logical connections explained above. 
Note that we previously derived the classical speed limit and thermodynamic uncertainty relation in a unified way 
in Refs.~\cite{Vo:2020:TURCSLPRE,Vo:2022:UKTUR}.
However, Refs.~\cite{Vo:2020:TURCSLPRE,Vo:2022:UKTUR} derived the classical speed limit as a
short time limit of the thermodynamic uncertainty relation,
whereas the derivation here does not use such a distinct setting for the speed limit. 

Thus far, we have considered a time-independent Markov process,
meaning that $H_\mathrm{sys}$ and $L_m$ are not depend on time. 
Here, we examine a time-dependent case with the time-dependent operators $H_\mathrm{sys}(s)$ and $L_m(s)$.
It is possible to introduce a time-dependent analogue of $\ket{\Psi(t)}$ introduced in Eq.~\eqref{eq:Psi_U_def}.
Using the time-dependent representation, we can derive speed limits and thermodynamic uncertainty relations similar to Eqs.~\eqref{eq:Lps_bound} and \eqref{eq:cTUR},
where the dynamical activity is replaced by
the generalized dynamical activity (Supplementary Note \SUPPTimeDependent{}).

We have considered geometric speed-limit relations in the bulk space. 
As shown in Eq.~\eqref{eq:Psi_U_def}, since the time evolution of
the composite system comprising the system and the quantum field
admits closed quantum dynamics, any relation that holds 
in the closed system should hold for the composite system as well. 
We here consider a consequence of the Heisenberg uncertainty relation \cite{Heisenberg:1927:UR,Robertson:1929:UncRel},
which is the most fundamental uncertainty relation in
quantum mechanics, in Eq.~\eqref{eq:Psi_U_def}. 
It can be shown that the Heisenberg uncertainty relation reduces to the thermodynamic uncertainty relation 
(Supplementary Note \SUPPHeisenberg{}).
It should also be noted that this correspondence is a consequence of the relation between
the Cram\'er--Rao inequality and the Heisenberg uncertainty relation as reported by Ref.~\cite{Frowis:2015:TQUR}. 
The Heisenberg uncertainty relation is a fundamental inequality to derive the Mandelstam-Tamm quantum speed limit \cite{Mandelstam:1945:QSL}.
Our result shows that the Heisenberg uncertainty relation
also plays a fundamental role in the thermodynamic uncertainty relation
when considering the bulk-boundary correspondence of the Markov process.

Thermodynamic uncertainty relations were originally derived as the inequality between current fluctuations and entropy production \cite{Barato:2015:UncRel,Gingrich:2016:TUP}. 
As such it might be possible to obtain a unified bulk/boundary approach for speed limit and thermodynamic uncertainty relations for which the thermodynamic cost involves solely entropy production. 
However, it is difficult to derive a unified bound for entropy production. 
To derive the bound, we should introduced another scaled continuous matrix product state that provides the same information regarding the number of jump events and the system state as the original dynamics while the Fisher information yielding entropy production.

In this paper, 
we studied the consequences of considering the bulk-boundary correspondence
in classical and quantum Markov processes.
These investigations could possibly be extended to employ
refined Heisenberg uncertainty relations, as shown in Ref.~\cite{Maccone:2014:UR}, as an example. 
Since any uncertainty relation that holds in closed quantum dynamics 
should hold in the time evolution shown by Eq.~\eqref{eq:Psi_U_def},
it can be anticipated that other uncertainty relations can be derived using the technique demonstrated herein. 

\section*{Methods}

\subsection*{Geometric bound}

We employ the geometric bounds given by Eqs.~\eqref{eq:classical_bound} and \eqref{eq:quantum_bound} to obtain speed limits and thermodynamic uncertainty relations. 

In Eq.~\eqref{eq:classical_bound},
the left hand side gives the path length corresponding to
the dynamics parametrized by $t \in [0,\tau]$
that connects the two states
under the Fisher information metric,
while the right hand side of Eq.~\eqref{eq:classical_bound} 
corresponds to the geodesic distance between
the two states \cite{Wootters:1981:StatDist}. 
Similarly, in Eq.~\eqref{eq:quantum_bound}, the left hand side gives the path length 
of the dynamics $\ket{\Psi(t)}$ under the Fubini-Study metric,
while the righthand side of Eq.~\eqref{eq:quantum_bound} 
is the geodesic distance between
the initial $\ket{\Psi(t_1)}$ and final $\ket{\Psi(t_2)}$ states under this metric. 

It is helpful here to assess the uniqueness of the metrics. 
In probability space, 
except for a constant factor,
the Fisher information metric is known to correspond to the unique contractive Riemannian metric.
In the case that a metric in the density operator space is considered, an infinite number of metrics is possible.
The geodesic distance can be analytically calculated for several metrics, such as the quantum Fisher information metric and the Wigner-Yanase information metric, both of which fall into the Fubini-Study metric for pure states. The continuous matrix product state is pure and so the Fubini-Study metric $\mathcal{J}(t)$ gives a unique metric \cite{Pires:2016:GQSL}.

\subsection*{Number operator and observables}

In the main text, we consider the observable $\mathfrak{C}(\Gamma)$ defined in Eq.~\eqref{eq:C_def}.
For the classical Markov process defined in Eq.~\eqref{eq:master_eq_def}, 
using the correspondence $m \to (\nu,\mu)$, Eq.~\eqref{eq:C_def} can be written as
\begin{equation}
    \mathfrak{C}(\Gamma)=\sum_{\nu,\mu,\nu\ne\mu}\alpha_{\nu\mu}\mathfrak{N}_{\nu\mu}(\Gamma).
    \label{eq:C_def2}
\end{equation}
As an example, when $\alpha_{\mu\nu} = -\alpha_{\nu\mu}$, $\mathfrak{C}(\Gamma)$ defines the time-integrated current
that is antisymmetric under time reversal.
In particular, the original thermodynamic uncertainty relation \cite{Barato:2015:UncRel,Gingrich:2016:TUP} states that the fluctuation of a time-integrated current such as this is bounded from below by the reciprocal of the entropy production.
In addition, if $\alpha_{\mu\nu} = -\alpha_{\nu\mu} = 1$ then $\mathfrak{C}(\Gamma)$ quantifies the amount of displacement, which can be used to quantify
the elapsed time on a Brownian clock \cite{Barato:2016:BrowClo}.

In the main text, we define $\mathcal{C}^\circ$ and $\mathcal{C}^{\bullet}$ in Eq.~\eqref{eq:Cbullet_Ccirc_def}. 
When we represent these observables as functions of a trajectory $\Gamma$ as was done in Eq.~\eqref{eq:C_def}, we have
\begin{align}
    \mathfrak{C}^{\circ}\equiv\sum_{m}\alpha_{m}\eta_{m}\left(\mathfrak{N}_{m}(\Gamma)\right),\,\,\,\mathfrak{C}^{\bullet}\equiv\sum_{m}\alpha_{m}\xi_{m}\left(\mathfrak{N}_{m}(\Gamma)\right),
    \label{eq:Cfrak_gen_def}
\end{align}
where the functions $\eta_m$ and $\xi_m$ are defined in Eqs.~\eqref{eq:Nm1_decomp_def} and \eqref{eq:Nm2_decomp_def}, respectively. 
Since
$\mathcal{C}^\circ$ and $\mathcal{C}^{\bullet}$ are generalizations of $\mathcal{C}$, they can recover $\mathcal{C}$ as a particular case. Moreover, they can express observables that are not covered by $\mathcal{C}$. 
An example of $\mathcal{C}^\circ$ that does not belong to $\mathcal{C}$ would be $\eta(n) = \mathrm{sgn}(n)$, where $\mathrm{sgn}$ is the sign function.
It gives a value of $1$ when there is more than one jump but a value of $0$ otherwise. 
We also note that $\mathcal{C}^\circ$ satisfies
$\braket{\mathcal{C}^\circ}_{t=0} = 0$ and $\dblbrace{\mathcal{C}^\circ}_{t=0} = 0$, which is an important property of $\mathcal{C}^\circ$ used in the derivation of Eq.~\eqref{eq:cTUR2}.

\subsection*{Derivation of speed limit relations}

We derive a classical speed limit relation from Eq.~\eqref{eq:classical_bound}. 
The Bhattacharyya coefficient satisfies monotonicity with
respect to any classical channel \cite{Liese:2006:Divergence}.
Since $P(\nu, t) = \sum_\Gamma \mathcal{P}(\Gamma,\nu, t)$, the monotonicity yields
\begin{equation}
    \mathrm{Bhat}(\mathcal{P}(\Gamma,\nu,t_{1}),\mathcal{P}(\Gamma,\nu,t_{2}))\le\mathrm{Bhat}(P(\nu,t_{1}),P(\nu,t_{2})).
    \label{eq:Bhatt_monotonic}
\end{equation}
Substituting Eqs.~\eqref{eq:Ic_and_A_relation} and \eqref{eq:Bhatt_monotonic} into Eq.~\eqref{eq:classical_bound}, we obtain the classical speed limit of Eq.~\eqref{eq:Lps_bound}. 

The quantum speed limit of Eq.~\eqref{eq:Ldo_bound} can be derived in a similar manner. 
The fidelity obeys
the monotonicity relation with respect to an arbitrary
completely positive and trace-preserving map \cite{Nielsen:2011:QuantumInfoBook}.
Since $\rho(t)=\mathrm{Tr}_{\mathrm{fld}}\left[\ket{\Psi(t)}\bra{\Psi(t)}\right]$ from Eq.~\eqref{eq:rho_Psi_Phi},
the following relation holds:
\begin{equation}
    \mathrm{Fid}(\ket{\Psi(t_{1})},\ket{\Psi(t_{2})})\le\mathrm{Fid}(\rho(t_{1}),\rho(t_{2})).
    \label{eq:monotonicity_fidelity}
\end{equation}
Using Eqs.~\eqref{eq:monotonicity_fidelity} and the quantum dynamical activity $\mathcal{B}(t)$ [Eq.~\eqref{eq:B_activity_def}], 
we obtain Eq.~\eqref{eq:Ldo_bound}.

\subsection*{Derivation of thermodynamic uncertainty relations}
Here, we derive classical thermodynamic uncertainty relations from
Eq.~\eqref{eq:Lps_bound}.
Let us consider the Hellinger distance between two
probability distributions $p_1(x)$ and $p_2(x)$:
\begin{align}
    \mathrm{Hel}^{2}(p_{1}(x),p_{2}(x))&\equiv\frac{1}{2}\sum_{x}\left(\sqrt{p_{1}(x)}-\sqrt{p_{2}(x)}\right)^{2}\nonumber\\&=1-\mathrm{Bhat}(p_{1}(x),p_{2}(x)).
    \label{eq:Hellinger_def}
\end{align}
We can assume that the probability distributions $p_1(x)$ and $p_2(x)$ are defined for a set of real values. 
We can define the mean and standard deviation of the distributions by
$\chi_{i}\equiv\sum_{x}x p_{i}(x)$ and $\sigma_{i}\equiv\sqrt{\sum_{x}x^{2}p_{i}(x)-\chi_{i}^{2}}$, respectively. 
Given the mean and standard deviation of $p_1(x)$ and $p_2(x)$, 
the lower bound of the Hellinger distance is
given by \cite{Nishiyama:2020:HellingerBound}:
\begin{equation}
    \mathrm{Hel}^{2}(p_{1}(x),p_{2}(x))\ge1-\left[\left(\frac{\chi_{1}-\chi_{2}}{\sigma_{1}+\sigma_{2}}\right)^{2}+1\right]^{-\frac{1}{2}}.
    \label{eq:Hellinger_lower_bound}
\end{equation}
We previously used Eq.~\eqref{eq:Hellinger_lower_bound} to
derive a quantum thermodynamic uncertainty relation in Ref.~\cite{Hasegawa:2021:QTURLEPRL}.
Knowing the entire trajectory $\Gamma$, we can compute the 
statistics of the number of jump events. 
As an example, for $\Gamma = [(s_1,m_1),(s_2,m_2),(s_3,m_3)]$, we know that 
there are three jump events at $s_1$, $s_2$ and $s_3$ during the time interval $[0,\tau]$.
Therefore, according to the monotonicity of the Bhattacharyya coefficient and Eq.~\eqref{eq:Hellinger_lower_bound}, we have
\begin{align}
    &\mathrm{Bhat}(\mathcal{P}(\Gamma,t_{1}),\mathcal{P}(\Gamma,t_{2}))\nonumber\\&\le\left[\left(\frac{\braket{\mathcal{C}^{\bullet}}_{t_{1}}-\braket{\mathcal{C}^{\bullet}}_{t_{2}}}{\dblbrace{\mathcal{C}^{\bullet}}_{t_{1}}+\dblbrace{\mathcal{C}^{\bullet}}_{t_{2}}}\right)^{2}+1\right]^{-\frac{1}{2}},
    \label{eq:Batt_mean_var_ineq}
\end{align}
where $\braket{\mathcal{C}^{\bullet}}_t$ and $\dblbrace{\mathcal{C}^{\bullet}}_t$ 
are defined in Eqs.~\eqref{eq:C_braket_def} and \eqref{eq:C_dblbrace_def}, respectively. 
For $0\le(1/2)\int_{t_{1}}^{t_{2}}dt\sqrt{\mathcal{I}(t)}\le\pi/2$, Eq.~\eqref{eq:classical_bound} yields
\begin{equation}
    \cos\left[\frac{1}{2}\int_{t_{1}}^{t_{2}}dt\sqrt{\mathcal{I}(t)}\right]\le\mathrm{Bhat}\left(\mathcal{P}(\Gamma,\nu,t_{1}),\mathcal{P}(\Gamma,\nu,t_{2})\right).
    \label{eq:Lps_bound2}
\end{equation}
Combining Eqs.~\eqref{eq:Batt_mean_var_ineq} and \eqref{eq:Lps_bound2}, we obtain Eq.~\eqref{eq:cTUR}.

Similarly, we can derive quantum thermodynamic uncertainty relations from
Eq.~\eqref{eq:quantum_bound}.
Regarding the quantum fidelity, 
a series of inequalities holds, as were employed in Ref.~\cite{Hasegawa:2021:QTURLEPRL}:
\begin{align}
    |\braket{\Psi(t_{2})|\Psi(t_{1})}|&\le\sum_{\Gamma}|\braket{\Psi(t_{2})|\Gamma}\braket{\Gamma|\Psi(t_{1})}|\nonumber\\&\le\sum_{\Gamma}\sqrt{\mathcal{P}(\Gamma,t_{1})\mathcal{P}(\Gamma,t_{2})}\nonumber\\&=\mathrm{Bhat}\left(\mathcal{P}(\Gamma,t_{1}),\mathcal{P}(\Gamma,t_{2})\right).
    \label{eq:LE_Hellinger_ineq}
\end{align}
The triangle inequality is used in the first line while 
the Cauchy-Schwarz inequality is employed in the first to second lines. 
From Eq.~\eqref{eq:quantum_bound}, for $0\le(1/2)\int_{t_{1}}^{t_{2}}dt\,\sqrt{\mathcal{J}(t)}\le\pi/2$, we have
\begin{equation}
    \cos\left[\frac{1}{2}\int_{t_{1}}^{t_{2}}dt\,\sqrt{\mathcal{J}(t)}\right]\le\left|\braket{\Psi(t_{2})|\Psi(t_{1})}\right|.
    \label{eq:quantum_bound2}
\end{equation}
Combining Eqs.~\eqref{eq:Batt_mean_var_ineq}, \eqref{eq:LE_Hellinger_ineq} and \eqref{eq:quantum_bound2}, 
we can derive Eq.~\eqref{eq:qTUR}.

Next, we derive the conventional thermodynamic uncertainty relation,
which was derived in Ref.~\cite{Terlizzi:2019:KUR}, from Eq.~\eqref{eq:cTUR}. 
We consider a time interval $[\tau-\varepsilon,\tau]$ for Eq.~\eqref{eq:cTUR},
where $\varepsilon > 0$ is an infinitesimally small parameter. 
Then we obtain
\begin{equation}
    \left(\frac{\dblbrace{\mathcal{C}^{\bullet}}_{\tau}+\dblbrace{\mathcal{C}^{\bullet}}_{\tau-\varepsilon}}{\braket{\mathcal{C}^{\bullet}}_{\tau}-\braket{\mathcal{C}^{\bullet}}_{\tau-\varepsilon}}\right)^{2}\ge\frac{1}{\tan\left[\frac{1}{2}\int_{\tau-\varepsilon}^{\tau}\frac{\sqrt{\mathcal{A}(t)}}{t}dt\right]^{2}}.
    \label{eq:cTUR_varepsilon}
\end{equation}
Since $\varepsilon$ is sufficiently small, we have
\begin{equation}
    \frac{d\braket{\mathcal{C}^{\bullet}}_{t}}{dt}=\frac{\braket{\mathcal{C}^{\bullet}}_{t}-\braket{\mathcal{C}^{\bullet}}_{t-\varepsilon}}{\varepsilon}.
    \label{eq:dCdt_def}
\end{equation}
Moreover, we consider a perturbation expansion for $\dblbrace{\mathcal{C}^{\bullet}}_{\tau-\varepsilon}$:
\begin{equation}
    \dblbrace{\mathcal{C}^{\bullet}}_{\tau-\varepsilon}=\dblbrace{\mathcal{C}^{\bullet}}_{\tau}+\varepsilon b_{1}+\varepsilon^{2}b_{2}+\cdots,
    \label{eq:Cstdev_expansion}
\end{equation}where $b_i \in \mathbb{R}$ are expansion coefficients.
Since $\varepsilon \ll 1$, considering the Taylor
expansion $(\tan x)^2= x^2 + O(x^3)$, we obtain
\begin{align}
\tan\left[\frac{1}{2}\int_{\tau-\varepsilon}^{\tau}\frac{\sqrt{\mathcal{A}(t)}}{t}dt\right]^{2}&\simeq\left(\frac{1}{2}\int_{\tau-\varepsilon}^{\tau}\frac{\sqrt{\mathcal{A}(t)}}{t}dt\right)^{2}\nonumber\\&=\frac{\mathcal{A}(\tau)}{4\tau^{2}}\varepsilon^{2}.
\label{eq:tan_taylor_exp}
\end{align}
Substituting Eqs.~\eqref{eq:dCdt_def}--\eqref{eq:tan_taylor_exp}
into Eq.~\eqref{eq:cTUR_varepsilon},
we obtain
\begin{equation}
    \left(\frac{2\dblbrace{\mathcal{C}^{\bullet}}_{\tau}+\varepsilon b_{1}+\varepsilon^{2}b_{2}+\cdots}{\varepsilon\partial_{\tau}\braket{\mathcal{C}^{\bullet}}_{\tau}}\right)^{2}\ge\frac{4\tau^{2}}{\mathcal{A}(\tau)\varepsilon^{2}}.
    \label{eq:cTUR_varep2}
\end{equation}
Taking a limit of $\varepsilon \to 0$, we obtain Eq.~\eqref{eq:cTUR_dCdt}.
We can repeat the same calculation for
the quantum dynamical activity $\mathcal{B}(t)$. 

\subsection*{Speed limit relation as minimum evolution time}

Speed limit relations are often provided as bounds for the minimum evolution time. 
As detailed in Ref.~\cite{Mirkin:2016:OpenQSL}, from speed limit relations shown in Eqs.~\eqref{eq:Lps_bound} and \eqref{eq:Ldo_bound}, 
we can introduce two types of minimum evolution time.
These can be explained using the quantum bound [Eq.~\eqref{eq:Ldo_bound}] because Ref.~\cite{Mirkin:2016:OpenQSL} addressed a quantum speed limit relation. 
The first type of minimum evolution time $\tau_{\mathrm{min}}$ can be implicitly defined by
\begin{align}
    \frac{1}{2}\int_{0}^{\tau_{\mathrm{min}}}\frac{\sqrt{\mathcal{B}(t)}}{t}dt=\mathcal{L}_{D}(\rho(0),\rho(\tau)).
    \label{eq:tau_min_impl_def}
\end{align}
Here, $\tau_\mathrm{min}$ is the time required to reach the geodesic length between $\rho(0)$ and $\rho(\tau)$ traveling along the actual evolution path. 

The second type of minimum evolution time can be obtained directly from Eq.~\eqref{eq:Ldo_bound}.
Let us define the average evolution speed as follows:
\begin{align}
    \mathcal{V}_{\mathrm{av}}\equiv\frac{1}{\tau}\int_{0}^{\tau}dt\frac{\sqrt{\mathcal{B}(t)}}{2t}.
    \label{eq:V_av_classical_def}
\end{align}
Using $\mathcal{V}_\mathrm{av}$, we obtain the bound:
\begin{align}
    \tau\ge\tau_{\mathrm{av}}\equiv\frac{\mathcal{L}_{D}\left(\rho(0),\rho(\tau)\right)}{\mathcal{V}_{\mathrm{av}}}.
    \label{eq:tau_av_QSL_def}
\end{align}
Note that the evaluation of Eq.~\eqref{eq:tau_av_QSL_def} requires information regarding $\tau$ because $\mathcal{V}_\mathrm{av}$ is typically dependent on $\tau$.
When considering a unitary evolution induced by a time-independent Hamiltonian and pure states, $\tau_\mathrm{min} = \tau_\mathrm{av}$ holds but they do not agree in general dynamics. Note that $\tau_{\mathrm{min}}$ and $\tau_{\mathrm{av}}$ can be defined in the classical bound [Eq.~\eqref{eq:Lps_bound}] in the same manner.

\section*{Data availability}
The data generated in this study are provided in the Source Data file. 

\section*{Code availability}
All codes used in this study are available from 
https://github.com/yoshihiko-hasegawa/BulkBoundaryBounds.

\section*{Acknowledgments}
The fruitful comments of Tan Van Vu are greatly appreciated.
This work was supported by JSPS KAKENHI Grant Numbers JP19K12153 and JP22H03659.

\section*{Author contributions}
This work was carried out by Y.H..

\section*{Competing interests}
The author declares no competing interests.

\begin{thebibliography}{75}%
\makeatletter
\providecommand \@ifxundefined [1]{%
 \@ifx{#1\undefined}
}%
\providecommand \@ifnum [1]{%
 \ifnum #1\expandafter \@firstoftwo
 \else \expandafter \@secondoftwo
 \fi
}%
\providecommand \@ifx [1]{%
 \ifx #1\expandafter \@firstoftwo
 \else \expandafter \@secondoftwo
 \fi
}%
\providecommand \natexlab [1]{#1}%
\providecommand \enquote  [1]{``#1''}%
\providecommand \bibnamefont  [1]{#1}%
\providecommand \bibfnamefont [1]{#1}%
\providecommand \citenamefont [1]{#1}%
\providecommand \href@noop [0]{\@secondoftwo}%
\providecommand \href [0]{\begingroup \@sanitize@url \@href}%
\providecommand \@href[1]{\@@startlink{#1}\@@href}%
\providecommand \@@href[1]{\endgroup#1\@@endlink}%
\providecommand \@sanitize@url [0]{\catcode `\\12\catcode `\$12\catcode
  `\&12\catcode `\#12\catcode `\^12\catcode `\_12\catcode `\%12\relax}%
\providecommand \@@startlink[1]{}%
\providecommand \@@endlink[0]{}%
\providecommand \url  [0]{\begingroup\@sanitize@url \@url }%
\providecommand \@url [1]{\endgroup\@href {#1}{\urlprefix }}%
\providecommand \urlprefix  [0]{URL }%
\providecommand \Eprint [0]{\href }%
\providecommand \doibase [0]{https://doi.org/}%
\providecommand \selectlanguage [0]{\@gobble}%
\providecommand \bibinfo  [0]{\@secondoftwo}%
\providecommand \bibfield  [0]{\@secondoftwo}%
\providecommand \translation [1]{[#1]}%
\providecommand \BibitemOpen [0]{}%
\providecommand \bibitemStop [0]{}%
\providecommand \bibitemNoStop [0]{.\EOS\space}%
\providecommand \EOS [0]{\spacefactor3000\relax}%
\providecommand \BibitemShut  [1]{\csname bibitem#1\endcsname}%
\let\auto@bib@innerbib\@empty
%</preamble>
\bibitem [{\citenamefont {Bousso}(2002)}]{Bousso:2002:HoloReview}%
  \BibitemOpen
  \bibfield  {author} {\bibinfo {author} {\bibfnamefont {R.}~\bibnamefont
  {Bousso}},\ }\bibfield  {title} {\bibinfo {title} {The holographic
  principle},\ }\href {https://doi.org/10.1103/RevModPhys.74.825} {\bibfield
  {journal} {\bibinfo  {journal} {Rev. Mod. Phys.}\ }\textbf {\bibinfo {volume}
  {74}},\ \bibinfo {pages} {825} (\bibinfo {year} {2002})}\BibitemShut
  {NoStop}%
\bibitem [{\citenamefont {Ammon}\ and\ \citenamefont
  {Erdmenger}(2015)}]{Ammon:2015:HolographyBook}%
  \BibitemOpen
  \bibfield  {author} {\bibinfo {author} {\bibfnamefont {M.}~\bibnamefont
  {Ammon}}\ and\ \bibinfo {author} {\bibfnamefont {J.}~\bibnamefont
  {Erdmenger}},\ }\href {https://doi.org/10.1017/CBO9780511846373} {\emph
  {\bibinfo {title} {Gauge/gravity duality: Foundations and applications}}}\
  (\bibinfo  {publisher} {Cambridge University Press},\ \bibinfo {year}
  {2015})\BibitemShut {NoStop}%
\bibitem [{\citenamefont {Baggioli}(2019)}]{Baggioli:2019:HoloBook}%
  \BibitemOpen
  \bibfield  {author} {\bibinfo {author} {\bibfnamefont {M.}~\bibnamefont
  {Baggioli}},\ }\href {https://doi.org/10.1007/978-3-030-35184-7} {\emph
  {\bibinfo {title} {Applied holography: a practical mini-course}}}\ (\bibinfo
  {publisher} {Springer},\ \bibinfo {year} {2019})\BibitemShut {NoStop}%
\bibitem [{\citenamefont {Policastro}\ \emph {et~al.}(2001)\citenamefont
  {Policastro}, \citenamefont {Son},\ and\ \citenamefont
  {Starinets}}]{Policastro:2001:HoloViscosity}%
  \BibitemOpen
  \bibfield  {author} {\bibinfo {author} {\bibfnamefont {G.}~\bibnamefont
  {Policastro}}, \bibinfo {author} {\bibfnamefont {D.~T.}\ \bibnamefont
  {Son}},\ and\ \bibinfo {author} {\bibfnamefont {A.~O.}\ \bibnamefont
  {Starinets}},\ }\bibfield  {title} {\bibinfo {title} {Shear viscosity of
  strongly coupled {$N\phantom{\rule{0ex}{0ex}}=\phantom{\rule{0ex}{0ex}}4$}
  supersymmetric yang-mills plasma},\ }\href
  {https://doi.org/10.1103/PhysRevLett.87.081601} {\bibfield  {journal}
  {\bibinfo  {journal} {Phys. Rev. Lett.}\ }\textbf {\bibinfo {volume} {87}},\
  \bibinfo {pages} {081601} (\bibinfo {year} {2001})}\BibitemShut {NoStop}%
\bibitem [{\citenamefont {Ryu}\ and\ \citenamefont
  {Takayanagi}(2006)}]{Ryu:2006:EntEnt}%
  \BibitemOpen
  \bibfield  {author} {\bibinfo {author} {\bibfnamefont {S.}~\bibnamefont
  {Ryu}}\ and\ \bibinfo {author} {\bibfnamefont {T.}~\bibnamefont
  {Takayanagi}},\ }\bibfield  {title} {\bibinfo {title} {Holographic derivation
  of entanglement entropy from the anti--de {Sitter} space/conformal field
  theory correspondence},\ }\href
  {https://doi.org/10.1103/PhysRevLett.96.181602} {\bibfield  {journal}
  {\bibinfo  {journal} {Phys. Rev. Lett.}\ }\textbf {\bibinfo {volume} {96}},\
  \bibinfo {pages} {181602} (\bibinfo {year} {2006})}\BibitemShut {NoStop}%
\bibitem [{\citenamefont {Hartnoll}\ \emph {et~al.}(2007)\citenamefont
  {Hartnoll}, \citenamefont {Kovtun}, \citenamefont {M\"uller},\ and\
  \citenamefont {Sachdev}}]{Hartnoll:2007:HoloNernst}%
  \BibitemOpen
  \bibfield  {author} {\bibinfo {author} {\bibfnamefont {S.~A.}\ \bibnamefont
  {Hartnoll}}, \bibinfo {author} {\bibfnamefont {P.~K.}\ \bibnamefont
  {Kovtun}}, \bibinfo {author} {\bibfnamefont {M.}~\bibnamefont {M\"uller}},\
  and\ \bibinfo {author} {\bibfnamefont {S.}~\bibnamefont {Sachdev}},\
  }\bibfield  {title} {\bibinfo {title} {Theory of the nernst effect near
  quantum phase transitions in condensed matter and in dyonic black holes},\
  }\href {https://doi.org/10.1103/PhysRevB.76.144502} {\bibfield  {journal}
  {\bibinfo  {journal} {Phys. Rev. B}\ }\textbf {\bibinfo {volume} {76}},\
  \bibinfo {pages} {144502} (\bibinfo {year} {2007})}\BibitemShut {NoStop}%
\bibitem [{\citenamefont {Seifert}(2012)}]{Seifert:2012:FTReview}%
  \BibitemOpen
  \bibfield  {author} {\bibinfo {author} {\bibfnamefont {U.}~\bibnamefont
  {Seifert}},\ }\bibfield  {title} {\bibinfo {title} {Stochastic
  thermodynamics, fluctuation theorems and molecular machines},\ }\href
  {http://stacks.iop.org/0034-4885/75/i=12/a=126001} {\bibfield  {journal}
  {\bibinfo  {journal} {Rep. Prog. Phys.}\ }\textbf {\bibinfo {volume} {75}},\
  \bibinfo {pages} {126001} (\bibinfo {year} {2012})}\BibitemShut {NoStop}%
\bibitem [{\citenamefont {Van~den Broeck}\ and\ \citenamefont
  {Esposito}(2015)}]{VandenBroeck:2015:Review}%
  \BibitemOpen
  \bibfield  {author} {\bibinfo {author} {\bibfnamefont {C.}~\bibnamefont
  {Van~den Broeck}}\ and\ \bibinfo {author} {\bibfnamefont {M.}~\bibnamefont
  {Esposito}},\ }\bibfield  {title} {\bibinfo {title} {Ensemble and trajectory
  thermodynamics: A brief introduction},\ }\href
  {https://doi.org/10.1016/j.physa.2014.04.035} {\bibfield  {journal} {\bibinfo
   {journal} {Physica A}\ }\textbf {\bibinfo {volume} {418}},\ \bibinfo {pages}
  {6} (\bibinfo {year} {2015})}\BibitemShut {NoStop}%
\bibitem [{\citenamefont {Funo}\ \emph {et~al.}(2018)\citenamefont {Funo},
  \citenamefont {Ueda},\ and\ \citenamefont {Sagawa}}]{Funo:2018:QFT}%
  \BibitemOpen
  \bibfield  {author} {\bibinfo {author} {\bibfnamefont {K.}~\bibnamefont
  {Funo}}, \bibinfo {author} {\bibfnamefont {M.}~\bibnamefont {Ueda}},\ and\
  \bibinfo {author} {\bibfnamefont {T.}~\bibnamefont {Sagawa}},\ }\bibfield
  {title} {\bibinfo {title} {Quantum fluctuation theorems},\ }in\ \href
  {https://doi.org/10.1007/978-3-319-99046-0_10} {\emph {\bibinfo {booktitle}
  {Thermodynamics in the Quantum Regime: Fundamental Aspects and New
  Directions}}},\ \bibinfo {editor} {edited by\ \bibinfo {editor}
  {\bibfnamefont {F.}~\bibnamefont {Binder}}, \bibinfo {editor} {\bibfnamefont
  {L.~A.}\ \bibnamefont {Correa}}, \bibinfo {editor} {\bibfnamefont
  {C.}~\bibnamefont {Gogolin}}, \bibinfo {editor} {\bibfnamefont
  {J.}~\bibnamefont {Anders}},\ and\ \bibinfo {editor} {\bibfnamefont
  {G.}~\bibnamefont {Adesso}}}\ (\bibinfo  {publisher} {Springer International
  Publishing},\ \bibinfo {year} {2018})\ pp.\ \bibinfo {pages}
  {249--273}\BibitemShut {NoStop}%
\bibitem [{\citenamefont {Manzano}\ and\ \citenamefont
  {Zambrini}(2021)}]{Manzano:2021:QThermo}%
  \BibitemOpen
  \bibfield  {author} {\bibinfo {author} {\bibfnamefont {G.}~\bibnamefont
  {Manzano}}\ and\ \bibinfo {author} {\bibfnamefont {R.}~\bibnamefont
  {Zambrini}},\ }\bibfield  {title} {\bibinfo {title} {Quantum thermodynamics
  under continuous monitoring: a general framework},\ }\href
  {https://doi.org/10.48550/arXiv.2112.02019} {\bibfield  {journal} {\bibinfo
  {journal} {arXiv:2112.02019}\ } (\bibinfo {year} {2021})}\BibitemShut
  {NoStop}%
\bibitem [{\citenamefont {Verstraete}\ and\ \citenamefont
  {Cirac}(2010)}]{Verstraete:2010:cMPS}%
  \BibitemOpen
  \bibfield  {author} {\bibinfo {author} {\bibfnamefont {F.}~\bibnamefont
  {Verstraete}}\ and\ \bibinfo {author} {\bibfnamefont {J.~I.}\ \bibnamefont
  {Cirac}},\ }\bibfield  {title} {\bibinfo {title} {Continuous matrix product
  states for quantum fields},\ }\href
  {https://doi.org/10.1103/PhysRevLett.104.190405} {\bibfield  {journal}
  {\bibinfo  {journal} {Phys. Rev. Lett.}\ }\textbf {\bibinfo {volume} {104}},\
  \bibinfo {pages} {190405} (\bibinfo {year} {2010})}\BibitemShut {NoStop}%
\bibitem [{\citenamefont {Osborne}\ \emph {et~al.}(2010)\citenamefont
  {Osborne}, \citenamefont {Eisert},\ and\ \citenamefont
  {Verstraete}}]{Osborne:2010:Holography}%
  \BibitemOpen
  \bibfield  {author} {\bibinfo {author} {\bibfnamefont {T.~J.}\ \bibnamefont
  {Osborne}}, \bibinfo {author} {\bibfnamefont {J.}~\bibnamefont {Eisert}},\
  and\ \bibinfo {author} {\bibfnamefont {F.}~\bibnamefont {Verstraete}},\
  }\bibfield  {title} {\bibinfo {title} {Holographic quantum states},\ }\href
  {https://doi.org/10.1103/PhysRevLett.105.260401} {\bibfield  {journal}
  {\bibinfo  {journal} {Phys. Rev. Lett.}\ }\textbf {\bibinfo {volume} {105}},\
  \bibinfo {pages} {260401} (\bibinfo {year} {2010})}\BibitemShut {NoStop}%
\bibitem [{\citenamefont {Garrahan}\ and\ \citenamefont
  {Lesanovsky}(2010)}]{Garrahan:2010:QJ}%
  \BibitemOpen
  \bibfield  {author} {\bibinfo {author} {\bibfnamefont {J.~P.}\ \bibnamefont
  {Garrahan}}\ and\ \bibinfo {author} {\bibfnamefont {I.}~\bibnamefont
  {Lesanovsky}},\ }\bibfield  {title} {\bibinfo {title} {Thermodynamics of
  quantum jump trajectories},\ }\href
  {https://link.aps.org/doi/10.1103/PhysRevLett.104.160601} {\bibfield
  {journal} {\bibinfo  {journal} {Phys. Rev. Lett.}\ }\textbf {\bibinfo
  {volume} {104}},\ \bibinfo {pages} {160601} (\bibinfo {year}
  {2010})}\BibitemShut {NoStop}%
\bibitem [{\citenamefont {Lesanovsky}\ \emph {et~al.}(2013)\citenamefont
  {Lesanovsky}, \citenamefont {van Horssen}, \citenamefont {Gu{\c{t}}{\u{a}}},\
  and\ \citenamefont {Garrahan}}]{Lesanovsky:2013:PhaseTrans}%
  \BibitemOpen
  \bibfield  {author} {\bibinfo {author} {\bibfnamefont {I.}~\bibnamefont
  {Lesanovsky}}, \bibinfo {author} {\bibfnamefont {M.}~\bibnamefont {van
  Horssen}}, \bibinfo {author} {\bibfnamefont {M.}~\bibnamefont
  {Gu{\c{t}}{\u{a}}}},\ and\ \bibinfo {author} {\bibfnamefont {J.~P.}\
  \bibnamefont {Garrahan}},\ }\bibfield  {title} {\bibinfo {title}
  {Characterization of dynamical phase transitions in quantum jump trajectories
  beyond the properties of the stationary state},\ }\href
  {https://doi.org/10.1103/PhysRevLett.110.150401} {\bibfield  {journal}
  {\bibinfo  {journal} {Phys. Rev. Lett.}\ }\textbf {\bibinfo {volume} {110}},\
  \bibinfo {pages} {150401} (\bibinfo {year} {2013})}\BibitemShut {NoStop}%
\bibitem [{\citenamefont {Garrahan}(2016)}]{Garrahan:2016:cMPS}%
  \BibitemOpen
  \bibfield  {author} {\bibinfo {author} {\bibfnamefont {J.~P.}\ \bibnamefont
  {Garrahan}},\ }\bibfield  {title} {\bibinfo {title} {Classical stochastic
  dynamics and continuous matrix product states: gauge transformations,
  conditioned and driven processes, and equivalence of trajectory ensembles},\
  }\href {https://doi.org/10.1088/1742-5468/2016/07/073208} {\bibfield
  {journal} {\bibinfo  {journal} {J. Stat. Mech: Theory Exp.}\ }\textbf
  {\bibinfo {volume} {2016}},\ \bibinfo {pages} {073208} (\bibinfo {year}
  {2016})}\BibitemShut {NoStop}%
\bibitem [{\citenamefont {Hasegawa}(2020)}]{Hasegawa:2020:QTURPRL}%
  \BibitemOpen
  \bibfield  {author} {\bibinfo {author} {\bibfnamefont {Y.}~\bibnamefont
  {Hasegawa}},\ }\bibfield  {title} {\bibinfo {title} {Quantum thermodynamic
  uncertainty relation for continuous measurement},\ }\href
  {https://doi.org/10.1103/PhysRevLett.125.050601} {\bibfield  {journal}
  {\bibinfo  {journal} {Phys. Rev. Lett.}\ }\textbf {\bibinfo {volume} {125}},\
  \bibinfo {pages} {050601} (\bibinfo {year} {2020})}\BibitemShut {NoStop}%
\bibitem [{\citenamefont
  {Hasegawa}(2021{\natexlab{a}})}]{Hasegawa:2021:QTURLEPRL}%
  \BibitemOpen
  \bibfield  {author} {\bibinfo {author} {\bibfnamefont {Y.}~\bibnamefont
  {Hasegawa}},\ }\bibfield  {title} {\bibinfo {title} {Irreversibility,
  {Loschmidt} echo, and thermodynamic uncertainty relation},\ }\href
  {https://doi.org/10.1103/PhysRevLett.127.240602} {\bibfield  {journal}
  {\bibinfo  {journal} {Phys. Rev. Lett.}\ }\textbf {\bibinfo {volume} {127}},\
  \bibinfo {pages} {240602} (\bibinfo {year} {2021}{\natexlab{a}})}\BibitemShut
  {NoStop}%
\bibitem [{\citenamefont
  {Hasegawa}(2021{\natexlab{b}})}]{Hasegawa:2021:FPTTUR}%
  \BibitemOpen
  \bibfield  {author} {\bibinfo {author} {\bibfnamefont {Y.}~\bibnamefont
  {Hasegawa}},\ }\bibfield  {title} {\bibinfo {title} {Thermodynamic
  uncertainty relation for quantum first passage process},\ }\href
  {https://doi.org/10.48550/arXiv.2106.09870} {\bibfield  {journal} {\bibinfo
  {journal} {arXiv.2106.09870}\ } (\bibinfo {year}
  {2021}{\natexlab{b}})}\BibitemShut {NoStop}%
\bibitem [{\citenamefont {Barato}\ and\ \citenamefont
  {Seifert}(2015)}]{Barato:2015:UncRel}%
  \BibitemOpen
  \bibfield  {author} {\bibinfo {author} {\bibfnamefont {A.~C.}\ \bibnamefont
  {Barato}}\ and\ \bibinfo {author} {\bibfnamefont {U.}~\bibnamefont
  {Seifert}},\ }\bibfield  {title} {\bibinfo {title} {Thermodynamic uncertainty
  relation for biomolecular processes},\ }\href
  {https://doi.org/10.1103/PhysRevLett.114.158101} {\bibfield  {journal}
  {\bibinfo  {journal} {Phys. Rev. Lett.}\ }\textbf {\bibinfo {volume} {114}},\
  \bibinfo {pages} {158101} (\bibinfo {year} {2015})}\BibitemShut {NoStop}%
\bibitem [{\citenamefont {Gingrich}\ \emph {et~al.}(2016)\citenamefont
  {Gingrich}, \citenamefont {Horowitz}, \citenamefont {Perunov},\ and\
  \citenamefont {England}}]{Gingrich:2016:TUP}%
  \BibitemOpen
  \bibfield  {author} {\bibinfo {author} {\bibfnamefont {T.~R.}\ \bibnamefont
  {Gingrich}}, \bibinfo {author} {\bibfnamefont {J.~M.}\ \bibnamefont
  {Horowitz}}, \bibinfo {author} {\bibfnamefont {N.}~\bibnamefont {Perunov}},\
  and\ \bibinfo {author} {\bibfnamefont {J.~L.}\ \bibnamefont {England}},\
  }\bibfield  {title} {\bibinfo {title} {Dissipation bounds all steady-state
  current fluctuations},\ }\href
  {https://doi.org/10.1103/PhysRevLett.116.120601} {\bibfield  {journal}
  {\bibinfo  {journal} {Phys. Rev. Lett.}\ }\textbf {\bibinfo {volume} {116}},\
  \bibinfo {pages} {120601} (\bibinfo {year} {2016})}\BibitemShut {NoStop}%
\bibitem [{\citenamefont {Garrahan}(2017)}]{Garrahan:2017:TUR}%
  \BibitemOpen
  \bibfield  {author} {\bibinfo {author} {\bibfnamefont {J.~P.}\ \bibnamefont
  {Garrahan}},\ }\bibfield  {title} {\bibinfo {title} {Simple bounds on
  fluctuations and uncertainty relations for first-passage times of counting
  observables},\ }\href {https://doi.org/10.1103/PhysRevE.95.032134} {\bibfield
   {journal} {\bibinfo  {journal} {Phys. Rev. E}\ }\textbf {\bibinfo {volume}
  {95}},\ \bibinfo {pages} {032134} (\bibinfo {year} {2017})}\BibitemShut
  {NoStop}%
\bibitem [{\citenamefont {Dechant}\ and\ \citenamefont
  {Sasa}(2018)}]{Dechant:2018:TUR}%
  \BibitemOpen
  \bibfield  {author} {\bibinfo {author} {\bibfnamefont {A.}~\bibnamefont
  {Dechant}}\ and\ \bibinfo {author} {\bibfnamefont {S.-i.}\ \bibnamefont
  {Sasa}},\ }\bibfield  {title} {\bibinfo {title} {Current fluctuations and
  transport efficiency for general {Langevin} systems},\ }\href
  {https://doi.org/10.1088/1742-5468/aac91a} {\bibfield  {journal} {\bibinfo
  {journal} {J. Stat. Mech: Theory Exp.}\ }\textbf {\bibinfo {volume} {2018}},\
  \bibinfo {pages} {063209} (\bibinfo {year} {2018})}\BibitemShut {NoStop}%
\bibitem [{\citenamefont {{Di Terlizzi}}\ and\ \citenamefont
  {Baiesi}(2019)}]{Terlizzi:2019:KUR}%
  \BibitemOpen
  \bibfield  {author} {\bibinfo {author} {\bibfnamefont {I.}~\bibnamefont {{Di
  Terlizzi}}}\ and\ \bibinfo {author} {\bibfnamefont {M.}~\bibnamefont
  {Baiesi}},\ }\bibfield  {title} {\bibinfo {title} {Kinetic uncertainty
  relation},\ }\href {https://doi.org/10.1088/1751-8121/aaee34} {\bibfield
  {journal} {\bibinfo  {journal} {J. Phys. A: Math. Theor.}\ }\textbf {\bibinfo
  {volume} {52}},\ \bibinfo {pages} {02LT03} (\bibinfo {year}
  {2019})}\BibitemShut {NoStop}%
\bibitem [{\citenamefont {Hasegawa}\ and\ \citenamefont
  {Van~Vu}(2019{\natexlab{a}})}]{Hasegawa:2019:CRI}%
  \BibitemOpen
  \bibfield  {author} {\bibinfo {author} {\bibfnamefont {Y.}~\bibnamefont
  {Hasegawa}}\ and\ \bibinfo {author} {\bibfnamefont {T.}~\bibnamefont
  {Van~Vu}},\ }\bibfield  {title} {\bibinfo {title} {Uncertainty relations in
  stochastic processes: An information inequality approach},\ }\href
  {https://doi.org/10.1103/PhysRevE.99.062126} {\bibfield  {journal} {\bibinfo
  {journal} {Phys. Rev. E}\ }\textbf {\bibinfo {volume} {99}},\ \bibinfo
  {pages} {062126} (\bibinfo {year} {2019}{\natexlab{a}})}\BibitemShut
  {NoStop}%
\bibitem [{\citenamefont {Hasegawa}\ and\ \citenamefont
  {Van~Vu}(2019{\natexlab{b}})}]{Hasegawa:2019:FTUR}%
  \BibitemOpen
  \bibfield  {author} {\bibinfo {author} {\bibfnamefont {Y.}~\bibnamefont
  {Hasegawa}}\ and\ \bibinfo {author} {\bibfnamefont {T.}~\bibnamefont
  {Van~Vu}},\ }\bibfield  {title} {\bibinfo {title} {Fluctuation theorem
  uncertainty relation},\ }\href
  {https://doi.org/10.1103/PhysRevLett.123.110602} {\bibfield  {journal}
  {\bibinfo  {journal} {Phys. Rev. Lett.}\ }\textbf {\bibinfo {volume} {123}},\
  \bibinfo {pages} {110602} (\bibinfo {year} {2019}{\natexlab{b}})}\BibitemShut
  {NoStop}%
\bibitem [{\citenamefont {Van~Vu}\ and\ \citenamefont
  {Hasegawa}(2019)}]{Vu:2019:UTURPRE}%
  \BibitemOpen
  \bibfield  {author} {\bibinfo {author} {\bibfnamefont {T.}~\bibnamefont
  {Van~Vu}}\ and\ \bibinfo {author} {\bibfnamefont {Y.}~\bibnamefont
  {Hasegawa}},\ }\bibfield  {title} {\bibinfo {title} {Uncertainty relations
  for underdamped {Langevin} dynamics},\ }\href
  {https://doi.org/10.1103/PhysRevE.100.032130} {\bibfield  {journal} {\bibinfo
   {journal} {Phys. Rev. E}\ }\textbf {\bibinfo {volume} {100}},\ \bibinfo
  {pages} {032130} (\bibinfo {year} {2019})}\BibitemShut {NoStop}%
\bibitem [{\citenamefont {Dechant}\ and\ \citenamefont
  {Sasa}(2020)}]{Dechant:2020:FRIPNAS}%
  \BibitemOpen
  \bibfield  {author} {\bibinfo {author} {\bibfnamefont {A.}~\bibnamefont
  {Dechant}}\ and\ \bibinfo {author} {\bibfnamefont {S.-i.}\ \bibnamefont
  {Sasa}},\ }\bibfield  {title} {\bibinfo {title} {Fluctuation--response
  inequality out of equilibrium},\ }\href
  {https://doi.org/10.1073/pnas.1918386117} {\bibfield  {journal} {\bibinfo
  {journal} {Proc. Natl. Acad. Sci. U.S.A.}\ }\textbf {\bibinfo {volume}
  {117}},\ \bibinfo {pages} {6430} (\bibinfo {year} {2020})}\BibitemShut
  {NoStop}%
\bibitem [{\citenamefont {Vo}\ \emph {et~al.}(2020)\citenamefont {Vo},
  \citenamefont {Van~Vu},\ and\ \citenamefont {Hasegawa}}]{Vo:2020:TURCSLPRE}%
  \BibitemOpen
  \bibfield  {author} {\bibinfo {author} {\bibfnamefont {V.~T.}\ \bibnamefont
  {Vo}}, \bibinfo {author} {\bibfnamefont {T.}~\bibnamefont {Van~Vu}},\ and\
  \bibinfo {author} {\bibfnamefont {Y.}~\bibnamefont {Hasegawa}},\ }\bibfield
  {title} {\bibinfo {title} {Unified approach to classical speed limit and
  thermodynamic uncertainty relation},\ }\href
  {https://doi.org/10.1103/PhysRevE.102.062132} {\bibfield  {journal} {\bibinfo
   {journal} {Phys. Rev. E}\ }\textbf {\bibinfo {volume} {102}},\ \bibinfo
  {pages} {062132} (\bibinfo {year} {2020})}\BibitemShut {NoStop}%
\bibitem [{\citenamefont {Koyuk}\ and\ \citenamefont
  {Seifert}(2020)}]{Koyuk:2020:TUR}%
  \BibitemOpen
  \bibfield  {author} {\bibinfo {author} {\bibfnamefont {T.}~\bibnamefont
  {Koyuk}}\ and\ \bibinfo {author} {\bibfnamefont {U.}~\bibnamefont
  {Seifert}},\ }\bibfield  {title} {\bibinfo {title} {Thermodynamic uncertainty
  relation for time-dependent driving},\ }\href
  {https://doi.org/10.1103/PhysRevLett.125.260604} {\bibfield  {journal}
  {\bibinfo  {journal} {Phys. Rev. Lett.}\ }\textbf {\bibinfo {volume} {125}},\
  \bibinfo {pages} {260604} (\bibinfo {year} {2020})}\BibitemShut {NoStop}%
\bibitem [{\citenamefont {Pietzonka}(2022)}]{Pietzonka:2021:PendulumTURPRL}%
  \BibitemOpen
  \bibfield  {author} {\bibinfo {author} {\bibfnamefont {P.}~\bibnamefont
  {Pietzonka}},\ }\bibfield  {title} {\bibinfo {title} {Classical pendulum
  clocks break the thermodynamic uncertainty relation},\ }\href
  {https://link.aps.org/doi/10.1103/PhysRevLett.128.130606} {\bibfield
  {journal} {\bibinfo  {journal} {Phys. Rev. Lett.}\ }\textbf {\bibinfo
  {volume} {128}},\ \bibinfo {pages} {130606} (\bibinfo {year}
  {2022})}\BibitemShut {NoStop}%
\bibitem [{\citenamefont {Erker}\ \emph {et~al.}(2017)\citenamefont {Erker},
  \citenamefont {Mitchison}, \citenamefont {Silva}, \citenamefont {Woods},
  \citenamefont {Brunner},\ and\ \citenamefont {Huber}}]{Erker:2017:QClockTUR}%
  \BibitemOpen
  \bibfield  {author} {\bibinfo {author} {\bibfnamefont {P.}~\bibnamefont
  {Erker}}, \bibinfo {author} {\bibfnamefont {M.~T.}\ \bibnamefont
  {Mitchison}}, \bibinfo {author} {\bibfnamefont {R.}~\bibnamefont {Silva}},
  \bibinfo {author} {\bibfnamefont {M.~P.}\ \bibnamefont {Woods}}, \bibinfo
  {author} {\bibfnamefont {N.}~\bibnamefont {Brunner}},\ and\ \bibinfo {author}
  {\bibfnamefont {M.}~\bibnamefont {Huber}},\ }\bibfield  {title} {\bibinfo
  {title} {Autonomous quantum clocks: Does thermodynamics limit our ability to
  measure time?},\ }\href {https://doi.org/10.1103/PhysRevX.7.031022}
  {\bibfield  {journal} {\bibinfo  {journal} {Phys. Rev. X}\ }\textbf {\bibinfo
  {volume} {7}},\ \bibinfo {pages} {031022} (\bibinfo {year}
  {2017})}\BibitemShut {NoStop}%
\bibitem [{\citenamefont {Brandner}\ \emph {et~al.}(2018)\citenamefont
  {Brandner}, \citenamefont {Hanazato},\ and\ \citenamefont
  {Saito}}]{Brandner:2018:Transport}%
  \BibitemOpen
  \bibfield  {author} {\bibinfo {author} {\bibfnamefont {K.}~\bibnamefont
  {Brandner}}, \bibinfo {author} {\bibfnamefont {T.}~\bibnamefont {Hanazato}},\
  and\ \bibinfo {author} {\bibfnamefont {K.}~\bibnamefont {Saito}},\ }\bibfield
   {title} {\bibinfo {title} {Thermodynamic bounds on precision in ballistic
  multiterminal transport},\ }\href
  {https://doi.org/10.1103/PhysRevLett.120.090601} {\bibfield  {journal}
  {\bibinfo  {journal} {Phys. Rev. Lett.}\ }\textbf {\bibinfo {volume} {120}},\
  \bibinfo {pages} {090601} (\bibinfo {year} {2018})}\BibitemShut {NoStop}%
\bibitem [{\citenamefont {Carollo}\ \emph {et~al.}(2019)\citenamefont
  {Carollo}, \citenamefont {Jack},\ and\ \citenamefont
  {Garrahan}}]{Carollo:2019:QuantumLDP}%
  \BibitemOpen
  \bibfield  {author} {\bibinfo {author} {\bibfnamefont {F.}~\bibnamefont
  {Carollo}}, \bibinfo {author} {\bibfnamefont {R.~L.}\ \bibnamefont {Jack}},\
  and\ \bibinfo {author} {\bibfnamefont {J.~P.}\ \bibnamefont {Garrahan}},\
  }\bibfield  {title} {\bibinfo {title} {Unraveling the large deviation
  statistics of {Markovian} open quantum systems},\ }\href
  {https://doi.org/10.1103/PhysRevLett.122.130605} {\bibfield  {journal}
  {\bibinfo  {journal} {Phys. Rev. Lett.}\ }\textbf {\bibinfo {volume} {122}},\
  \bibinfo {pages} {130605} (\bibinfo {year} {2019})}\BibitemShut {NoStop}%
\bibitem [{\citenamefont {Liu}\ and\ \citenamefont
  {Segal}(2019)}]{Liu:2019:QTUR}%
  \BibitemOpen
  \bibfield  {author} {\bibinfo {author} {\bibfnamefont {J.}~\bibnamefont
  {Liu}}\ and\ \bibinfo {author} {\bibfnamefont {D.}~\bibnamefont {Segal}},\
  }\bibfield  {title} {\bibinfo {title} {Thermodynamic uncertainty relation in
  quantum thermoelectric junctions},\ }\href
  {https://doi.org/10.1103/PhysRevE.99.062141} {\bibfield  {journal} {\bibinfo
  {journal} {Phys. Rev. E}\ }\textbf {\bibinfo {volume} {99}},\ \bibinfo
  {pages} {062141} (\bibinfo {year} {2019})}\BibitemShut {NoStop}%
\bibitem [{\citenamefont {Guarnieri}\ \emph {et~al.}(2019)\citenamefont
  {Guarnieri}, \citenamefont {Landi}, \citenamefont {Clark},\ and\
  \citenamefont {Goold}}]{Guarnieri:2019:QTURPRR}%
  \BibitemOpen
  \bibfield  {author} {\bibinfo {author} {\bibfnamefont {G.}~\bibnamefont
  {Guarnieri}}, \bibinfo {author} {\bibfnamefont {G.~T.}\ \bibnamefont
  {Landi}}, \bibinfo {author} {\bibfnamefont {S.~R.}\ \bibnamefont {Clark}},\
  and\ \bibinfo {author} {\bibfnamefont {J.}~\bibnamefont {Goold}},\ }\bibfield
   {title} {\bibinfo {title} {Thermodynamics of precision in quantum
  nonequilibrium steady states},\ }\href
  {https://doi.org/10.1103/PhysRevResearch.1.033021} {\bibfield  {journal}
  {\bibinfo  {journal} {Phys. Rev. Research}\ }\textbf {\bibinfo {volume}
  {1}},\ \bibinfo {pages} {033021} (\bibinfo {year} {2019})}\BibitemShut
  {NoStop}%
\bibitem [{\citenamefont {Saryal}\ \emph {et~al.}(2019)\citenamefont {Saryal},
  \citenamefont {Friedman}, \citenamefont {Segal},\ and\ \citenamefont
  {Agarwalla}}]{Saryal:2019:TUR}%
  \BibitemOpen
  \bibfield  {author} {\bibinfo {author} {\bibfnamefont {S.}~\bibnamefont
  {Saryal}}, \bibinfo {author} {\bibfnamefont {H.~M.}\ \bibnamefont
  {Friedman}}, \bibinfo {author} {\bibfnamefont {D.}~\bibnamefont {Segal}},\
  and\ \bibinfo {author} {\bibfnamefont {B.~K.}\ \bibnamefont {Agarwalla}},\
  }\bibfield  {title} {\bibinfo {title} {Thermodynamic uncertainty relation in
  thermal transport},\ }\href
  {https://link.aps.org/doi/10.1103/PhysRevE.100.042101} {\bibfield  {journal}
  {\bibinfo  {journal} {Phys. Rev. E}\ }\textbf {\bibinfo {volume} {100}},\
  \bibinfo {pages} {042101} (\bibinfo {year} {2019})}\BibitemShut {NoStop}%
\bibitem [{\citenamefont
  {Hasegawa}(2021{\natexlab{c}})}]{Hasegawa:2020:TUROQS}%
  \BibitemOpen
  \bibfield  {author} {\bibinfo {author} {\bibfnamefont {Y.}~\bibnamefont
  {Hasegawa}},\ }\bibfield  {title} {\bibinfo {title} {Thermodynamic
  uncertainty relation for general open quantum systems},\ }\href
  {https://doi.org/10.1103/PhysRevLett.126.010602} {\bibfield  {journal}
  {\bibinfo  {journal} {Phys. Rev. Lett.}\ }\textbf {\bibinfo {volume} {126}},\
  \bibinfo {pages} {010602} (\bibinfo {year} {2021}{\natexlab{c}})}\BibitemShut
  {NoStop}%
\bibitem [{\citenamefont {Sacchi}(2021)}]{Sacchi:2021:BosonicTUR}%
  \BibitemOpen
  \bibfield  {author} {\bibinfo {author} {\bibfnamefont {M.~F.}\ \bibnamefont
  {Sacchi}},\ }\bibfield  {title} {\bibinfo {title} {Thermodynamic uncertainty
  relations for bosonic {Otto} engines},\ }\href
  {https://doi.org/10.1103/PhysRevE.103.012111} {\bibfield  {journal} {\bibinfo
   {journal} {Phys. Rev. E}\ }\textbf {\bibinfo {volume} {103}},\ \bibinfo
  {pages} {012111} (\bibinfo {year} {2021})}\BibitemShut {NoStop}%
\bibitem [{\citenamefont {Kalaee}\ \emph {et~al.}(2021)\citenamefont {Kalaee},
  \citenamefont {Wacker},\ and\ \citenamefont {Potts}}]{Kalaee:2021:QTURPRE}%
  \BibitemOpen
  \bibfield  {author} {\bibinfo {author} {\bibfnamefont {A.~A.~S.}\
  \bibnamefont {Kalaee}}, \bibinfo {author} {\bibfnamefont {A.}~\bibnamefont
  {Wacker}},\ and\ \bibinfo {author} {\bibfnamefont {P.~P.}\ \bibnamefont
  {Potts}},\ }\bibfield  {title} {\bibinfo {title} {Violating the thermodynamic
  uncertainty relation in the three-level maser},\ }\href
  {https://doi.org/10.1103/PhysRevE.104.L012103} {\bibfield  {journal}
  {\bibinfo  {journal} {Phys. Rev. E}\ }\textbf {\bibinfo {volume} {104}},\
  \bibinfo {pages} {L012103} (\bibinfo {year} {2021})}\BibitemShut {NoStop}%
\bibitem [{\citenamefont {Monnai}(2022)}]{Monnai:2022:QTUR}%
  \BibitemOpen
  \bibfield  {author} {\bibinfo {author} {\bibfnamefont {T.}~\bibnamefont
  {Monnai}},\ }\bibfield  {title} {\bibinfo {title} {Thermodynamic uncertainty
  relation for quantum work distribution: Exact case study for a perturbed
  oscillator},\ }\href {https://doi.org/10.1103/PhysRevE.105.034115} {\bibfield
   {journal} {\bibinfo  {journal} {Phys. Rev. E}\ }\textbf {\bibinfo {volume}
  {105}},\ \bibinfo {pages} {034115} (\bibinfo {year} {2022})}\BibitemShut
  {NoStop}%
\bibitem [{\citenamefont {Horowitz}\ and\ \citenamefont
  {Gingrich}(2019)}]{Horowitz:2019:TURReview}%
  \BibitemOpen
  \bibfield  {author} {\bibinfo {author} {\bibfnamefont {J.~M.}\ \bibnamefont
  {Horowitz}}\ and\ \bibinfo {author} {\bibfnamefont {T.~R.}\ \bibnamefont
  {Gingrich}},\ }\bibfield  {title} {\bibinfo {title} {Thermodynamic
  uncertainty relations constrain non-equilibrium fluctuations},\ }\href
  {https://doi.org/10.1038/s41567-019-0702-6} {\bibfield  {journal} {\bibinfo
  {journal} {Nat. Phys.}\ } (\bibinfo {year} {2019})}\BibitemShut {NoStop}%
\bibitem [{\citenamefont {Mandelstam}\ and\ \citenamefont
  {Tamm}(1945)}]{Mandelstam:1945:QSL}%
  \BibitemOpen
  \bibfield  {author} {\bibinfo {author} {\bibfnamefont {L.}~\bibnamefont
  {Mandelstam}}\ and\ \bibinfo {author} {\bibfnamefont {I.}~\bibnamefont
  {Tamm}},\ }\bibfield  {title} {\bibinfo {title} {The uncertainty relation
  between energy and time in non-relativistic quantum mechanics},\ }\href
  {https://doi.org/10.1007/978-3-642-74626-0_8} {\bibfield  {journal} {\bibinfo
   {journal} {J. Phys. USSR}\ }\textbf {\bibinfo {volume} {9}},\ \bibinfo
  {pages} {249} (\bibinfo {year} {1945})}\BibitemShut {NoStop}%
\bibitem [{\citenamefont {Margolus}\ and\ \citenamefont
  {Levitin}(1998)}]{Margolus:1998:QSL}%
  \BibitemOpen
  \bibfield  {author} {\bibinfo {author} {\bibfnamefont {N.}~\bibnamefont
  {Margolus}}\ and\ \bibinfo {author} {\bibfnamefont {L.~B.}\ \bibnamefont
  {Levitin}},\ }\bibfield  {title} {\bibinfo {title} {The maximum speed of
  dynamical evolution},\ }\href
  {http://www.sciencedirect.com/science/article/pii/S0167278998000542}
  {\bibfield  {journal} {\bibinfo  {journal} {Physica D: Nonlinear Phenomena}\
  }\textbf {\bibinfo {volume} {120}},\ \bibinfo {pages} {188 } (\bibinfo {year}
  {1998})}\BibitemShut {NoStop}%
\bibitem [{\citenamefont {Deffner}\ and\ \citenamefont
  {Lutz}(2010)}]{Deffner:2010:GenClausius}%
  \BibitemOpen
  \bibfield  {author} {\bibinfo {author} {\bibfnamefont {S.}~\bibnamefont
  {Deffner}}\ and\ \bibinfo {author} {\bibfnamefont {E.}~\bibnamefont {Lutz}},\
  }\bibfield  {title} {\bibinfo {title} {Generalized {Clausius} inequality for
  nonequilibrium quantum processes},\ }\href
  {https://doi.org/10.1103/PhysRevLett.105.170402} {\bibfield  {journal}
  {\bibinfo  {journal} {Phys. Rev. Lett.}\ }\textbf {\bibinfo {volume} {105}},\
  \bibinfo {pages} {170402} (\bibinfo {year} {2010})}\BibitemShut {NoStop}%
\bibitem [{\citenamefont {Taddei}\ \emph {et~al.}(2013)\citenamefont {Taddei},
  \citenamefont {Escher}, \citenamefont {Davidovich},\ and\ \citenamefont
  {de~Matos~Filho}}]{Taddei:2013:QSL}%
  \BibitemOpen
  \bibfield  {author} {\bibinfo {author} {\bibfnamefont {M.~M.}\ \bibnamefont
  {Taddei}}, \bibinfo {author} {\bibfnamefont {B.~M.}\ \bibnamefont {Escher}},
  \bibinfo {author} {\bibfnamefont {L.}~\bibnamefont {Davidovich}},\ and\
  \bibinfo {author} {\bibfnamefont {R.~L.}\ \bibnamefont {de~Matos~Filho}},\
  }\bibfield  {title} {\bibinfo {title} {Quantum speed limit for physical
  processes},\ }\href {https://link.aps.org/doi/10.1103/PhysRevLett.110.050402}
  {\bibfield  {journal} {\bibinfo  {journal} {Phys. Rev. Lett.}\ }\textbf
  {\bibinfo {volume} {110}},\ \bibinfo {pages} {050402} (\bibinfo {year}
  {2013})}\BibitemShut {NoStop}%
\bibitem [{\citenamefont {del Campo}\ \emph {et~al.}(2013)\citenamefont {del
  Campo}, \citenamefont {Egusquiza}, \citenamefont {Plenio},\ and\
  \citenamefont {Huelga}}]{DelCampo:2013:OpenQSL}%
  \BibitemOpen
  \bibfield  {author} {\bibinfo {author} {\bibfnamefont {A.}~\bibnamefont {del
  Campo}}, \bibinfo {author} {\bibfnamefont {I.~L.}\ \bibnamefont {Egusquiza}},
  \bibinfo {author} {\bibfnamefont {M.~B.}\ \bibnamefont {Plenio}},\ and\
  \bibinfo {author} {\bibfnamefont {S.~F.}\ \bibnamefont {Huelga}},\ }\bibfield
   {title} {\bibinfo {title} {Quantum speed limits in open system dynamics},\
  }\href {https://doi.org/10.1103/PhysRevLett.110.050403} {\bibfield  {journal}
  {\bibinfo  {journal} {Phys. Rev. Lett.}\ }\textbf {\bibinfo {volume} {110}},\
  \bibinfo {pages} {050403} (\bibinfo {year} {2013})}\BibitemShut {NoStop}%
\bibitem [{\citenamefont {Deffner}\ and\ \citenamefont
  {Lutz}(2013)}]{Deffner:2013:DrivenQSL}%
  \BibitemOpen
  \bibfield  {author} {\bibinfo {author} {\bibfnamefont {S.}~\bibnamefont
  {Deffner}}\ and\ \bibinfo {author} {\bibfnamefont {E.}~\bibnamefont {Lutz}},\
  }\bibfield  {title} {\bibinfo {title} {Energy-time uncertainty relation for
  driven quantum systems},\ }\href
  {https://doi.org/10.1088/1751-8113/46/33/335302} {\bibfield  {journal}
  {\bibinfo  {journal} {J. Phys. A: Math. Theor.}\ }\textbf {\bibinfo {volume}
  {46}},\ \bibinfo {pages} {335302} (\bibinfo {year} {2013})}\BibitemShut
  {NoStop}%
\bibitem [{\citenamefont {Pires}\ \emph {et~al.}(2016)\citenamefont {Pires},
  \citenamefont {Cianciaruso}, \citenamefont {C\'eleri}, \citenamefont
  {Adesso},\ and\ \citenamefont {Soares-Pinto}}]{Pires:2016:GQSL}%
  \BibitemOpen
  \bibfield  {author} {\bibinfo {author} {\bibfnamefont {D.~P.}\ \bibnamefont
  {Pires}}, \bibinfo {author} {\bibfnamefont {M.}~\bibnamefont {Cianciaruso}},
  \bibinfo {author} {\bibfnamefont {L.~C.}\ \bibnamefont {C\'eleri}}, \bibinfo
  {author} {\bibfnamefont {G.}~\bibnamefont {Adesso}},\ and\ \bibinfo {author}
  {\bibfnamefont {D.~O.}\ \bibnamefont {Soares-Pinto}},\ }\bibfield  {title}
  {\bibinfo {title} {Generalized geometric quantum speed limits},\ }\href
  {https://doi.org/10.1103/PhysRevX.6.021031} {\bibfield  {journal} {\bibinfo
  {journal} {Phys. Rev. X}\ }\textbf {\bibinfo {volume} {6}},\ \bibinfo {pages}
  {021031} (\bibinfo {year} {2016})}\BibitemShut {NoStop}%
\bibitem [{\citenamefont {O'Connor}\ \emph {et~al.}(2021)\citenamefont
  {O'Connor}, \citenamefont {Guarnieri},\ and\ \citenamefont
  {Campbell}}]{OConnor:2021:ActionSL}%
  \BibitemOpen
  \bibfield  {author} {\bibinfo {author} {\bibfnamefont {E.}~\bibnamefont
  {O'Connor}}, \bibinfo {author} {\bibfnamefont {G.}~\bibnamefont
  {Guarnieri}},\ and\ \bibinfo {author} {\bibfnamefont {S.}~\bibnamefont
  {Campbell}},\ }\bibfield  {title} {\bibinfo {title} {Action quantum speed
  limits},\ }\href {https://link.aps.org/doi/10.1103/PhysRevA.103.022210}
  {\bibfield  {journal} {\bibinfo  {journal} {Phys. Rev. A}\ }\textbf {\bibinfo
  {volume} {103}},\ \bibinfo {pages} {022210} (\bibinfo {year}
  {2021})}\BibitemShut {NoStop}%
\bibitem [{\citenamefont {Shiraishi}\ \emph {et~al.}(2018)\citenamefont
  {Shiraishi}, \citenamefont {Funo},\ and\ \citenamefont
  {Saito}}]{Shiraishi:2018:SpeedLimit}%
  \BibitemOpen
  \bibfield  {author} {\bibinfo {author} {\bibfnamefont {N.}~\bibnamefont
  {Shiraishi}}, \bibinfo {author} {\bibfnamefont {K.}~\bibnamefont {Funo}},\
  and\ \bibinfo {author} {\bibfnamefont {K.}~\bibnamefont {Saito}},\ }\bibfield
   {title} {\bibinfo {title} {Speed limit for classical stochastic processes},\
  }\href {https://link.aps.org/doi/10.1103/PhysRevLett.121.070601} {\bibfield
  {journal} {\bibinfo  {journal} {Phys. Rev. Lett.}\ }\textbf {\bibinfo
  {volume} {121}},\ \bibinfo {pages} {070601} (\bibinfo {year}
  {2018})}\BibitemShut {NoStop}%
\bibitem [{\citenamefont {Ito}(2018)}]{Ito:2018:InfoGeo}%
  \BibitemOpen
  \bibfield  {author} {\bibinfo {author} {\bibfnamefont {S.}~\bibnamefont
  {Ito}},\ }\bibfield  {title} {\bibinfo {title} {Stochastic thermodynamic
  interpretation of information geometry},\ }\href
  {https://doi.org/10.1103/PhysRevLett.121.030605} {\bibfield  {journal}
  {\bibinfo  {journal} {Phys. Rev. Lett.}\ }\textbf {\bibinfo {volume} {121}},\
  \bibinfo {pages} {030605} (\bibinfo {year} {2018})}\BibitemShut {NoStop}%
\bibitem [{\citenamefont {Ito}\ and\ \citenamefont
  {Dechant}(2020)}]{Ito:2020:TimeTURPRX}%
  \BibitemOpen
  \bibfield  {author} {\bibinfo {author} {\bibfnamefont {S.}~\bibnamefont
  {Ito}}\ and\ \bibinfo {author} {\bibfnamefont {A.}~\bibnamefont {Dechant}},\
  }\bibfield  {title} {\bibinfo {title} {Stochastic time evolution, information
  geometry, and the {Cram\'er}-{Rao} bound},\ }\href
  {https://doi.org/10.1103/PhysRevX.10.021056} {\bibfield  {journal} {\bibinfo
  {journal} {Phys. Rev. X}\ }\textbf {\bibinfo {volume} {10}},\ \bibinfo
  {pages} {021056} (\bibinfo {year} {2020})}\BibitemShut {NoStop}%
\bibitem [{\citenamefont {Nicholson}\ \emph {et~al.}(2020)\citenamefont
  {Nicholson}, \citenamefont {Garcia-Pintos}, \citenamefont {del Campo},\ and\
  \citenamefont {Green}}]{Nicholson:2020:TIUncRel}%
  \BibitemOpen
  \bibfield  {author} {\bibinfo {author} {\bibfnamefont {S.~B.}\ \bibnamefont
  {Nicholson}}, \bibinfo {author} {\bibfnamefont {L.~P.}\ \bibnamefont
  {Garcia-Pintos}}, \bibinfo {author} {\bibfnamefont {A.}~\bibnamefont {del
  Campo}},\ and\ \bibinfo {author} {\bibfnamefont {J.~R.}\ \bibnamefont
  {Green}},\ }\bibfield  {title} {\bibinfo {title} {Time-information
  uncertainty relations in thermodynamics},\ }\href
  {https://doi.org/10.1038/s41567-020-0981-y} {\bibfield  {journal} {\bibinfo
  {journal} {Nat. Phys.}\ }\textbf {\bibinfo {volume} {16}},\ \bibinfo {pages}
  {1211} (\bibinfo {year} {2020})}\BibitemShut {NoStop}%
\bibitem [{\citenamefont {Van~Vu}\ and\ \citenamefont
  {Hasegawa}(2021)}]{Vu:2021:GeomBound}%
  \BibitemOpen
  \bibfield  {author} {\bibinfo {author} {\bibfnamefont {T.}~\bibnamefont
  {Van~Vu}}\ and\ \bibinfo {author} {\bibfnamefont {Y.}~\bibnamefont
  {Hasegawa}},\ }\bibfield  {title} {\bibinfo {title} {Geometrical bounds of
  the irreversibility in {Markovian} systems},\ }\href
  {https://doi.org/10.1103/PhysRevLett.126.010601} {\bibfield  {journal}
  {\bibinfo  {journal} {Phys. Rev. Lett.}\ }\textbf {\bibinfo {volume} {126}},\
  \bibinfo {pages} {010601} (\bibinfo {year} {2021})}\BibitemShut {NoStop}%
\bibitem [{\citenamefont {Deffner}\ and\ \citenamefont
  {Campbell}(2017)}]{Deffner:2017:QSLReview}%
  \BibitemOpen
  \bibfield  {author} {\bibinfo {author} {\bibfnamefont {S.}~\bibnamefont
  {Deffner}}\ and\ \bibinfo {author} {\bibfnamefont {S.}~\bibnamefont
  {Campbell}},\ }\bibfield  {title} {\bibinfo {title} {Quantum speed limits:
  from {Heisenberg}'s uncertainty principle to optimal quantum control},\
  }\href {https://doi.org/10.1088/1751-8121/aa86c6} {\bibfield  {journal}
  {\bibinfo  {journal} {J. Phys. A: Math. Theor.}\ }\textbf {\bibinfo {volume}
  {50}},\ \bibinfo {pages} {453001} (\bibinfo {year} {2017})}\BibitemShut
  {NoStop}%
\bibitem [{\citenamefont {Li}\ \emph {et~al.}(2019)\citenamefont {Li},
  \citenamefont {Horowitz}, \citenamefont {Gingrich},\ and\ \citenamefont
  {Fakhri}}]{Li:2019:EPInference}%
  \BibitemOpen
  \bibfield  {author} {\bibinfo {author} {\bibfnamefont {J.}~\bibnamefont
  {Li}}, \bibinfo {author} {\bibfnamefont {J.~M.}\ \bibnamefont {Horowitz}},
  \bibinfo {author} {\bibfnamefont {T.~R.}\ \bibnamefont {Gingrich}},\ and\
  \bibinfo {author} {\bibfnamefont {N.}~\bibnamefont {Fakhri}},\ }\bibfield
  {title} {\bibinfo {title} {Quantifying dissipation using fluctuating
  currents},\ }\href {https://doi.org/10.1038/s41467-019-09631-x} {\bibfield
  {journal} {\bibinfo  {journal} {Nat. Commun.}\ }\textbf {\bibinfo {volume}
  {10}},\ \bibinfo {pages} {1666} (\bibinfo {year} {2019})}\BibitemShut
  {NoStop}%
\bibitem [{\citenamefont {Manikandan}\ \emph {et~al.}(2020)\citenamefont
  {Manikandan}, \citenamefont {Gupta},\ and\ \citenamefont
  {Krishnamurthy}}]{Manikandan:2019:InferEPPRL}%
  \BibitemOpen
  \bibfield  {author} {\bibinfo {author} {\bibfnamefont {S.~K.}\ \bibnamefont
  {Manikandan}}, \bibinfo {author} {\bibfnamefont {D.}~\bibnamefont {Gupta}},\
  and\ \bibinfo {author} {\bibfnamefont {S.}~\bibnamefont {Krishnamurthy}},\
  }\bibfield  {title} {\bibinfo {title} {Inferring entropy production from
  short experiments},\ }\href
  {https://link.aps.org/doi/10.1103/PhysRevLett.124.120603} {\bibfield
  {journal} {\bibinfo  {journal} {Phys. Rev. Lett.}\ }\textbf {\bibinfo
  {volume} {124}},\ \bibinfo {pages} {120603} (\bibinfo {year}
  {2020})}\BibitemShut {NoStop}%
\bibitem [{\citenamefont {Van~Vu}\ \emph {et~al.}(2020)\citenamefont {Van~Vu},
  \citenamefont {Vo},\ and\ \citenamefont {Hasegawa}}]{Vu:2020:EPInferPRE}%
  \BibitemOpen
  \bibfield  {author} {\bibinfo {author} {\bibfnamefont {T.}~\bibnamefont
  {Van~Vu}}, \bibinfo {author} {\bibfnamefont {V.~T.}\ \bibnamefont {Vo}},\
  and\ \bibinfo {author} {\bibfnamefont {Y.}~\bibnamefont {Hasegawa}},\
  }\bibfield  {title} {\bibinfo {title} {Entropy production estimation with
  optimal current},\ }\href
  {https://link.aps.org/doi/10.1103/PhysRevE.101.042138} {\bibfield  {journal}
  {\bibinfo  {journal} {Phys. Rev. E}\ }\textbf {\bibinfo {volume} {101}},\
  \bibinfo {pages} {042138} (\bibinfo {year} {2020})}\BibitemShut {NoStop}%
\bibitem [{\citenamefont {Otsubo}\ \emph {et~al.}(2020)\citenamefont {Otsubo},
  \citenamefont {Ito}, \citenamefont {Dechant},\ and\ \citenamefont
  {Sagawa}}]{Otsubo:2020:EPInferPRE}%
  \BibitemOpen
  \bibfield  {author} {\bibinfo {author} {\bibfnamefont {S.}~\bibnamefont
  {Otsubo}}, \bibinfo {author} {\bibfnamefont {S.}~\bibnamefont {Ito}},
  \bibinfo {author} {\bibfnamefont {A.}~\bibnamefont {Dechant}},\ and\ \bibinfo
  {author} {\bibfnamefont {T.}~\bibnamefont {Sagawa}},\ }\bibfield  {title}
  {\bibinfo {title} {Estimating entropy production by machine learning of
  short-time fluctuating currents},\ }\href
  {https://doi.org/10.1103/PhysRevE.101.062106} {\bibfield  {journal} {\bibinfo
   {journal} {Phys. Rev. E}\ }\textbf {\bibinfo {volume} {101}},\ \bibinfo
  {pages} {062106} (\bibinfo {year} {2020})}\BibitemShut {NoStop}%
\bibitem [{\citenamefont {Rold{\'{a}}n}\ \emph {et~al.}(2021)\citenamefont
  {Rold{\'{a}}n}, \citenamefont {Barral}, \citenamefont {Martin}, \citenamefont
  {Parrondo},\ and\ \citenamefont {J{\"u}licher}}]{Roldan:2021:EPInfer}%
  \BibitemOpen
  \bibfield  {author} {\bibinfo {author} {\bibfnamefont {{\'{E}}.}~\bibnamefont
  {Rold{\'{a}}n}}, \bibinfo {author} {\bibfnamefont {J.}~\bibnamefont
  {Barral}}, \bibinfo {author} {\bibfnamefont {P.}~\bibnamefont {Martin}},
  \bibinfo {author} {\bibfnamefont {J.~M.~R.}\ \bibnamefont {Parrondo}},\ and\
  \bibinfo {author} {\bibfnamefont {F.}~\bibnamefont {J{\"u}licher}},\
  }\bibfield  {title} {\bibinfo {title} {Quantifying entropy production in
  active fluctuations of the hair-cell bundle from time irreversibility and
  uncertainty relations},\ }\href {https://doi.org/10.1088/1367-2630/ac0f18}
  {\bibfield  {journal} {\bibinfo  {journal} {New J. Phys.}\ }\textbf {\bibinfo
  {volume} {23}},\ \bibinfo {pages} {083013} (\bibinfo {year}
  {2021})}\BibitemShut {NoStop}%
\bibitem [{\citenamefont {Wootters}(1981)}]{Wootters:1981:StatDist}%
  \BibitemOpen
  \bibfield  {author} {\bibinfo {author} {\bibfnamefont {W.~K.}\ \bibnamefont
  {Wootters}},\ }\bibfield  {title} {\bibinfo {title} {Statistical distance and
  {Hilbert} space},\ }\href {https://doi.org/10.1103/PhysRevD.23.357}
  {\bibfield  {journal} {\bibinfo  {journal} {Phys. Rev. D}\ }\textbf {\bibinfo
  {volume} {23}},\ \bibinfo {pages} {357} (\bibinfo {year} {1981})}\BibitemShut
  {NoStop}%
\bibitem [{\citenamefont {Maes}(2020)}]{Maes:2020:FrenesyPR}%
  \BibitemOpen
  \bibfield  {author} {\bibinfo {author} {\bibfnamefont {C.}~\bibnamefont
  {Maes}},\ }\bibfield  {title} {\bibinfo {title} {Frenesy: Time-symmetric
  dynamical activity in nonequilibria},\ }\href
  {https://doi.org/10.1016/j.physrep.2020.01.002} {\bibfield  {journal}
  {\bibinfo  {journal} {Phys. Rep.}\ }\textbf {\bibinfo {volume} {850}},\
  \bibinfo {pages} {1} (\bibinfo {year} {2020})}\BibitemShut {NoStop}%
\bibitem [{\citenamefont {Uhlmann}(1992)}]{Uhlmann:1992:BuresGeodesic}%
  \BibitemOpen
  \bibfield  {author} {\bibinfo {author} {\bibfnamefont {A.}~\bibnamefont
  {Uhlmann}},\ }\bibinfo {title} {The metric of {Bures} and the geometric
  phase},\ in\ \href {https://doi.org/10.1007/978-94-011-2801-8_23} {\emph
  {\bibinfo {booktitle} {Groups and Related Topics: Proceedings of the First
  Max Born Symposium}}},\ \bibinfo {editor} {edited by\ \bibinfo {editor}
  {\bibfnamefont {R.}~\bibnamefont {Gielerak}}, \bibinfo {editor}
  {\bibfnamefont {J.}~\bibnamefont {Lukierski}},\ and\ \bibinfo {editor}
  {\bibfnamefont {Z.}~\bibnamefont {Popowicz}}}\ (\bibinfo  {publisher}
  {Springer Netherlands},\ \bibinfo {address} {Dordrecht},\ \bibinfo {year}
  {1992})\ pp.\ \bibinfo {pages} {267--274}\BibitemShut {NoStop}%
\bibitem [{\citenamefont {Meyer}(2021)}]{Meyer:2021:QFI}%
  \BibitemOpen
  \bibfield  {author} {\bibinfo {author} {\bibfnamefont {J.~J.}\ \bibnamefont
  {Meyer}},\ }\bibfield  {title} {\bibinfo {title} {{Fisher} information in
  noisy intermediate-scale quantum applications},\ }\href
  {https://doi.org/10.22331/q-2021-09-09-539} {\bibfield  {journal} {\bibinfo
  {journal} {{Quantum}}\ }\textbf {\bibinfo {volume} {5}},\ \bibinfo {pages}
  {539} (\bibinfo {year} {2021})}\BibitemShut {NoStop}%
\bibitem [{\citenamefont {Nielsen}\ and\ \citenamefont
  {Chuang}(2011)}]{Nielsen:2011:QuantumInfoBook}%
  \BibitemOpen
  \bibfield  {author} {\bibinfo {author} {\bibfnamefont {M.~A.}\ \bibnamefont
  {Nielsen}}\ and\ \bibinfo {author} {\bibfnamefont {I.~L.}\ \bibnamefont
  {Chuang}},\ }\href@noop {} {\emph {\bibinfo {title} {Quantum Computation and
  Quantum Information}}}\ (\bibinfo  {publisher} {Cambridge University Press},\
  \bibinfo {address} {New York, NY, USA},\ \bibinfo {year} {2011})\BibitemShut
  {NoStop}%
\bibitem [{\citenamefont {Gammelmark}\ and\ \citenamefont
  {M\o{}lmer}(2014)}]{Gammelmark:2014:QCRB}%
  \BibitemOpen
  \bibfield  {author} {\bibinfo {author} {\bibfnamefont {S.}~\bibnamefont
  {Gammelmark}}\ and\ \bibinfo {author} {\bibfnamefont {K.}~\bibnamefont
  {M\o{}lmer}},\ }\bibfield  {title} {\bibinfo {title} {{Fisher} information
  and the quantum {Cram\'er}-{Rao} sensitivity limit of continuous
  measurements},\ }\href {https://doi.org/10.1103/PhysRevLett.112.170401}
  {\bibfield  {journal} {\bibinfo  {journal} {Phys. Rev. Lett.}\ }\textbf
  {\bibinfo {volume} {112}},\ \bibinfo {pages} {170401} (\bibinfo {year}
  {2014})}\BibitemShut {NoStop}%
\bibitem [{\citenamefont {Vo}\ \emph {et~al.}(2022)\citenamefont {Vo},
  \citenamefont {Vu},\ and\ \citenamefont {Hasegawa}}]{Vo:2022:UKTUR}%
  \BibitemOpen
  \bibfield  {author} {\bibinfo {author} {\bibfnamefont {V.~T.}\ \bibnamefont
  {Vo}}, \bibinfo {author} {\bibfnamefont {T.~V.}\ \bibnamefont {Vu}},\ and\
  \bibinfo {author} {\bibfnamefont {Y.}~\bibnamefont {Hasegawa}},\ }\bibfield
  {title} {\bibinfo {title} {Unified thermodynamic kinetic uncertainty
  relation},\ }\href {https://arxiv.org/abs/2203.11501} {\bibfield  {journal}
  {\bibinfo  {journal} {arXiv:2203.11501}\ } (\bibinfo {year}
  {2022})}\BibitemShut {NoStop}%
\bibitem [{\citenamefont {Heisenberg}(1927)}]{Heisenberg:1927:UR}%
  \BibitemOpen
  \bibfield  {author} {\bibinfo {author} {\bibfnamefont {W.}~\bibnamefont
  {Heisenberg}},\ }\bibfield  {title} {\bibinfo {title} {{\"U}ber den
  anschaulichen inhalt der quantentheoretischen kinematik und mechanik},\
  }\href {https://doi.org/10.1007/BF01397280} {\bibfield  {journal} {\bibinfo
  {journal} {Z. Phys.}\ }\textbf {\bibinfo {volume} {43}},\ \bibinfo {pages}
  {172} (\bibinfo {year} {1927})}\BibitemShut {NoStop}%
\bibitem [{\citenamefont {Robertson}(1929)}]{Robertson:1929:UncRel}%
  \BibitemOpen
  \bibfield  {author} {\bibinfo {author} {\bibfnamefont {H.~P.}\ \bibnamefont
  {Robertson}},\ }\bibfield  {title} {\bibinfo {title} {The uncertainty
  principle},\ }\href {https://doi.org/10.1103/PhysRev.34.163} {\bibfield
  {journal} {\bibinfo  {journal} {Phys. Rev.}\ }\textbf {\bibinfo {volume}
  {34}},\ \bibinfo {pages} {163} (\bibinfo {year} {1929})}\BibitemShut
  {NoStop}%
\bibitem [{\citenamefont {Fr\"owis}\ \emph {et~al.}(2015)\citenamefont
  {Fr\"owis}, \citenamefont {Schmied},\ and\ \citenamefont
  {Gisin}}]{Frowis:2015:TQUR}%
  \BibitemOpen
  \bibfield  {author} {\bibinfo {author} {\bibfnamefont {F.}~\bibnamefont
  {Fr\"owis}}, \bibinfo {author} {\bibfnamefont {R.}~\bibnamefont {Schmied}},\
  and\ \bibinfo {author} {\bibfnamefont {N.}~\bibnamefont {Gisin}},\ }\bibfield
   {title} {\bibinfo {title} {Tighter quantum uncertainty relations following
  from a general probabilistic bound},\ }\href
  {https://doi.org/10.1103/PhysRevA.92.012102} {\bibfield  {journal} {\bibinfo
  {journal} {Phys. Rev. A}\ }\textbf {\bibinfo {volume} {92}},\ \bibinfo
  {pages} {012102} (\bibinfo {year} {2015})}\BibitemShut {NoStop}%
\bibitem [{\citenamefont {Maccone}\ and\ \citenamefont
  {Pati}(2014)}]{Maccone:2014:UR}%
  \BibitemOpen
  \bibfield  {author} {\bibinfo {author} {\bibfnamefont {L.}~\bibnamefont
  {Maccone}}\ and\ \bibinfo {author} {\bibfnamefont {A.~K.}\ \bibnamefont
  {Pati}},\ }\bibfield  {title} {\bibinfo {title} {Stronger uncertainty
  relations for all incompatible observables},\ }\href
  {https://doi.org/10.1103/PhysRevLett.113.260401} {\bibfield  {journal}
  {\bibinfo  {journal} {Phys. Rev. Lett.}\ }\textbf {\bibinfo {volume} {113}},\
  \bibinfo {pages} {260401} (\bibinfo {year} {2014})}\BibitemShut {NoStop}%
\bibitem [{\citenamefont {Barato}\ and\ \citenamefont
  {Seifert}(2016)}]{Barato:2016:BrowClo}%
  \BibitemOpen
  \bibfield  {author} {\bibinfo {author} {\bibfnamefont {A.~C.}\ \bibnamefont
  {Barato}}\ and\ \bibinfo {author} {\bibfnamefont {U.}~\bibnamefont
  {Seifert}},\ }\bibfield  {title} {\bibinfo {title} {Cost and precision of
  {Brownian} clocks},\ }\href {https://doi.org/10.1103/PhysRevX.6.041053}
  {\bibfield  {journal} {\bibinfo  {journal} {Phys. Rev. X}\ }\textbf {\bibinfo
  {volume} {6}},\ \bibinfo {pages} {041053} (\bibinfo {year}
  {2016})}\BibitemShut {NoStop}%
\bibitem [{\citenamefont {Liese}\ and\ \citenamefont
  {Vajda}(2006)}]{Liese:2006:Divergence}%
  \BibitemOpen
  \bibfield  {author} {\bibinfo {author} {\bibfnamefont {F.}~\bibnamefont
  {Liese}}\ and\ \bibinfo {author} {\bibfnamefont {I.}~\bibnamefont {Vajda}},\
  }\bibfield  {title} {\bibinfo {title} {On divergences and informations in
  statistics and information theory},\ }\href
  {https://doi.org/10.1109/TIT.2006.881731} {\bibfield  {journal} {\bibinfo
  {journal} {IEEE Trans. Inf. Theor.}\ }\textbf {\bibinfo {volume} {52}},\
  \bibinfo {pages} {4394} (\bibinfo {year} {2006})}\BibitemShut {NoStop}%
\bibitem [{\citenamefont {Nishiyama}(2020)}]{Nishiyama:2020:HellingerBound}%
  \BibitemOpen
  \bibfield  {author} {\bibinfo {author} {\bibfnamefont {T.}~\bibnamefont
  {Nishiyama}},\ }\bibfield  {title} {\bibinfo {title} {A tight lower bound for
  the {Hellinger} distance with given means and variances},\ }\href
  {https://arxiv.org/abs/2010.13548} {\bibfield  {journal} {\bibinfo  {journal}
  {arXiv:2010.13548}\ } (\bibinfo {year} {2020})}\BibitemShut {NoStop}%
\bibitem [{\citenamefont {Mirkin}\ \emph {et~al.}(2016)\citenamefont {Mirkin},
  \citenamefont {Toscano},\ and\ \citenamefont
  {Wisniacki}}]{Mirkin:2016:OpenQSL}%
  \BibitemOpen
  \bibfield  {author} {\bibinfo {author} {\bibfnamefont {N.}~\bibnamefont
  {Mirkin}}, \bibinfo {author} {\bibfnamefont {F.}~\bibnamefont {Toscano}},\
  and\ \bibinfo {author} {\bibfnamefont {D.~A.}\ \bibnamefont {Wisniacki}},\
  }\bibfield  {title} {\bibinfo {title} {Quantum-speed-limit bounds in an open
  quantum evolution},\ }\href
  {https://link.aps.org/doi/10.1103/PhysRevA.94.052125} {\bibfield  {journal}
  {\bibinfo  {journal} {Phys. Rev. A}\ }\textbf {\bibinfo {volume} {94}},\
  \bibinfo {pages} {052125} (\bibinfo {year} {2016})}\BibitemShut {NoStop}%
\end{thebibliography}
\end{document}

\title{Supplementary Information for\\
``Unifying Speed Limit, Thermodynamic Uncertainty Relation and Heisenberg Principle via Bulk-Boundary Correspondence''
}
\author{Yoshihiko Hasegawa${}^1$}
\email{hasegawa@biom.t.u-tokyo.ac.jp}
\affiliation{${}^1$Department of Information and Communication Engineering, Graduate
School of Information Science and Technology, The University of Tokyo,
Tokyo 113-8656, Japan}

\maketitle

\refstepcounter{toinum}
\section*{Supplementary Note \thetoinum: Continuous matrix product state}

This subsection explains the calculations associated with the continuous matrix product state.
For notational convenience, we define
\begin{align}
    Q(t)&\equiv\mathfrak{U}\left(t;H_{\mathrm{sys}},\{L_{m}\}\right)\nonumber\\&=\mathbb{T}\exp\left[-i\int_{0}^{t}ds\,\left\{ H_{\mathrm{sys}}\otimes\mathbb{I}_{\mathrm{fld}}+\sum_{m}\left(iL_{m}\otimes\phi_{m}^{\dagger}(s)-iL_{m}^{\dagger}\otimes\phi_{m}(s)\right)\right\} \right],
    \label{eq:U_unitary_def}
\end{align}
where $\mathfrak{U}$ is shown in Eq.~\UfrakUdef{} in the main text.
The continuous matrix product state is given by [Eq.~\cMPSUdefII{} in the main text]
\begin{equation}
    \ket{\Phi(t)}=Q(t)\ket{\psi(0)}\otimes\ket{\mathrm{vac}}.
    \label{eq:Phi_cMPS_def}
\end{equation}
From Eq.~\eqref{eq:U_unitary_def}, we have
\begin{align}
    dQ(t)&=Q(t+dt)-Q(t)\nonumber\\&=\left[\exp\left[-i\left\{ H_{\mathrm{sys}}dt\otimes\mathbb{I}_{\mathrm{fld}}+\sum_{m}\left(iL_{m}\otimes d\phi_{m}^{\dagger}(t)-iL_{m}^{\dagger}\otimes d\phi_{m}(t)\right)\right\} \right]-1\right]Q(t)\nonumber\\&=\sum_{n=1}^{\infty}\frac{1}{n!}\left[-i\left\{ H_{\mathrm{sys}}dt\otimes\mathbb{I}_{\mathrm{fld}}+\sum_{m}\left(iL_{m}\otimes d\phi_{m}^{\dagger}(t)-iL_{m}^{\dagger}\otimes d\phi_{m}(t)\right)\right\} \right]^{n}Q(t),
    \label{eq:dU_def}
\end{align}
where
\begin{align}
    d\phi_{m}(t)&\equiv\int_{t}^{t+dt}\phi_{m}(s)ds,\label{eq:dphi_def}\\
    d\phi_{m}^{\dagger}(t)&\equiv\int_{t}^{t+dt}\phi_{m}^{\dagger}(s)ds.\label{eq:dphi_dag_def}
\end{align}
Because the terms with $n \ge 3$ vanish, Eq.~\eqref{eq:dU_def} yields
\begin{equation}
    d\ket{\Phi(t)}=\left[-iH_{\mathrm{sys}}dt\otimes\mathbb{I}_{\mathrm{fld}}+\sum_{m}L_{m}\otimes d\phi_{m}^{\dagger}(t)-\frac{1}{2}\sum_{m}L_{m}^{\dagger}L_{m}dt\otimes\mathbb{I}_{\mathrm{fld}}\right]\ket{\Phi(t)},
    \label{eq:dPhi_def}
\end{equation}
in which the Ito rule is used for $\phi_m(s)$.
Then $\ket{\Phi(t)}$ is represented by
\begin{equation}
    \ket{\Phi(t)}=\mathbb{T}\exp\left[-i\int_{0}^{t}ds\left(H_{\mathrm{eff}}\otimes\mathbb{I}_{\mathrm{fld}}+\sum_{m}iL_{m}\otimes\phi_{m}^{\dagger}(s)\right)\right]\ket{\psi(0)}\otimes\ket{\mathrm{vac}},
    \label{eq:Phi_alt_def}
\end{equation}where $H_\mathrm{eff}$ is the effective Hamiltonian, defined by
\begin{equation}
H_{\mathrm{eff}}\equiv H_{\mathrm{sys}}-\frac{i}{2}\sum_{m=1}^{M}L_{m}^{\dagger}L_{m}.
\label{eq:Heff_def}
\end{equation}

We now consider an alternative representation of $\ket{\Psi(t)}$. Based on Eq.~\eqref{eq:Phi_alt_def},
$\ket{\Psi(\tau)}$ can be expressed by
\begin{equation}
    \ket{\Psi(\tau)}=\mathbb{T}\exp\left[-i\int_{0}^{\tau}ds\,\left(H_{\mathrm{eff}}\otimes\mathbb{I}_{\mathrm{fld}}+\sum_{m}iL_{m}\otimes\phi_{m}^{\dagger}(s)\right)\right]\ket{\psi(0)}\otimes\ket{\mathrm{vac}}.
    \label{eq:cMPS_alt}
\end{equation}
Let us consider a discretization from $0$ to $\Delta s \equiv \tau / N$,
where $N$ is a sufficiently large natural number. 
Then the one-step evolution of Eq.~\eqref{eq:cMPS_alt} is given by
\begin{equation}
    \ket{\Psi(\tau)}_{\Delta s}=\left[\left(\mathbb{I}-i\Delta s H_{\mathrm{eff}}\otimes\mathbb{I}_{\mathrm{fld}}\right)+\sum_{m}\left(L_{m}\otimes\Delta\phi_{m}^{\dagger}\right)\right]\ket{\psi(0)}\otimes\ket{\mathrm{vac}}.
    \label{eq:cMPS_disc}
\end{equation}
Using $\Delta \phi_m \Delta \phi_{m'}^\dagger = \delta_{mm'}\Delta s$,
which can be derived from the canonical commutation relation,
we have
\begin{align}
    \mathrm{Tr}_{\mathrm{fld}}\left[\ket{\Psi(\tau)}_{\Delta s}\bra{\Psi(\tau)}_{\Delta s}\right]&=\left(\mathbb{I}_{\mathrm{sys}}-i\Delta sH_{\mathrm{eff}}\right)\ket{\psi(0)}\bra{\psi(0)}\left(\mathbb{I}_{\mathrm{sys}}+i\Delta sH_{\mathrm{eff}}^{\dagger}\right)+\sum_{m}\left(L_{m}\sqrt{\Delta s}\right)\ket{\psi(0)}\bra{\psi(0)}\left(L_{m}^{\dagger}\sqrt{\Delta s}\right)\nonumber\\&=\sum_{m=0}^{M}V_{m}\ket{\psi(0)}\bra{\psi(0)}V_{m}^{\dagger},
    \label{eq:Kraus_repr}
\end{align}where $V_m$ comprises the Kraus operators. These are defined by
\begin{align}
    V_{0}&\equiv\mathbb{I}_{\mathrm{sys}}-i\Delta sH_{\mathrm{eff}},\label{eq:V0_def}\\V_{m}&\equiv\sqrt{\Delta s}L_{m}\,\,\,\,\,(1\le m\le M).\label{eq:Vm_def}
\end{align}
Generalizing Eq.~\eqref{eq:Kraus_repr}, it is possible to compute $\mathrm{Tr}_{\mathrm{fld}}\left[\ket{\Psi(t_{1})}\bra{\Psi(t_{2})}\right]$ as
\begin{align}
    \mathrm{Tr}_{\mathrm{fld}}\left[\ket{\Psi(t_{1})}\bra{\Psi(t_{2})}\right]=\sum_{m_{N-1}}\cdots\sum_{m_{0}}V_{m_{N-1}}(t_{1})\cdots V_{m_{0}}(t_{1})\ket{\psi(0)}\bra{\psi(0)}V_{m_{0}}^{\dagger}(t_{2})\cdots V_{m_{N-1}}^{\dagger}(t_{2}),
    \label{eq:two_sided_Kraus}
\end{align}
where $V_m(t)$ are defined by
\begin{align}
    V_{0}(t)&\equiv\mathbb{I}_{\mathrm{sys}}-i\Delta s\frac{t}{\tau}H_{\mathrm{eff}},\label{eq:V0t_def}\\
    V_{m}(t)&\equiv\sqrt{\Delta s}\sqrt{\frac{t}{\tau}}L_{m}\,\,\,\,\,(1\le m\le M).\label{eq:Vmt_def}
\end{align}
For $N\to \infty$, Eq.~\eqref{eq:two_sided_Kraus} yields the two-sided Lindblad equation [cf. Eq.~\eqref{eq:two_side_Lindblad_def}].

\refstepcounter{toinum}
\section*{Supplementary Note \thetoinum: Classical Fisher information\label{sec:CFI_calculation}}

\begin{figure*}
\includegraphics[width=12cm]{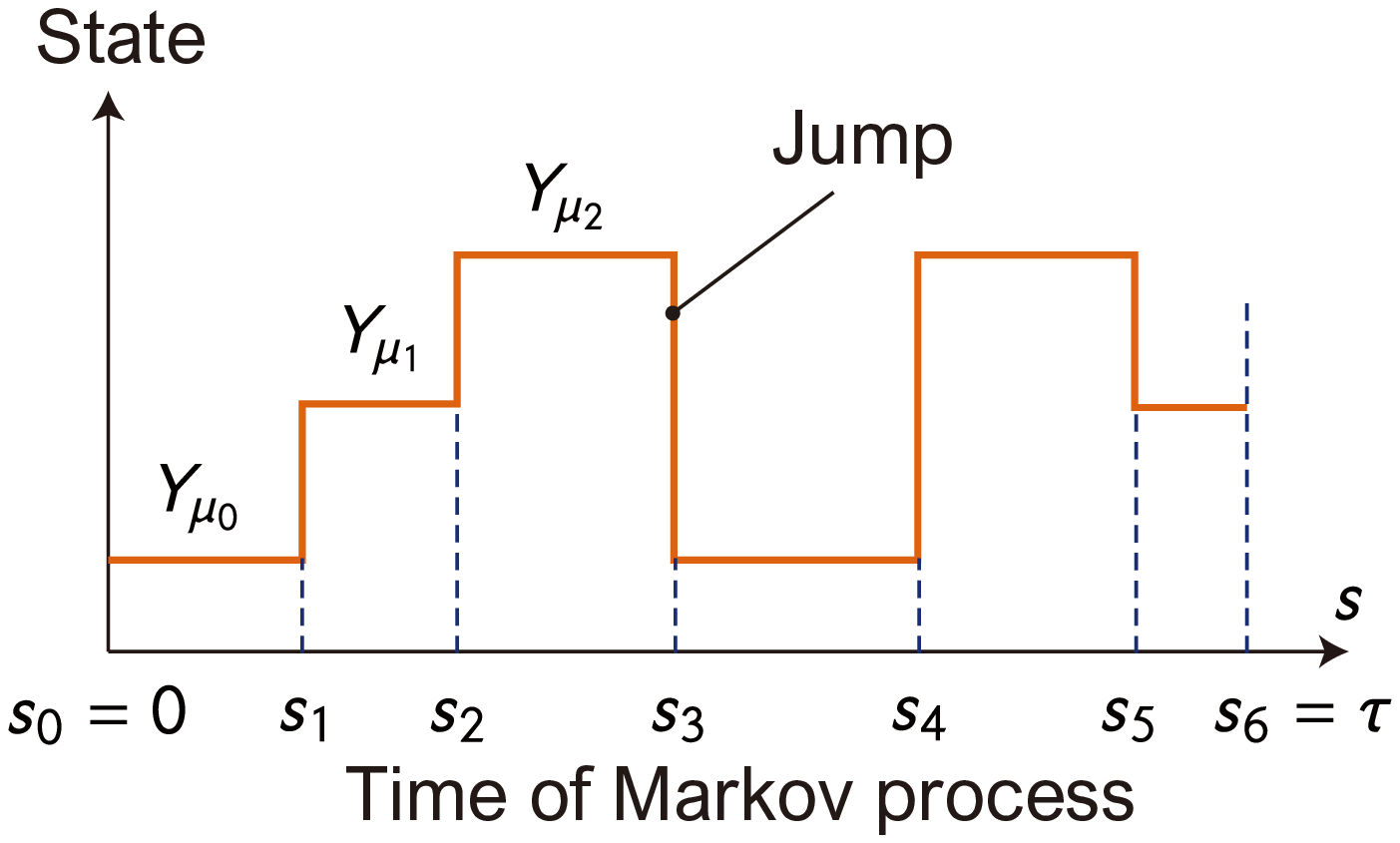} 
\caption{
\textbf{Example of a classical Markov process. }
A diagram of a process that starts from $s=0$ and ends at $s=\tau$ while
undergoing $K=5$ jump events.
Here, $s_l$ denotes the time stamp for the $l$th jump event,
where $s_0 \equiv 0$ and $s_{K+1} = s_6 = \tau$. In addition,
$Y_{\mu_l}$ denotes the state after the $l$th jump event. 
\label{fig:jump_example}}
\end{figure*}

In this section, we calculate the classical Fisher information for a classical Markov process. 
We first consider the general parameter inference in a classical Markov process. 
Let $\theta \in \mathbb{R}$ be a parameter of interest. 
Suppose that there are $K$ jump events in a trajectory,
and let $s_l$ ($l\in\{0,1,2,\cdots,K\}$) be the time stamp for the
$l$th jump event with $s_0=0$ and $s_{K+1}=\tau$.
Let $Y_{\mu_l}$ be the state after the $l$th jump event, where 
$Y_{\mu_0}$ is the initial state of the process and 
$W_{\mu \nu}(\theta)$ is the transition rate from $Y_\nu$ to $Y_\mu$,
which depends on the parameter $\theta$, and
$P(\mu_0;\theta)$ is the initial distribution of the Markov process,
which is also dependent on $\theta$.
From the path integral representation \cite{Seifert:2012:FTReview}, 
the probability of the trajectory $\Gamma$ with
the initial state $Y_{\mu_0}$ is
\begin{equation}
    \mathcal{P}(\mu_{0},\Gamma;\theta)=P(\mu_{0};\theta)\mathcal{P}(\Gamma|\mu_{0};\theta),
    \label{eq:P_traj_def}
\end{equation}where $\mathcal{P}(\Gamma|\mu_0;\theta)$
is the conditional probability of $\Gamma$ given
the initial state $Y_{\mu_0}$.
$\mathcal{P}(\Gamma|\mu_0;\theta)$ is given by
\begin{equation}
    \ln\mathcal{P}(\Gamma|\mu_{0};\theta)=\sum_{l=1}^{K}\ln W_{\mu_{l}\mu_{l-1}}(\theta)-\sum_{l=0}^{K}\int_{s_{l}}^{s_{l+1}}ds\,R(\mu_{l};\theta),
    \label{eq:P_cond_traj_def}
\end{equation}
where $R(\mu;\theta)$ is the escape rate defined by
\begin{equation}
    R(\mu;\theta)\equiv\sum_{\nu(\ne\mu)}W_{\nu\mu}(\theta).
    \label{eq:R_def}
\end{equation}
Using Eq.~\eqref{eq:P_cond_traj_def}, we can evaluate the Fisher information:
\begin{equation}
    -\left\langle \frac{\partial^{2}}{\partial\theta^{2}}\ln\mathcal{P}(\mu_{0},\Gamma;\theta)\right\rangle =-\left\langle \frac{\partial^{2}}{\partial\theta^{2}}\ln P(\mu_{0};\theta)\right\rangle -\left\langle \sum_{l=1}^{K}\frac{\partial^{2}}{\partial\theta^{2}}\ln W_{\mu_{l}\mu_{l-1}}(\theta)\right\rangle +\left\langle \sum_{l=0}^{K}\int_{s_{l}}^{s_{l+1}}ds\,\frac{\partial^{2}}{\partial\theta^{2}}R(\mu_{l};\theta)\right\rangle ,
    \label{eq:P_traj_dd}
\end{equation}
where the bracket denotes the expectation. 
The second term on the right hand side of Eq.~\eqref{eq:P_traj_dd} becomes
\begin{align}
    \left\langle \sum_{l=1}^{K}\frac{\partial^{2}}{\partial\theta^{2}}\ln W_{\mu_{l}\mu_{l-1}}(\theta)\right\rangle &=\int_{0}^{\tau}ds\sum_{\mu,\nu,\mu\ne\nu}P(\mu,s;\theta)W_{\nu\mu}(\theta)\frac{\partial^{2}}{\partial\theta^{2}}\ln W_{\nu\mu}(\theta)\nonumber\\&=\int_{0}^{\tau}ds\sum_{\mu,\nu,\mu\ne\nu}P(\mu,s;\theta)W_{\nu\mu}(\theta)\left[-\frac{\left(\partial_{\theta}W_{\nu\mu}(\theta)\right)^{2}}{W_{\nu\mu}(\theta)^{2}}+\frac{\partial_{\theta}^{2}W_{\nu\mu}(\theta)}{W_{\nu\mu}(\theta)}\right]\nonumber\\&=\int_{0}^{\tau}ds\sum_{\mu,\nu,\mu\ne\nu}P(\mu,s;\theta)\left[-\frac{\left(\partial_{\theta}W_{\nu\mu}(\theta)\right)^{2}}{W_{\nu\mu}(\theta)}+\frac{\partial^{2}}{\partial\theta^{2}}W_{\nu\mu}(\theta)\right].
    \label{eq:second_term}
\end{align}
Similarly, the third term on the right hand side of Eq.~\eqref{eq:P_traj_dd} becomes
\begin{equation}
    \left\langle \sum_{l=0}^{K}\int_{s_{l}}^{s_{l+1}}ds\,\frac{\partial^{2}}{\partial\theta^{2}}R(\mu_{l};\theta)\right\rangle =\int_{0}^{\tau}ds\,\sum_{\mu,\nu,\mu\ne\nu}P(\mu,s;\theta)\frac{\partial^{2}}{\partial\theta^{2}}W_{\nu\mu}(\theta).
    \label{eq:third_term}
\end{equation}
Combining Eqs.~\eqref{eq:second_term} and \eqref{eq:third_term},
the Fisher information is
\begin{equation}
    \mathcal{I}(\theta)=-\left\langle \frac{\partial^{2}}{\partial\theta^{2}}\ln P(\mu_{0};\theta)\right\rangle +\int_{0}^{\tau}ds\sum_{\mu,\nu,\mu\ne\nu}P(\mu,s;\theta)\frac{\left(\partial_{\theta}W_{\nu\mu}(\theta)\right)^{2}}{W_{\nu\mu}(\theta)}.
    \label{eq:Fisher_info_theta}
\end{equation}

In the main text, it is demonstrated that the classical Fisher information [Eq.~\CFIUdef{} in the main text] is given by
\begin{equation}
    \mathcal{I}(t)=\sum_{\Gamma,\nu}\mathcal{P}(\Gamma,\nu,t)\left(-\frac{\partial^{2}}{\partial t^{2}}\ln\mathcal{P}(\Gamma,\nu,t)\right),
    \label{eq:Fisher_def}
\end{equation}
where $\mathcal{P}(\Gamma,\nu,t)$ is the probability of measuring a trajectory $\Gamma$ and $Y_\nu$ at the end time.
Knowing a specific trajectory $\Gamma$ and the final state $Y_\nu$, we can uniquely specify the state change of the dynamics. Therefore, Eq.~\eqref{eq:Fisher_def} can be rewritten as
\begin{equation}
    \mathcal{I}(t)=\sum_{\mu_{0},\Gamma}\mathcal{P}(\mu_{0},\Gamma,t)\left(-\frac{\partial^{2}}{\partial t^{2}}\ln\mathcal{P}(\mu_{0},\Gamma,t)\right),
\end{equation}
where $\mathcal{P}(\mu_{0},\Gamma,t)$ is the probability of measuring a trajectory $\Gamma$ given the initial state $Y_{\mu_0}$.
Equations~\eqref{eq:P_traj_def}--\eqref{eq:Fisher_info_theta}
can then be used to calculate the Fisher information. 
Let $P(\nu,s;\boldsymbol{W})$ be the probability of being $Y_\nu$ at time $s$,
the dynamics of which is governed by a classical Markov process
with the transition rate $\boldsymbol{W}\equiv \{W_{\mu\nu}\}_{\mu,\nu}$.
From Eq.~\eqref{eq:Fisher_info_theta}, we obtain
\begin{align}
    \mathcal{I}(t)&=\int_{0}^{\tau}ds\sum_{\nu,\mu,\nu\ne\mu}\frac{P(\nu,s;\frac{t}{\tau}\boldsymbol{W})}{\frac{t}{\tau}W_{\mu\nu}}\left(\frac{\partial\frac{t}{\tau}W_{\mu\nu}}{\partial t}\right)^{2}\nonumber\\&=\int_{0}^{\tau}ds\,\sum_{\nu,\mu,\nu\ne\mu}\frac{1}{t\tau}P\left(\nu,s;\frac{t}{\tau}\boldsymbol{W}\right)W_{\mu\nu}.
    \label{eq:Fisher2}
\end{align}
In Eq.~\eqref{eq:Fisher2}, the boundary term $P\left(\nu,s=0;\frac{t}{\tau}\boldsymbol{W}\right)$
does not come into play since
the initial state does not depend on $t$.
We define the dynamical activity from $0$ to $t$ as follows:
\begin{equation}
    \mathcal{A}(t;\boldsymbol{W})\equiv\int_{0}^{t}ds\sum_{\nu,\mu,\nu\ne\mu}P(\mu,s;\boldsymbol{W})W_{\nu\mu},
    \label{eq:A_da_def}
\end{equation}which
quantifies the average number of jumps in $[0,t]$.
Since the transition rate is time-independent,
we have the relation:
\begin{equation}
    P\left(\nu,s;\frac{t}{\tau}\boldsymbol{W}\right)=P\left(\nu,\frac{t}{\tau}s;\boldsymbol{W}\right).
    \label{eq:Pn_time_relation}
\end{equation}
Substituting Eq.~\eqref{eq:Pn_time_relation} into Eq.~\eqref{eq:Fisher2},
we obtain
\begin{equation}
    \mathcal{I}(t)=\frac{\mathcal{A}\left(\tau;\frac{t}{\tau}\boldsymbol{W}\right)}{t^{2}}=\frac{\mathcal{A}(t;\boldsymbol{W})}{t^{2}},
    \label{eq:Fisher3}
\end{equation}
which is Eq.~\IcUandUAUrelation{} in the main text. 

\refstepcounter{toinum}
\section*{Supplementary Note \thetoinum: Quantum Fisher information\label{sec:QFI}}

The geometric bound given by Eq.~\quantumUbound{} in the main text concerns the calculation of the quantum Fisher information $\mathcal{J}(t)$,
which is expressed by Eq.~\QFIUdef{} in the main text. 
Specifically, we have
\begin{align}
    \braket{\Psi(t_{2})|\Psi(t_{1})}=\mathrm{Tr}_{\mathrm{sys,fld}}\left[\ket{\Psi(t_{1})}\bra{\Psi(t_{2})}\right]=\mathrm{Tr}_{\mathrm{sys}}\left[\zeta(\tau;t_{1},t_{2})\right],
    \label{eq:two_side_def}
\end{align}
where $\zeta(\tau;t_{1},t_{2})\equiv\mathrm{Tr}_{\mathrm{fld}}\left[\ket{\Psi(t_{1})}\bra{\Psi(t_{2})}\right]$. From Eq.~\eqref{eq:two_sided_Kraus},
$\zeta(s;t_1,t_2)$ obeys the two-sided Lindblad equation \cite{Gammelmark:2014:QCRB}:
\begin{align}
    \frac{d}{ds}\zeta(s;t_{1},t_{2})=-iH_{1,\mathrm{sys}}\zeta+i\zeta H_{2,\mathrm{sys}}+\sum_{m}L_{1,m}\zeta L_{2,m}^{\dagger}-\frac{1}{2}\sum_{m}\left[L_{1,m}^{\dagger}L_{1,m}\zeta+\zeta L_{2,m}^{\dagger}L_{2,m}\right],
    \label{eq:two_side_Lindblad_def}
\end{align}
where $H_{1,\mathrm{sys}}\equiv(t_{1}/\tau)H_{\mathrm{sys}}$ and 
$L_{1,m}\equiv\sqrt{t_{1}/\tau}L_{m}$ ($H_{2,\mathrm{sys}}$ and $L_{2,m}$
are defined in a similar manner). 
Note that $\zeta(s;t_1,t_2)$ is not a density operator because
$\mathrm{Tr}_\mathrm{sys}[\zeta(s;t_1,t_2)] \ne 1$ in general. 
Calculating Eq.~\eqref{eq:two_side_Lindblad_def} with the initial state $\zeta(0;t_1,t_2) = \ket{\psi(0)}\bra{\psi(0)}$ to $s=\tau$, the fidelity can be computed. 
Practically, we compute $\mathcal{J}(t)$ based on the relationship: 
\begin{align}
    \mathcal{J}(t)=\frac{8}{dt^{2}}\left[1-\left|\braket{\Psi(t)|\Psi(t+dt)}\right|\right],
    \label{eq:Jt_by_Fidelity}
\end{align}
where $dt$ is taken to be a sufficiently small increment. 
Equation~\eqref{eq:Jt_by_Fidelity} can be shown by using $\ket{\partial_t \Psi(t)}dt = \ket{\Psi(t+dt)}-\ket{\Psi(t)}$. 
Since $\ket{\Psi(t)}$ can be calculated by Eq.~\eqref{eq:two_side_Lindblad_def}, we can calculate the quantum Fisher information via Eq.~\eqref{eq:Jt_by_Fidelity}.

\refstepcounter{toinum}
\section*{Supplementary Note \thetoinum: Initially mixed state case\label{eq:mixed_state_case}}

The calculation in the main text assumes an initially pure state,
i.e., $\rho(0)=\ket{\psi(0)}\bra{\psi(0)}$ but 
here we consider an initially mixed-state case. 
We can introduce an ancilla that purifies the initial state:
\begin{equation}
    \rho(0) = \mathrm{Tr}_\mathrm{anc}[\ket{\tilde{\psi}(0)}\bra{\tilde{\psi}(0)}],
    \label{eq:purification_def}
\end{equation}
where $\ket{\tilde{\psi}(0)}$ is a purification of $\rho(0)$ and $\mathrm{Tr}_\mathrm{anc}$ is the trace operation with respect to the ancilla.
We define the
scaled
continuous matrix product state for the purified state as follows:
\begin{align}
    \ket{\tilde{\Psi}(t)}&=\tilde{\mathfrak{U}}\left(\tau;\frac{t}{\tau}H_{\mathrm{sys}},\left\{ \sqrt{\frac{t}{\tau}}L_{m}\right\} \right)\ket{\tilde{\psi}(0)}\otimes\ket{\mathrm{vac}},\label{eq:Psi_tilde_def}\\
    \tilde{\mathfrak{U}}\left(\tau;H_{\mathrm{sys}},\left\{ L_{m}\right\} \right)&\equiv \mathbb{T}\exp\left[-i\int_{0}^{\tau}ds\left(H_{\mathrm{eff}}\otimes\mathbb{I}_{\mathrm{anc}}\otimes\mathbb{I}_{\mathrm{fld}}+\sum_{m}iL_{m}\otimes\mathbb{I}_{\mathrm{anc}}\otimes\phi_{m}^{\dagger}(s)\right)\right],\nonumber
\end{align}
where $\mathbb{I}_\mathrm{anc}$ is the identity operator in the ancilla.
In Eq.~\quantumUbound{} in the main text,
we consider the relation:
\begin{equation}
    \frac{1}{2}\int_{t_{1}}^{t_{2}}dt\,\sqrt{\mathcal{J}(t)}\ge\arccos\left[\left|\braket{\Psi(t_{2})|\Psi(t_{1})}\right|\right].
    \label{eq:quantum_bound}
\end{equation}
Using the purified state $\ket{\tilde{\Psi}(t)}$ in Eq.~\eqref{eq:Psi_tilde_def}, from Eq.~\eqref{eq:quantum_bound}, the following relation holds:
\begin{equation}
\frac{1}{2}\int_{t_{1}}^{t_{2}}dt\,\sqrt{\tilde{\mathcal{J}}(t)}\ge\arccos\left[\left|\braket{\tilde{\Psi}(t_{2})|\tilde{\Psi}(t_{1})}\right|\right],
    \label{eq:quantum_bound_puri}
\end{equation}where $\tilde{\mathcal{J}}(t)$ is the quantum Fisher information defined using the purified state:
\begin{equation}
    \tilde{\mathcal{J}}(t)\equiv4\left[\braket{\partial_{t}\tilde{\Psi}(t)|\partial_{t}\tilde{\Psi}(t)}-\left|\braket{\partial_{t}\tilde{\Psi}(t)|\tilde{\Psi}(t)}\right|^{2}\right].
    \label{eq:QFI_purified_def}
\end{equation}
Due to the monotonicity of the quantum fidelity, the right hand side of Eq.~\eqref{eq:quantum_bound_puri} can be evaluated in the same manner
as in the case of the initially pure state [cf. Eq.~\monotonicityUfidelity{} in the main text]. That is 
\begin{equation}
    \mathrm{Fid}(\rho(t_{1}),\rho(t_{2}))\ge\mathrm{Fid}(\ket{\tilde{\Psi}(t_{1})},\ket{\tilde{\Psi}(t_{2})}),
    \label{eq:monotonicity_puri}
\end{equation}
which indicates that Eq.~\LdoUbound{} in the main text should hold 
for the initially mixed state case. 
Similarly, Eq.~\LEUHellingerUineq{} in the main text is satisfied
for the initially mixed state case. 

Next, we consider the evaluation of $\tilde{\mathcal{J}}(t)$,
which appears on the left hand side of Eq.~\eqref{eq:quantum_bound_puri}
and can be computed through 
the fidelity $\braket{\tilde{\Psi(t_{2})}|\tilde{\Psi(t_{1})}}$.
Let us define $\tilde{V}_m$ as follows:
\begin{equation}
    \tilde{V}_{m}(t)\equiv V_{m}(t)\otimes\mathbb{I}_{\mathrm{anc}}\,\,\,\,(0\le m\le M),
    \label{eq:Vm_tilde_def}
\end{equation}
where $V_m(t)$ is defined by Eqs.~\eqref{eq:V0t_def} and \eqref{eq:Vmt_def}.
We thus obtain
\begin{align}
    \braket{\tilde{\Psi}(t_{2})|\tilde{\Psi}(t_{1})}&=\mathrm{Tr}_{\mathrm{sys,anc,fld}}\left[\ket{\tilde{\Psi}(t_{1})}\bra{\tilde{\Psi}(t_{2})}\right]\nonumber\\&=\mathrm{Tr}_{\mathrm{sys,anc}}\left[\sum_{m_{N-1}}\cdots\sum_{m_{0}}\tilde{V}_{m_{N-1}}(t_{1})\cdots\tilde{V}_{m_{0}}(t_{1})\ket{\tilde{\psi}(0)}\bra{\tilde{\psi}(0)}\tilde{V}_{m_{0}}^{\dagger}(t_{2})\cdots\tilde{V}_{m_{N-1}}^{\dagger}(t_{2})\right]\nonumber\\&=\mathrm{Tr}_{\mathrm{sys}}\left[\sum_{m_{N-1}}\cdots\sum_{m_{0}}V_{m_{N-1}}(t_{1})\cdots V_{m_{0}}(t_{1})\rho(0)V_{m_{0}}^{\dagger}(t_{2})\cdots V_{m_{N-1}}^{\dagger}(t_{2})\right].
    \label{eq:fidelity_puri_calc}
\end{align}
For $N\to \infty$, the last line of Eq.~\eqref{eq:fidelity_puri_calc} gives the two-sided Lindblad equation of Eq.~\eqref{eq:two_side_Lindblad_def}
with the initial density $\zeta(s=0;t_1,t_2)=\rho(0)$. 
Then, when we compute the quantum Fisher information using the two-sided Lindblad equation, Eqs.~\LdoUbound{} and \qTUR{} in the main text hold for the initially mixed state case.

\refstepcounter{toinum}
\section*{Supplementary Note \thetoinum: Numerical simulation\label{eq:simulation_detail}}

To better understand the bounds obtained from the present calculations, we consider paradigmatic classical and quantum models. 
For classical dynamics, we consider $N_S$ states Markov process, the dynamics of which is governed by Eq.~\masterUeqUdef{} in the main text. 
When calculating fluctuations of the observable $\mathcal{C}$
in the classical thermodynamic uncertainty relation, each trajectory is generated by the Gillespie algorithm \cite{Allen:2010:SPinBiology}. 
The dynamical activity $\mathcal{A}(t)$ is evaluated by performing the numerical integration to Eq.~\eqref{eq:A_da_def}, where the time-dependent probability distribution $P(\mu,s;\boldsymbol{W})$ is obtained by the matrix exponential. 

For quantum dynamics, 
we employ a two-level atom driven by a classical laser field. 
This dynamics is governed by the Lindblad equation [Eq.~\LindbladUdef{} in the main text], where the Hamiltonian and the jump operator are given by
\begin{align}
    H_{\mathrm{sys}}&=\Delta\ket{\epsilon_{e}}\bra{\epsilon_{e}}+\frac{\Omega}{2}\left(\ket{\epsilon_{e}}\bra{\epsilon_{g}}+\ket{\epsilon_{g}}\bra{\epsilon_{e}}\right),\label{eq:Hsys_two_level_atom}\\
    L&=\sqrt{\kappa}\ket{\epsilon_{g}}\bra{\epsilon_{e}}.
    \label{eq:L_two_level_atom}
\end{align}
Here,
$\ket{\epsilon_e}$ and $\ket{\epsilon_g}$ denote the excited and ground states, respectively, 
$\Delta$ is the extent of detuning between the laser field and the atomic transition frequencies, $\Omega$ is the Rabi-oscillation frequency and  $\kappa$ is the decay rate. 
The jump operator $L$ induces a jump from the excited state $\ket{\epsilon_e}$ to the ground state $\ket{\epsilon_g}$. 
To calculate the fluctuation of the observable $\mathcal{C}$, we generate quantum trajectories. 
Each trajectory obeys the stochastic Schr{\"o}dinger equation:
\begin{equation}
d\rho=-i[H_{\mathrm{sys}},\rho]ds+\rho\mathrm{Tr}_{\mathrm{sys}}\left[L\rho L^{\dagger}\right]ds-\frac{\left\{ L^{\dagger}L,\rho\right\} }{2}ds+\left(\frac{L\rho L^{\dagger}}{\mathrm{Tr}_{\mathrm{sys}}[L\rho L^{\dagger}]}-\rho\right)d\mathfrak{n},
\label{eq:SSE_sim_def}
\end{equation}
where $d\mathfrak{n}$ is a noise increment having a value of $1$ when a jump event (photon) is detected within $ds$ and otherwise has a value of $0$.
The conditional expectation of $d\mathfrak{n}$ is  $\mathrm{Tr}_{\mathrm{sys}}[L\rho(s)L^{\dagger}]ds$, where $\rho(s)$ is a solution of 
Eq.~\eqref{eq:SSE_sim_def}. 

\subsection*{Speed limit relations}

\begin{figure}
\includegraphics[width=16cm]{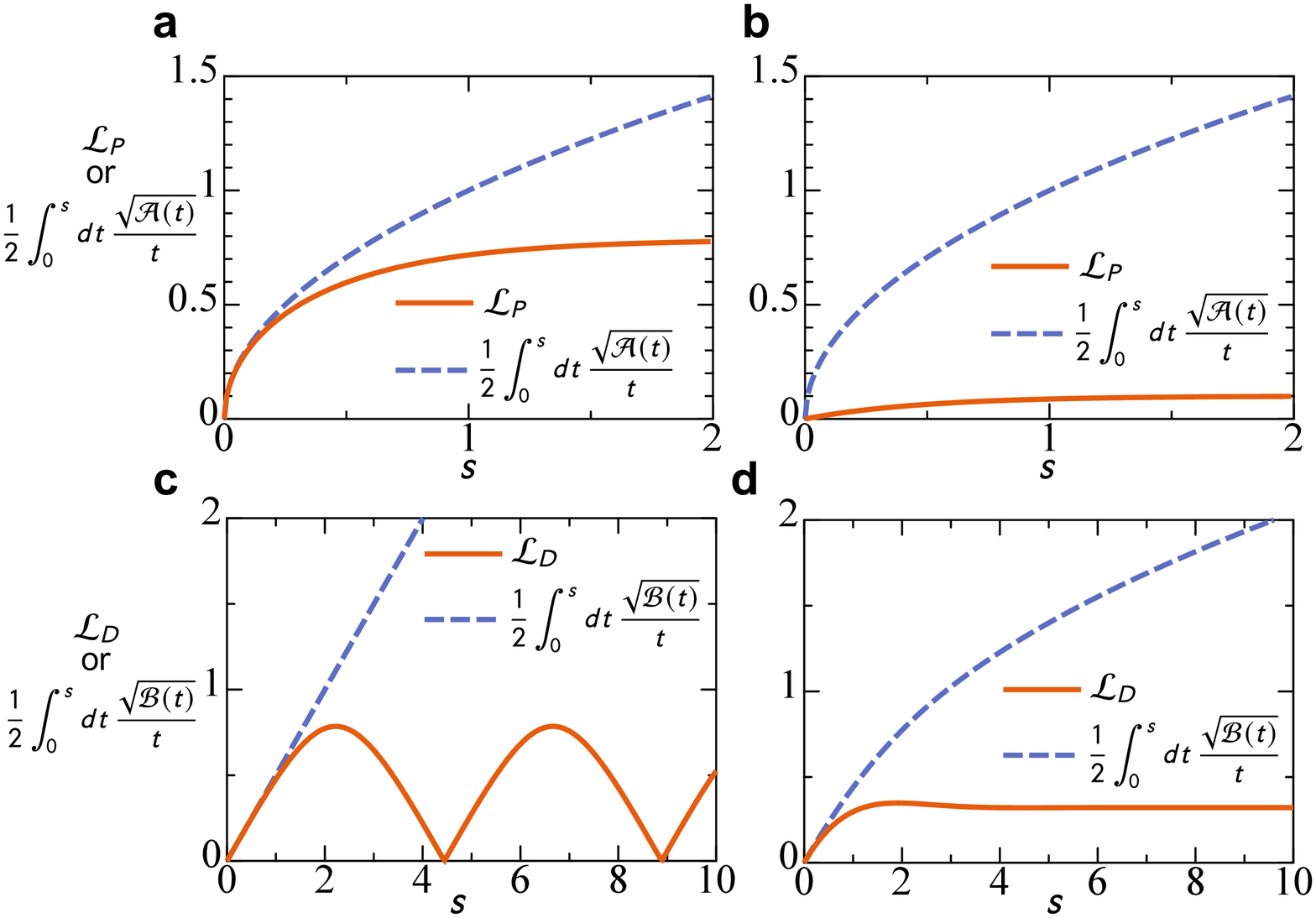} 
\caption{
\textbf{Numerical calculations of the classical and quantum speed limit relations.}
\textbf{a} and \textbf{b} Classical speed limit relation for the Markov process ($N_S=2$). 
$(1/2)\int_{0}^{s}\sqrt{\mathcal{A}(t)}/t\,dt$ and $\mathcal{L}_P(P(\nu,0),P(\nu,s))$ are plotted with the dashed and solid lines, respectively as functions of time $s$ for different initial probabilities: \textbf{a} $[P(\nu,0)]_{\nu=1,2}=[1, 0]$ and \textbf{b} $[P(\nu,0)]_{\nu=1,2}=[0.6, 0.4]$. The transition matrix $W$ is set to $W_{12} = W_{21} = 1$.
\textbf{c} and \textbf{d} Quantum speed limit relation for the two-level atom model. 
$(1/2)\int_{0}^{s}\sqrt{\mathcal{B}(t)}/t\,dt$ and $\mathcal{L}_D(\rho(0),\rho(s))$ are plotted with the dashed and solid lines, respectively, as functions of time $s$ for different $\kappa$ values: \textbf{c} $\kappa = 0$, corresponding to the closed quantum dynamics, and \textbf{d} $\kappa = 2$. The other parameters are $\Delta = 1$ and $\Omega =1$. 
\label{fig:csl_qsl}}
\end{figure}

Here, we first calculate the classical speed limit relation. 
We use a two-state Markov process ($N_S = 2$) and calculate $(1/2)\int_{0}^{s}\sqrt{\mathcal{A}(t)}/t\,dt$ and $\mathcal{L}_{P}(P(\nu,0),P(\nu,s))$ in Eq.~\LpsUbound{} in the main text.
These are shown by the dashed and solid lines, respectively, in Figs.~\ref{fig:csl_qsl}(a) and (b). 
Figures~\ref{fig:csl_qsl}(a) and (b) employ different initial distributions, (a) $[P(\nu,0)]_{\nu=1,2}=[1,0]$ and (b) $[P(\nu,0)]_{\nu=1,2}=[0.6,0.4]$, while the other settings are the same. 
From Fig.~\ref{fig:csl_qsl}(a), 
it is evident that $(1/2)\int_{0}^{s}\sqrt{\mathcal{A}(t)}/t\,dt$ and $\mathcal{L}_P(P(\nu,0),P(\nu,s))$ are almost equivalent for $s < 0.5$.
The classical speed limit concerns two inequalities. 
Since the first inequality is Eq.~\classicalUbound{} in the main text and saturates when the dynamics is the geodesic with respect to the
Fisher information metric, this inequality saturates when $s$ is sufficiently small.
The second inequality is Eq.~\BhattUmonotonic{} in the main text, which concerns the monotonicity with respect to stochastic maps. 
For Fig.~\ref{fig:csl_qsl}(b),
the difference between $(1/2)\int_{0}^{s}\sqrt{\mathcal{A}(t)}/t\,dt$ and $\mathcal{L}_P(P(\nu,0),P(\nu,s))$ is large even at an earlier time.
Therefore, the divergence in Fig.~\ref{fig:csl_qsl}(b) is caused by the second inequality. 
This result indicates that, if the initial distribution is $[P(\nu,0)]_{\nu=1,2}=[0.6,0.4]$, there is greater ambiguity between the trajectory information and the distribution $P(\nu,s)$. 

We now consider the quantum case. 
We use the two-level atom model to calculate $(1/2)\int_{0}^{s}\sqrt{\mathcal{B}(t)}/t\,dt$ and $\mathcal{L}_{D}(\rho(0),\rho(s))$ in Eq.~\LdoUbound{} in the main text,
as indicated by the dashed and solid lines, respectively, in Figs.~\ref{fig:csl_qsl}(c) and (d). 
Here, we consider two cases (c) $\kappa = 0$ and (d) $\kappa = 2$ while the other settings are the same. 
For $\kappa=0$, the jump operator $L$ vanishes and hence the system reduces to a closed quantum dynamics, meaning that Figs.~\ref{fig:csl_qsl}(c) and (d) highlight the differences between closed and open quantum dynamics. 
It is evident that, in both cases, $(1/2)\int_{0}^{s}\sqrt{\mathcal{B}(t)}/t\,dt$ is bounded from below by $\mathcal{L}_{D}(\rho(0),\rho(s))$ and 
the extent of saturation is greater for lower values of $s$, which is similar to the classical case.
Comparing Figs.~\ref{fig:csl_qsl}(c) and (d) shows that $(1/2)\int_{0}^{s}\sqrt{\mathcal{B}(t)}/t\,dt$ and $\mathcal{L}_{D}(\rho(0),\rho(s))$ are closer for (c), suggesting that Eq.~\LdoUbound{} in the main text is tighter for the closed dynamics. 

\subsection*{Thermodynamic uncertainty relations}

\begin{figure}
\includegraphics[width=17cm]{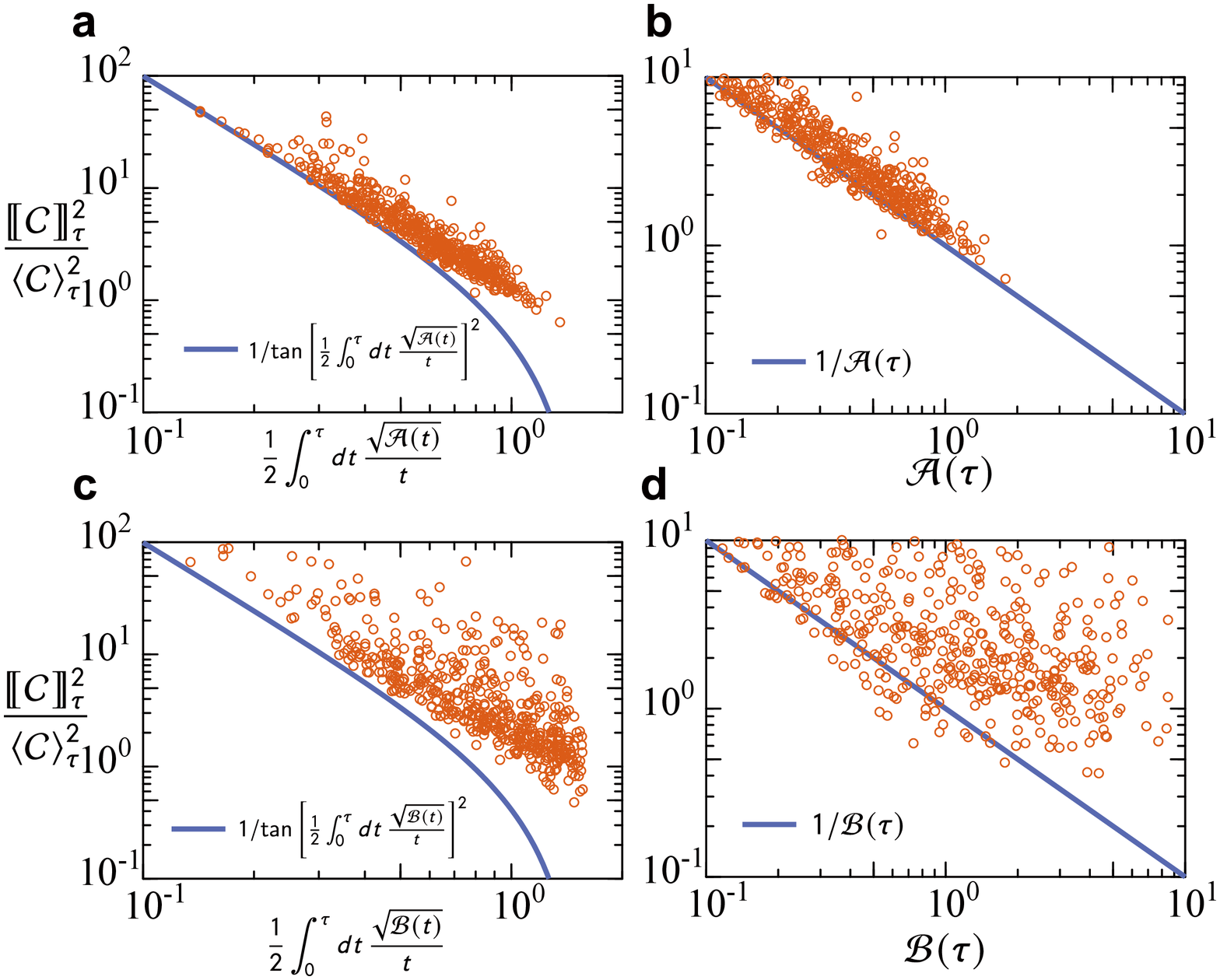} 
\caption{
\textbf{Numerical simulation of the classical and quantum thermodynamic uncertainty relations.}
\textbf{a} and \textbf{b} Precision $\dblbrace{\mathcal{C}}_{\tau}^{2}/\braket{\mathcal{C}}_{\tau}^{2}$ as a function of \textbf{a} $(1/2)\int_{0}^{\tau}\sqrt{\mathcal{A}(t)}/t\;dt$ or \textbf{b} $\mathcal{A}(\tau)$ for random realizations of the classical Markov chain. 
Random realizations are plotted by circles. In \textbf{a}, the solid line indicates the lower bound of Eq.~\cTURII{} in the main text. In \textbf{b}, the solid line is $1/\mathcal{A}(\tau)$, which is the lower bound of the steady-state classical thermodynamic uncertainty relation [Eq.~\convUcTUR{} in the main text]. 
In \textbf{a} and \textbf{b}, the parameter ranges are $\tau \in [0.1, 1]$, $W_{\mu\nu} \in [0,1]$, and $N_S \in \{2,3,\ldots, 5\}$. 
\textbf{c} and \textbf{d} Precision $\dblbrace{\mathcal{C}}_{\tau}^{2}/\braket{\mathcal{C}}_{\tau}^{2}$ as a function of \textbf{c} $(1/2)\int_{0}^{\tau}\sqrt{\mathcal{B}(t)}/t\;dt$ or \textbf{d} $\mathcal{B}(\tau)$ for random realizations of the quantum two-level atom model. 
Random realizations are plotted by circles. In \textbf{c}, the solid line indicates the lower bound of Eq.~\cTURII{} in the main text with $\mathcal{A}(t)$ replaced by $\mathcal{B}(t)$. In 
\textbf{d},
the solid line is $1/\mathcal{B}(\tau)$, which is the lower bound of the steady-state quantum thermodynamic uncertainty relation [Eq.~\convUcTUR{} in the main text with $\mathcal{A}(\tau)$ replaced by $\mathcal{B}(\tau)$]. 
In \textbf{c} and \textbf{d}, the parameter ranges are $\Delta \in [0.1, 3]$, $\Omega \in [0.1, 3]$, $\kappa \in [0.1, 3]$, and $\tau \in [0.1, 1]$. 
\label{fig:ctur_qtur}}
\end{figure}

We next focus on the classical thermodynamic uncertainty relations. 
We use a classical Markov process with $N_S$ states, where $N_S$ is determined at random.
After determining $N_S$, we decide the topology of the Markov process. The other model parameters, including the transition rate $W_{\mu\nu}$, are randomly selected (see the caption of Fig.~\ref{fig:ctur_qtur} for the parameter ranges). Moreover, we randomly generate the initial distribution $P(\nu,0)$. For each selected parameter set, we generate trajectories and calculate $\dblbrace{\mathcal{C}}_{\tau}^{2}/\braket{\mathcal{C}}_{\tau}^{2}$. Since we select the initial distribution randomly, the dynamics represents an out-of-steady state.  In Fig.~\ref{fig:ctur_qtur}(a), we plot $\dblbrace{\mathcal{C}}_{\tau}^{2}/\braket{\mathcal{C}}_{\tau}^{2}$ as a function of $(1/2)\int_{0}^{\tau}\sqrt{\mathcal{A}(t)}/t\,dt$ by circles, where the solid line shows the lower bound of Eq.~\cTURII{} in the main text. 
Because all realizations are above the solid line, it is apparent that Eq.~\cTURII{} in the main text has been numerically verified.
Since the conventional thermodynamic uncertainty relation \cite{Garrahan:2017:TUR,Terlizzi:2019:KUR}, shown by Eq.~\convUcTUR{} in the main text, holds when the system is in the steady state, 
we plot the realizations as a function of $\mathcal{A}(\tau)$ in Fig.~\ref{fig:ctur_qtur}(b), where the solid line now indicates $1/\mathcal{A}(\tau)$ showing the lower bound of Eq.~\convUcTUR{} in the main text. 
Some data points are seen to be below the solid line, confirming that Eq.~\convUcTUR{} in the main text does not hold for the out-of-steady state. 
The points below the lower bound tend to be found in the low dynamical activity region, because the initial state dependence is lost in the high dynamical activity region. 
Although the bound derived in Ref.~\cite{Terlizzi:2019:KUR} [Eq.~\cTURUdCdt{} in the main text] is applicable to any time-independent Markov process, the denominator on the left side of Eq.~\cTURUdCdt{} in the main text represents the time derivative of the average value of the observable rather than the time-integrated observable.

We also perform a similar numerical simulation for the quantum two-level atom dynamics.
Again, we randomly select the model parameters (see the caption of Fig.~\ref{fig:ctur_qtur}(c) for the parameter ranges) and randomly generate the initial density operator $\rho(0)$ to calculate the fluctuation $\dblbrace{\mathcal{C}}_{\tau}^{2}/\braket{\mathcal{C}}_{\tau}^{2}$.  
Figure~\ref{fig:ctur_qtur}(c) shows $\dblbrace{\mathcal{C}}_{\tau}^{2}/\braket{\mathcal{C}}_{\tau}^{2}$ as a function of $(1/2)\int_{0}^{\tau}\sqrt{\mathcal{B}(t)}/t\,dt$ by circles, while the solid line shows $1/\tan\left[\frac{1}{2}\int_{0}^{\tau}\sqrt{\mathcal{B}(t)}/t\,dt\right]^{2}$. 
This confirms that Eq.~\cTURII{} in the main text with $\mathcal{A}(t)$ replaced by $\mathcal{B}(t)$ holds for this out-of-steady state dynamics.
In addition, Fig.~\ref{fig:ctur_qtur}(d) plots $\dblbrace{\mathcal{C}}_{\tau}^{2}/\braket{\mathcal{C}}_{\tau}^{2}$ as a function of $\mathcal{B}(\tau)$,
where the solid line denotes $1/ \mathcal{B}(\tau)$. 
From Fig.~\ref{fig:ctur_qtur}(d), 
we confirm that the steady-state thermodynamic uncertainty relation derived in Ref.~\cite{Hasegawa:2020:QTURPRL},
which is Eq.~\convUcTUR{} in the main text with $\mathcal{A}(\tau)$ replaced with $\mathcal{B}(\tau)$,
does not hold for the out-of-steady state dynamics.

\refstepcounter{toinum}
\section*{Supplementary Note \thetoinum: Time-dependent case\label{sec:time_dependent_case}}

The main paper considers a time-independent Markov process, i.e., $H_\mathrm{sys}$ and $L_m$ are not dependent on time. 
Here, we consider a time-dependent case with the time-dependent operators 
$H_\mathrm{sys}(s)$ and $L_m(s)$.
The time-dependent version of
$\mathfrak{U}\left(\tau;(t/\tau)H_{\mathrm{sys}},\left\{ \sqrt{t/\tau}L_{m}\right\} \right)$ [Eq.~\PsiUUUdef{} in the main text] is given by
\begin{equation}
    \mathbb{T}\exp\left[-i\int_{0}^{\tau}ds\,\left(\frac{t}{\tau}H_{\mathrm{sys}}\left(\frac{t}{\tau}s\right)\otimes\mathbb{I}_{\mathrm{fld}}+\sum_{m}\left(i\sqrt{\frac{t}{\tau}}L_{m}\left(\frac{t}{\tau}s\right)\otimes\phi^{\dagger}(s)-i\sqrt{\frac{t}{\tau}}L_{m}^{\dagger}\left(\frac{t}{\tau}s\right)\otimes\phi(s)\right)\right)\right].
    \label{eq:U_time_def1}
\end{equation}
Again, the density operator $\rho(t)=\rho_{\mathrm{sys}}^{\Psi}(t)=\mathrm{Tr}_{\mathrm{fld}}\left[\ket{\Psi(t)}\bra{\Psi(t)}\right]$ yields
results consistent with the time-dependent Lindblad equation. 
Moreover, the observable that counts the number of jump events yields the same statistics as
the genuine continuous matrix product state $\ket{\Phi(t)}$. 
However, note that the classical Fisher information $\mathcal{I}(t)$
does not calculated into the dynamical activity $\mathcal{A}(t)$ for the time-dependent case.
The dynamical activity quantifies the extent of the dynamics of the Markov process.
When the system is time-independent, the state change is solely induced by
stochastic jumps. However, when the transition rate depends on time,
the transition rate variation also affects the dynamics, which
cannot be included in the definition of the dynamical activity. 
If we define the generalized dynamical activity via $\mathcal{A}_g(t) \equiv t^2 \mathcal{I}(t)$ as in Eq.~\BUactivityUdef{} in the main text, 
all the results presented in the manuscript hold for the time-dependent case. 
As noted in the main text, 
for the quantum case, all the relations hold
for the time-dependent case when using the quantum dynamical
activity $\mathcal{B}(\tau)$.

\refstepcounter{toinum}
\section*{Supplementary Note \thetoinum: Heisenberg uncertainty relation}

Here, we derive the 
quantum
thermodynamic uncertainty relation by considering the Heisenberg uncertainty relation in the bulk space. 
The Heisenberg uncertainty relation \cite{Heisenberg:1927:UR}, generalized by Robertson \cite{Robertson:1929:UncRel}, is given by
\begin{equation}
    \dblbrace{\mathcal{X}}_{t}\dblbrace{\mathcal{Y}}_{t}\ge\frac{1}{2}\left|\braket{[\mathcal{X},\mathcal{Y}]}_{t}\right|,
    \label{eq:Robertson_UR}
\end{equation}where $\mathcal{X}$ and $\mathcal{Y}$ are arbitrary Hermitian operators, and we define the mean and standard deviation as follows:
\begin{align}
    \braket{\mathcal{X}}_{t}&\equiv\braket{\Psi(t)|\mathcal{X}|\Psi(t)},\label{eq:X_braket_def}\\
    \dblbrace{\mathcal{X}}_{t}&\equiv\sqrt{\braket{\mathcal{X}^{2}}_{t}-\braket{\mathcal{X}}_{t}^{2}}.\label{eq:X_brace_def}
\end{align}
Equation~\eqref{eq:Robertson_UR} is a statement about the relation between the
precision of two observables and their incompatibility as quantified by
the commutation relation. 

For notational convenience, we define
\begin{align}
    U(t)\equiv\mathfrak{U}\left(\tau;\frac{t}{\tau}H_{\mathrm{sys}},\left\{ \sqrt{\frac{t}{\tau}}L_{m}\right\} \right).
    \label{eq:Ut_define}
\end{align}
Following Ref.~\cite{Escher:2011:NoisyQFI},
we consider the Hermitian operator:
\begin{equation}
    \mathcal{K}(t)\equiv-i\frac{dU^{\dagger}(t)}{dt}U(t).
    \label{eq:K_def}
\end{equation}
Since $U^\dagger(t)U(t) = \mathbb{I}$, the following relation holds:
\begin{equation}
    \frac{dU^{\dagger}(t)U(t)}{dt}=\frac{dU^{\dagger}(t)}{dt}U(t)+U^{\dagger}(t)\frac{dU(t)}{dt}=0,
    \label{eq:UU_condition}
\end{equation}which guarantees the hermicity of $\mathcal{K}(t)$:
\begin{equation}
    \mathcal{K}^{\dagger}(t)=iU^{\dagger}(t)\frac{dU(t)}{dt}=-i\frac{dU^{\dagger}(t)}{dt}U(t)=\mathcal{K}(t).
    \label{eq:K_dag}
\end{equation}
Let us consider a Heisenberg representation of the observable $\mathcal{X}$ considered in Eq.~\eqref{eq:Robertson_UR}:
\begin{equation}
    \mathcal{X}(t)=U^{\dagger}(t)\mathcal{X}U(t).
    \label{eq:Xt_def}
\end{equation}
We substitute $\mathcal{X} \leftarrow \mathcal{X}(t)$ and
$\mathcal{Y} \leftarrow \mathcal{K}(t)$ into Eq.~\eqref{eq:Robertson_UR} to obtain
\begin{align}
    \dblbrace{\mathcal{X}(t)}_{t=0}\dblbrace{\mathcal{K}(t)}_{t=0}&\ge\frac{1}{2}\left|\braket{[\mathcal{X}(t),\mathcal{K}(t)]}_{t=0}\right|,\nonumber\\
    &=\frac{1}{2}\left|\left\langle \frac{d\mathcal{X}(t)}{dt}\right\rangle _{t=0}\right|,
    \label{eq:Robertson_calc}
\end{align}
where we used the Heisenberg equation for the observable $\mathcal{X}$,  $\partial_{t}\mathcal{X}(t)=i[\mathcal{K}(t),\mathcal{X}(t)]$,
when deriving the last line of Eq.~\eqref{eq:Robertson_calc}. 
Apparently, $\dblbrace{\mathcal{X}(t)}_{t=0}=\dblbrace{\mathcal{X}}_{t}$ in Eq.~\eqref{eq:Robertson_calc}.
We evaluate $\dblbrace{\mathcal{K}(t)}_{t=0}$ in Eq.~\eqref{eq:Robertson_calc} as follows:
\begin{equation}
\dblbrace{\mathcal{K}(t)}_{t=0}^{2}=\braket{\mathcal{K}(t)^{2}}_{t=0}-\braket{\mathcal{K}(t)}_{t=0}^{2},
\label{eq:K_var_eval}
\end{equation}
where
\begin{align}
    \braket{\mathcal{K}(t)^{2}}_{t=0}&=\braket{\mathcal{K}(t)\mathcal{K}^{\dagger}(t)}_{t=0}\nonumber\\
    &=\braket{\Psi(0)|\frac{dU^{\dagger}(t)}{dt}\frac{dU(t)}{dt}|\Psi(0)}\nonumber\\
    &=\braket{\partial_{t}\Psi(t)|\partial_{t}\Psi(t)},
    \label{eq:K_var_calc}
\end{align}and
\begin{align}
    \braket{\mathcal{K}(t)}_{t=0}&=i\braket{\Psi(0)|U^{\dagger}(t)\frac{dU(t)}{dt}|\Psi(0)}\nonumber\\&=i\braket{\Psi(t)|\partial_{t}\Psi(t)}.
    \label{eq:K_mean_calc}
\end{align}
Equations~\eqref{eq:K_var_eval}--\eqref{eq:K_mean_calc} show that
\begin{equation}
    \mathcal{J}(t)=4\dblbrace{\mathcal{K}(t)}_{t=0}^{2},
    \label{eq:QFI_var_def}
\end{equation}
where the quantum Fisher information $\mathcal{J}(t)$
is defined by Eq.~\QFIUdef{} in the main text.
Considering the observable $\mathbb{I}_\mathrm{sys}\otimes\mathcal{C}^{\bullet}$
for $\mathcal{X}$ and using Eq.~\BUactivityUdef{} in the main text to convert from $\mathcal{J}(t)$ to $\mathcal{B}(t)$,
we derive
\begin{equation}
    \frac{\dblbrace{\mathcal{C}^{\bullet}}_{\tau}^{2}}{\tau^{2}\left(\partial_{\tau}\braket{\mathcal{C}^{\bullet}}_{\tau}\right)^{2}}\ge\frac{1}{\mathcal{B}(\tau)},
    \label{eq:qTUR_dCdt}
\end{equation}
which is the thermodynamic uncertainty relation given in Eq.~\cTURUdCdt{} in the main text with
$\mathcal{A}(\tau)$ replaced by $\mathcal{B}(\tau)$.

\refstepcounter{toinum}
\section*{Supplementary Note \thetoinum: Function and variable definition}

In this section, we provide the variable and function definitions used in Fig.~\FIGrelation{} in the main text.
The equation numbers in the following list correspond to those in the main text.

\begin{description}
\item [{$\mathcal{I}(t)$}] Fisher information [Eq.~\CFIUdef{}]
\item [{$\mathcal{J}(t)$}] Quantum Fisher information [Eq.~\QFIUdef{}]
\item [{$\Gamma$}] Trajectory [Eq.~\cMPSUtrajUdef{}]
\item [{$\mathcal{P}(\Gamma,\nu,t)$}] Probability of observing trajectory $\Gamma$ and being $Y_\nu$ at time $t$ [Eq.~\pathUprobUdef{}]
\item [{$\ket{\Psi(t)}$}] Scaled continuous matrix product state at time
$t$ [Eq.~\PsiUUUdef{}]
\item [{$P(\nu,t)$}] Probability of $\nu$ at time $t$ [Eq.~\masterUeqUdef{}]
\item [{$\rho(t)$}] Density matrix at time $t$ [Eq.~\LindbladUdef{}]
\item [{$\mathcal{A}(t)$}] Dynamical activity [Eq.~\activityUdef{}]
\item [{$\mathcal{B}(t)$}] Quantum dynamical activity [Eq.~\BUactivityUdef{}]
\item [{$\mathcal{L}_{P}$}] Bhattacharya angle [Eq.~\LPUdef{}]
\item [{$\mathcal{L}_{D}$}] Bures angle [Eq.~\LDUdef{}]
\item [{$\mathcal{C}$}] Weighted sum of the number operator $\mathcal{N}_m$ [Eqs.~\NmUdecompUdef{} and \countingUobsUdef{}]
\item [{$\mathcal{C}^{\circ}$}] Weighted sum of the number operator $\mathcal{N}_m^\circ$ [Eqs.~\NmIUdecompUdef{} and \CbulletUCcircUdef{}]
\item [{$\mathcal{C}^{\bullet}$}] Weighted sum of the number operator $\mathcal{N}_m^\bullet$ [Eqs.~\NmIIUdecompUdef{} and \CbulletUCcircUdef{}]
\end{description}